\documentclass[[aps,prb,showpacs,preprintnumbers,amsmath,amssymb]{revtex4}
\usepackage{natbib} 
\usepackage{graphicx}
\usepackage{epstopdf}
\usepackage{xcolor}
\usepackage{amsmath}

\usepackage[utf8x]{inputenc}

\begin{document}

\title{Is biosensing revolution approaching? \\Review on Biocompatible ODMR Techniques}

\author{G. Petrini$^{1,2}$, E. Moreva$^2$, E. Bernardi$^2$, P. Traina$^2$, G. Tomagra$^3$, V. Carabelli$^3$,I. P. Degiovanni$^2$, M. Genovese$^{2,4}$.\\
\textit{$^1$ Physics Department, University of Torino, Torino, Italy\\
$^2$ Istituto Nazionale di Ricerca Metrologica (INRIM), Strada delle cacce 91, Torino, Italy\\
$^3$Department of Drug and Science Technology and NIS Inter-departmental Centre Torino, University of Torino, Italy\\
$^4$Istituto Nazionale di Fisica Nucleare (INFN) Sez. Torino, Torino, Italy }}

\begin{abstract}
Understanding the human brain is one of the most significant challenges of the 21st century. As theoretical studies continue to improve the description of the complex mechanisms that regulate biological processes, in parallel numerous experiments are conducted to enrich or verify these theoretical predictions also with the aim of extrapolating more accurate models. 
In the field of magnetometry, among the various sensors proposed for biological application, NV centers are emerging as a promising solution due to their perfect biocompatibility and the possibility of being positioned in close proximity to the cell membrane, thus allowing a nanometric spatial resolution down to the nano-scale.
Still many issues must be overcome to obtain both spatial resolution and sensitivity capable of revealing the very weak electromagnetic fields generated by neurons, or other excitable cells, during their firing activity. 
However, over the last few years, significant improvements have been achieved in this direction, thanks to the use of innovative techniques. 
In this review, the new results regarding the application of NV centers will be analyzed and the main challenges that must be afforded for leading to practical applications will be discussed.
\end{abstract}
\newcommand{\ket}[1]{\mbox{\ensuremath{|#1\rangle}}}
\newcommand{\bra}[1]{\mbox{\ensuremath{\langle#1|}}}
\flushbottom
\maketitle
\thispagestyle{empty}

\section{Introduction}
Electromagnetic field sensing is of the utmost importance for several applications in current scientific research, fostering the search for novel high sensitivity sensors. Several innovative electromagnetic field sensors emerged in the last years \cite{mathias1990limitations,fejtl2006micro,spira2013multi,vengalattore2007high,faley2006new,baudenbacher2003monolithic,knappe2010cross,nakajima2016application,maestro2010cdse,brites2016lanthanides}, whose main goal is revealing less and less intense fields with an increased spatial resolution.
In particular high sensitivity sensing coupled to high resolution is of the utmost relevance in biological research, especially, for instance, in studies of human brain cell currents, which are typically extremely faint. 
Localized monitoring of neuronal fields \cite{hines1993neuron,hines1997neuron,santamaria2009hodgkin,arcas2003computation} would allow not only the investigation of brain currents during cognitive processes in order to improve neurological diagnostic systems, but also identifying the early stages of neurodegenerative disease, like Parkinson’s, Alzheimers's disease and other forms of dementia \cite{recording1995edited,waxman1998demyelinating}.\\
Furthermore, since localized temperature gradients and heat dissipation occur within cellular microdomains, the exploitation of sensing probes for multi-task applications would be very fruitful \cite{kiyatkin2018brain,maurer2013nanometer}.
Among the various sensing devices that have emerged over the years \cite{bai2016micro,wu2016diamond}, promising sensors for the detection of biological fields are color centers \cite{beveratos2001nonclassical,beveratos2002room} in diamond \cite{zaitsev2013optical,bernardi2020biocompatible,yu2005bright,barone2019long,specht2004ordered,schirhagl2014nitrogen,kucsko2013nanometre,jeske2016laser}. Color centers are impurities in the crystalline matrix that, when stimulated, emit fluorescence. In particular, the nitrogen-vacancy (NV) complex \cite{doherty2013nitrogen} is by far the most promising due to its level structure\cite{wolf2017diamond}. 
This dependence allows the realization of techniques for optical initialization and spin readout by means of the Optically Detected Magnetic Resonance (ODMR) technique \cite{gruber1997scanning}.
Furthermore, its spin energy levels \cite{doherty2012theory} are sensitive not only to electromagnetic fields \cite{balasubramanian2008nanoscale,pham2011magnetic,schloss2018simultaneous,taylor2008high,degen2008scanning,maze2008nanoscale,dolde2011electric}, but also to temperature variations \cite{acosta2010temperature}. 
These exceptional properties, together with their photostability at room-temperature and the non-toxicity of diamond \cite{guarina2018nanodiamonds}, promotes the NV complex as a very promising candidate for biological application \cite{le2013optical,davis2018mapping,barry2020sensitivity,simpson2017non,ermakova2017thermogenetic,fujiwara2020realtime}.\\\\
The paper is structured as follows: section \textbf{2} reports a brief description of the theory of quantum sensing, section \textbf{3} analyzes the magnetic field generated by mammalian neuronal cells and cardiac tissue, section \textbf{4} deals with experiments aimed at the detection of cells fields and finally in section \textbf{5} the experimental techniques, used to enhance the sensitivity of NV centers to be used as biosensors, are highlighted.
\section{The theory of quantum sensing with NV$^{-}$ centers}
\begin{figure}[h]\begin{center}
		a)\includegraphics[scale=0.24]{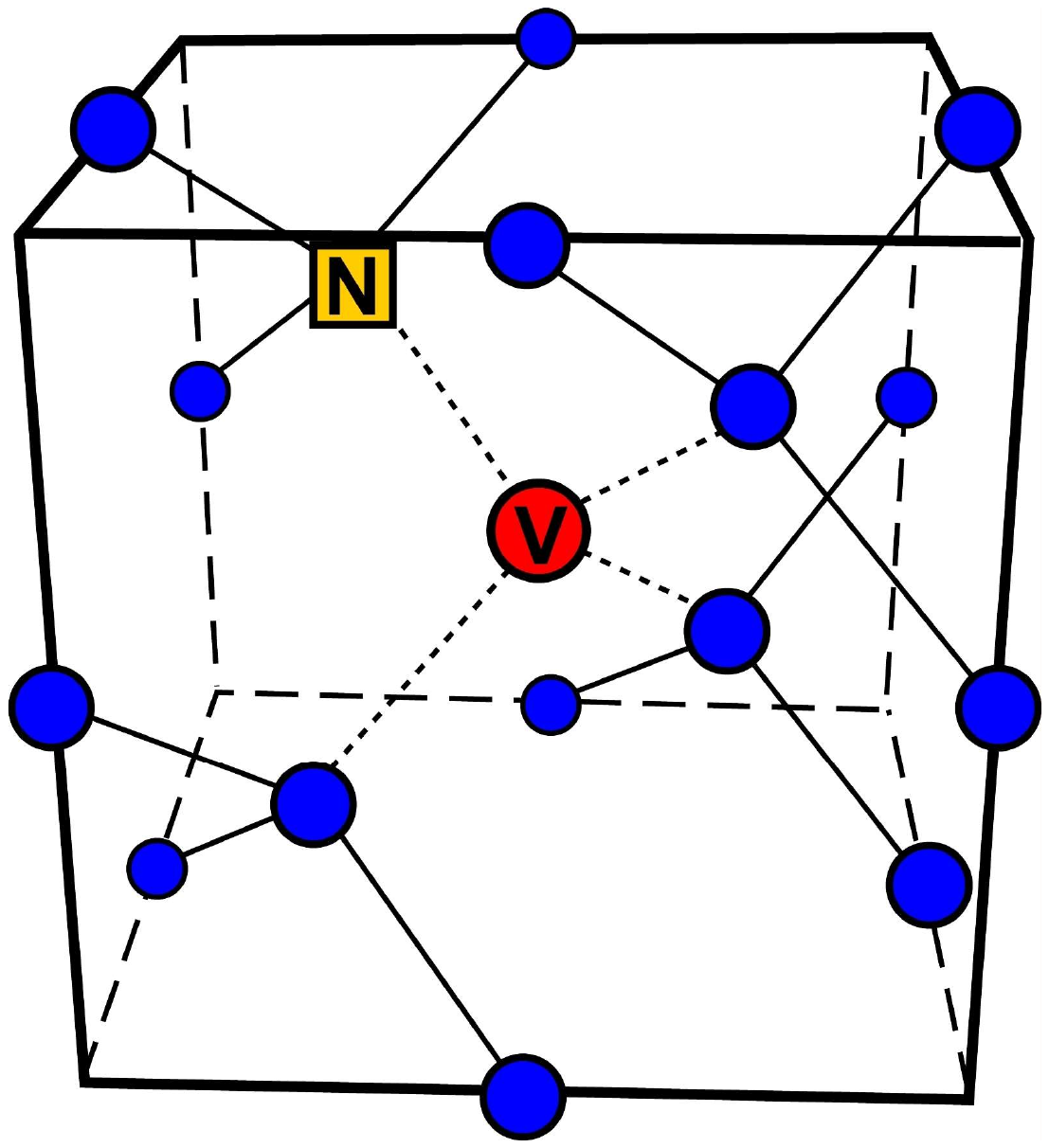} \quad b)\includegraphics[scale=0.2]{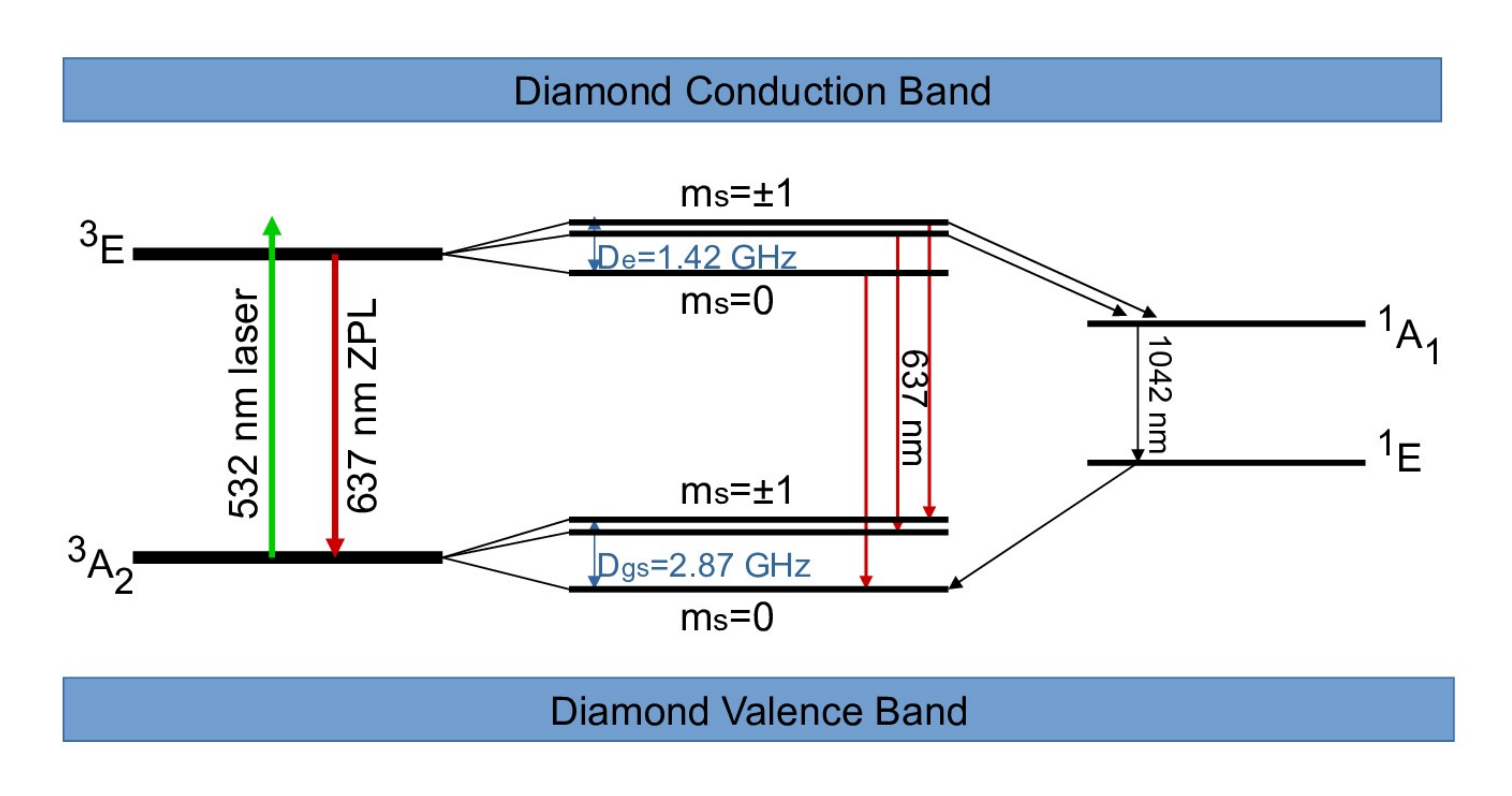}
	\end{center}
	\caption{a) Diamond crystalline structure with nitrogen-vacancy defect;
		b) NV$^-$ radiative state transitions that occur during laser pumping. Radiative optical transition $^3$E $\rightarrow$ $^3$A$_2$ with 637 nm zero phonon line (ZPL), and non optical transition $^1$E $\rightarrow$ $^1$A$_1$ with 1042 nm ZPL. Non radiative intersystem crossing (ISC) transitions subsist between $^3$E $\rightarrow$ $^1$A$_1$ and between $^1$E $\rightarrow$ $^3$A$_2$.}
	\label{fig:NV_levels}
\end{figure}

The nitrogen-vacancy (NV) defect is a natural complex of impurities in diamond crystalline matrix. This complex is composed of a substitution nitrogen atom and a vacancy-type defect, located in adjacent reticular sites \cite{doherty2013nitrogen}. This system has a pyramidal symmetry (C$_{3v}$) and it has, as axis of symmetry, the line that connects the nitrogen atom with the vacancy (see \textbf{Figure \ref{fig:NV_levels}a}). 
With respect to the tetrahedral structure of the diamond, there are 4 possible orientations of this defect, all equiprobable in conditions of conventional syntheses. Moreover, there are two charged states in which it is possible to find the nitrogen-vacancy defects, distinguished by the number of electrons involved. The 3 carbon atoms surrounding the vacancy contribute to sharing 1 electron each to the complex, while nitrogen contributes with 2. If, in total, only these 5 electrons are present in the system, the center is electrically neutral and it is referred to as NV$^0$, with total electronic spin $S$ = 1/2. 
Alternatively, the defect can trap 1 additional electron from the surrounding lattice, creating the NV$^-$ center. 
In this case the total electronic spin becomes $S$ = 1, with spin component along symmetry axis $\{|m_s =0>,|m_s=+1>,|m_s=−1>\}$. 
The most promising configuration for quantum sensing exploits the spin property of the NV$^-$ complex. 
Its sp$^3$ orbitals linearly combine to form 4 molecular orbitals: the lowest energy state of the ground configuration that is the orbital singlet, spin triplet state $^3$A$_2$, the electronic excited states that are orbital doublet, spin triplet $^3$E, spin singlet orbital singlet $^1$E and $^1$A$_1$.

As we will discuss, due to its electronic levels the NV$^-$ complex is the most promising configuration for quantum sensing.
By irradiating the complex with a 532 nm pump laser (\textbf{Figure \ref{fig:NV_levels}b}), the electronic state is excited in a non-resonant way and afterwards it relaxes to the fundamental state emitting a photon with wave lenght between 637 nm (zero phonon line) and 800 nm (phonon sideband).
While the optical excitation from the $|m_s= 0>$ state is spin preserving, the transition from $|m_s=\pm 1>$ has a finite branching ratio into the metastable singlet $^1$E, with a lifetime of 300 ns. This singlet state relaxes into $|m_s= 0>$ through a non-radiative processes and weak infrared emission peaking at 1042 nm, leading to a drop in fluorescence output up to 30$\%$ for a single NV$^-$, or 1-2$\%$ for a large NV$^-$ ensemble, with respect to the situation when the system is initialized in  $|m_s= 0>$. This allows the optical readout of the spin state.\\
Since the main focus of the present work is on sensing, from now on we will refer to NV$^-$ as NV for simplicity.
\subsection{NV ground electronic state}
The Hamiltonian of  $^3$A$_2$, the ground spin state of the NV system, can be written in the following form \cite{loubser1978electron,doherty2012theory}:
\begin{equation}
\frac{\hat{H}_{gs}}{h}= \hat{S}\textbf{D}\hat{S}+ \hat{S}\textbf{A}\hat{I}+ \hat{I}\textbf{Q}\hat{I}
\end{equation}
where $\hat{S}=(\hat{S_x},\hat{S_y},\hat{S_z})$ and $\hat{I}=(\hat{I_x},\hat{I_y},\hat{I_z})$ are the dimensionless electron and nitrogen nuclear spin operators, respectively. The first term represents the fine structure splitting due to the electronic spin-spin interaction, coupled by the fine structure tensor \textbf{D}. The second term is generated by the hyperfine interaction between NV electrons and the nitrogen nucleus ($I$ = 1 for a $^{14}$N nucleus, while $I$ = 1/2 for a $^{15}$N nucleus), with the hyperfine tensor \textbf{A}. Finally, the third term represents the nuclear electric quadrupole interaction, with the electric quadrupole tensor \textbf{Q}. It should be noted that, in this notation, the component \textit{z} coincides with the NV axis of symmetry. Due to the symmetry of the NV center, \textbf{D}, \textbf{A}, and \textbf{Q} are diagonal in the NV coordinate system \cite{bauch2018ultralong,shin2014optically} and, in terms of the natural spin-triplet basis $\{|m_s =0>,|m_s=+1>,|m_s=−1>\}$, the Hamiltonian can be written as:
\begin{equation}
\frac{\hat{H}_{gs}}{h}= \underbrace{ D_{gs} [\hat{S}_{z}^2-\hat{S}^2/3]}_{electronic\ spin-spin\ interaction} + \underbrace{A^{//}_{gs} \hat{S}_z \hat{I}_z+ A^{\perp}_{gs}[ \hat{S}_x \hat{I}_x+ \hat{S}_y \hat{I}_y ]}_{electron-nucleus\ spin\ interaction}+\underbrace{ Q_{gs} [\hat{I}_{z}^2-\hat{I}^2/3] }_{nuclear\ spin-spin\ interaction}
\end{equation}
where $D_{gs}$$\simeq$ 2.87 GHz is the zero field splitting, $Q_{gs}$ is the nuclear electric quadrupole parameter, $A^{//}_{gs}$ and $A^{\perp}_{gs}$ are the axial and non-axial magnetic hyperfine parameters \cite{felton2009hyperfine,he1993paramagnetic}. The parameters values are reported in \textbf{Table \ref{tab:Coupling_coefficinet}}.
\begin{table}[htbp]
	\begin{center}
		\begin{tabular}{ll}
		\hline 
		Hyperfine parameters          & Value     \\
		\hline 
		Zero field splitting    & $D_{gs}$ $\simeq$ 2.87 GHz                   \\
		
		 Axial hyperfine term   & $A^{//}_{gs,^{14}N}$  $\simeq$ -2.14 MHz   \\ &    $A^{//}_{gs,^{15}N}$  $\simeq$ 3.03 MHz            \\
		
		 Transverse hyperfine term & $A^{\perp}_{gs,^{14}N}$  $\simeq$ -2.70 MHz    \\&    $A^{\perp}_{gs,^{15}N}$  $\simeq$ 3.65 MHz          \\
		
		Nuclear electric quadrupole term & $Q_{gs}$ $\simeq$ -5 MHz                        \\
		\hline 
	\end{tabular} 
	\caption{Hyperfine parameters for the NV defect determined at room temperature.}
	\label{tab:Coupling_coefficinet}
\end{center}
\end{table} 
\subsection{The Optically Detected Magnetic Resonance technique}
\begin{figure}[h]\begin{center}
		a)\includegraphics[scale=0.1]{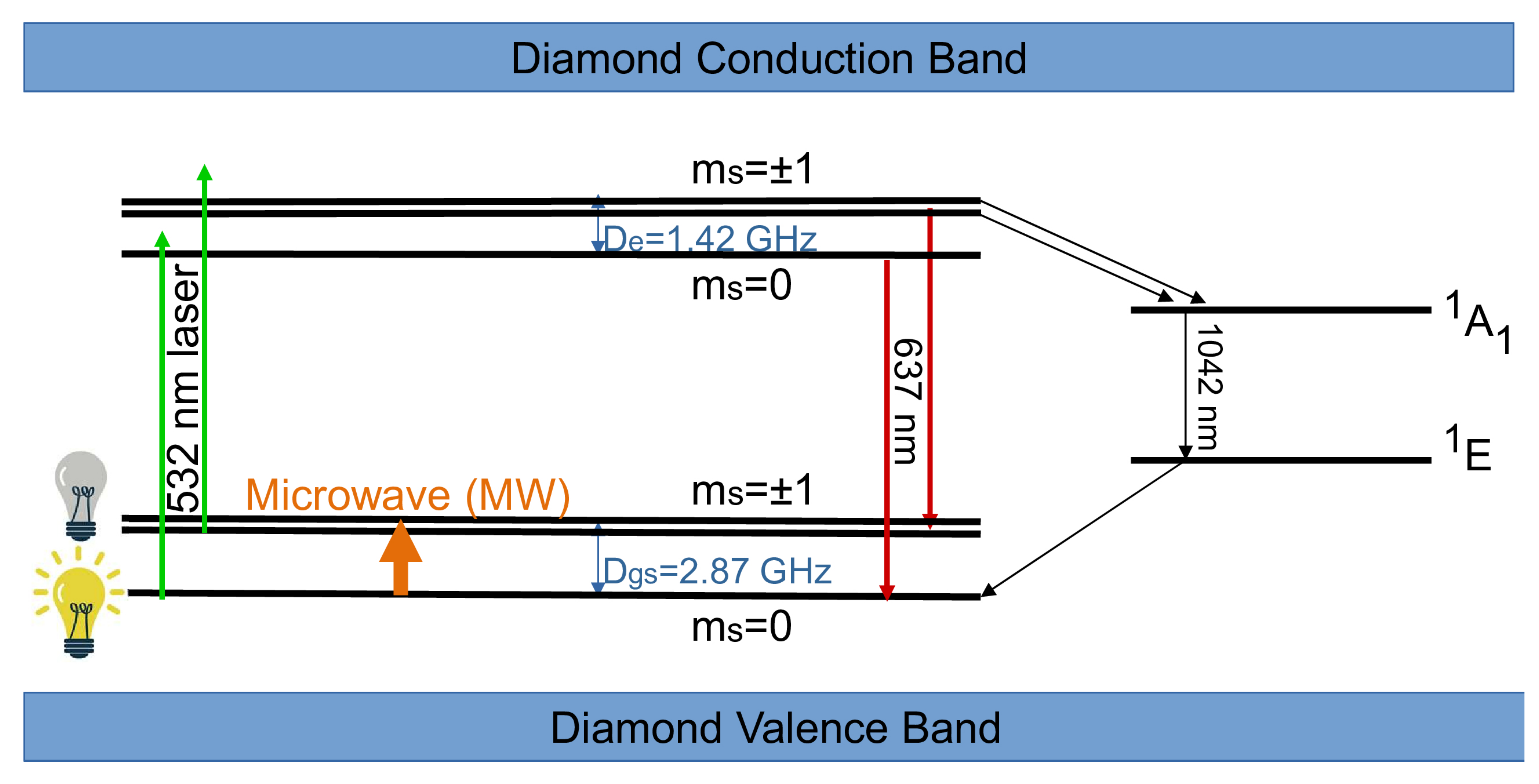} \quad b)\includegraphics[scale=0.2]{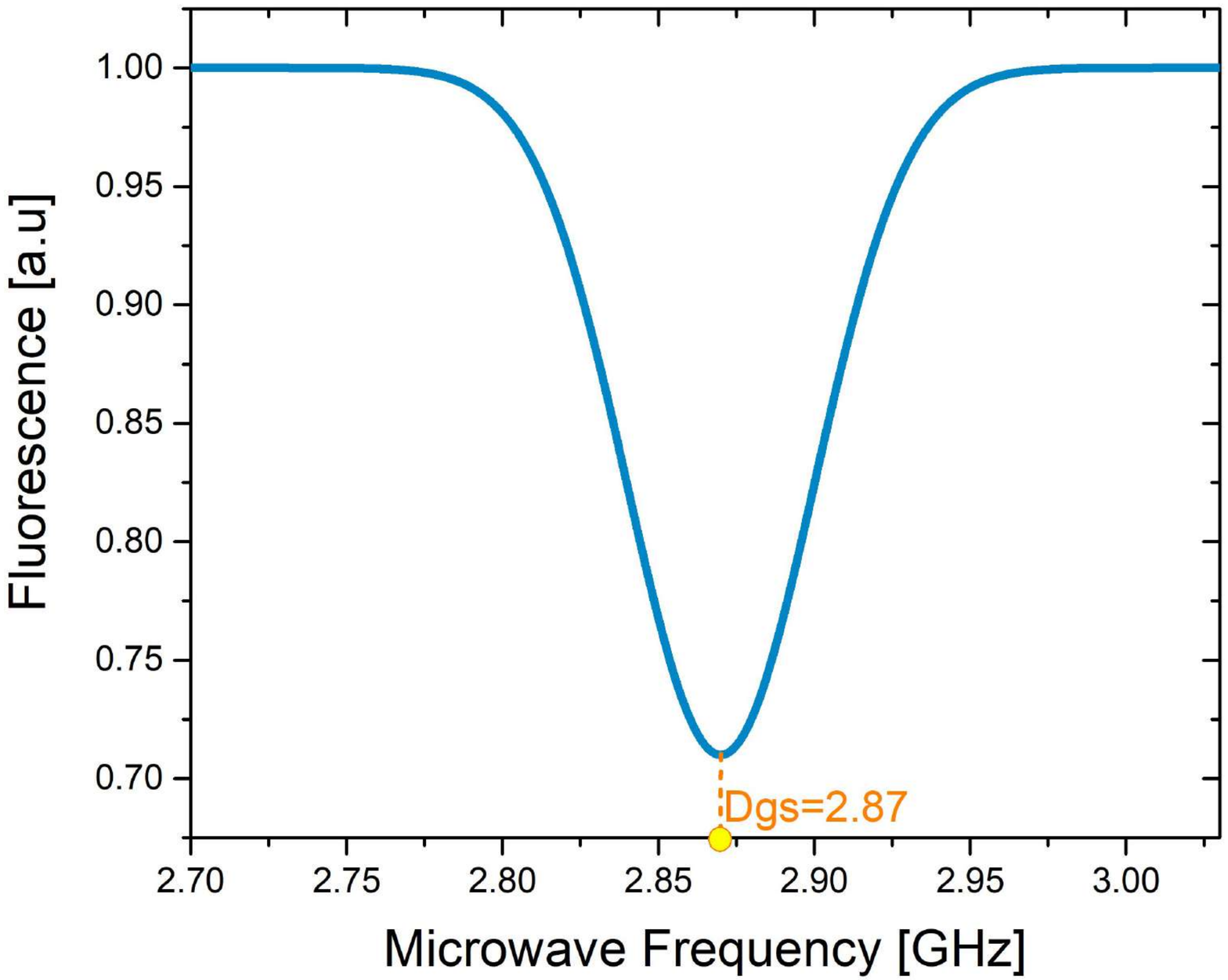}
	\end{center}
	\caption{a) NV radiative state transitions that occur during laser pumping and MW excitation. The coupling of the state $|m_s=\pm1>$ with the metastable level generates a statistically lower fluorescent emission than the $|m_s=0>$ initialized electronic state.
		b) Fluorescence collected from the NV center as a function of the MW frequency varies. A dip in correspondence of the zero field splitting D$_{gs}$ (resonance frequency of the undisturbed NV center, at room temperature) can be observed.}
	\label{fig:NVlevels_ODMR}
\end{figure}

One of the characteristics that makes NV centers so attractive and convenient as key element in various type of sensors is the possibility to discriminate the spin components of the electronic state. This is allowed by the different coupling of the $|m_s =0>$ state with a metastable level, compared to the $|m_s= \pm 1>$ states and results in a variation of the photoluminescence (PL) of the defect under laser non resonant excitation. Optically Detected Magnetic Resonance (ODMR) \cite{gruber1997scanning} consists in the application of a microwave field (MW) on the sample, simultaneously with its exposure to a non-resonant laser at a frequency higher than the resonant one, corresponding to the energy gap between the ground and the $^3$E level (e.g. 532 nm) (see \textbf{Figure \ref{fig:NVlevels_ODMR}a}). 
When the frequency of the MW reaches the ground state resonance $D_{gs}$ of the NVs, with a certain probability (depending on the MW power), those NV centers will be initialized in the states $|m_s=\pm1>$ rather than $|m_s=0>$. As mentioned, this corresponds to a reduction in photoluminescence of the NV centers, as it can be observed, e.g., in \textbf{Figure \ref{fig:NVlevels_ODMR}b}, where a typical ODMR spectrum is reported with the expected fluorescence dip at the zero field splitting frequency $D_{gs}$.

The coupling terms of a NV center with the electric, magnetic fields and local temperature variations will be analyzed in following.

\subsection{Magnetic field sensing}
\begin{figure}[h]\begin{center}
		a)\includegraphics[scale=0.105]{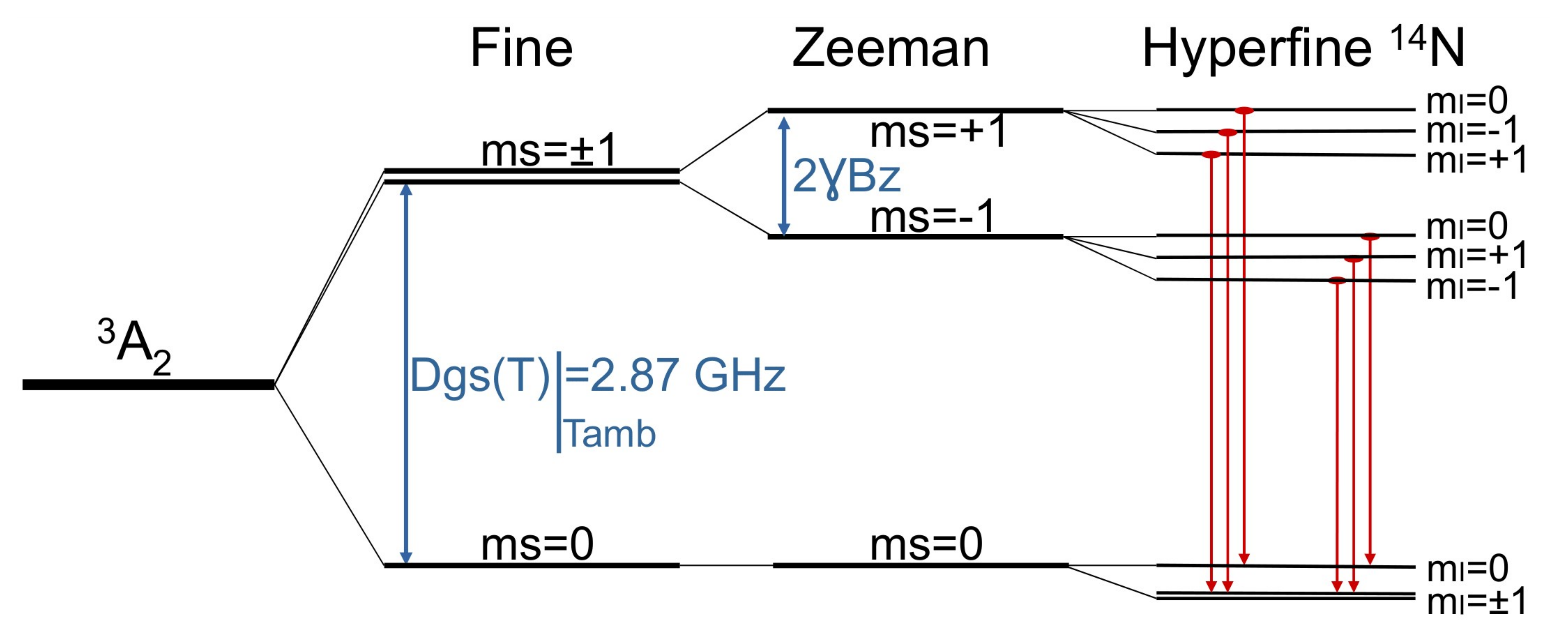} \quad b)\includegraphics[scale=0.105]{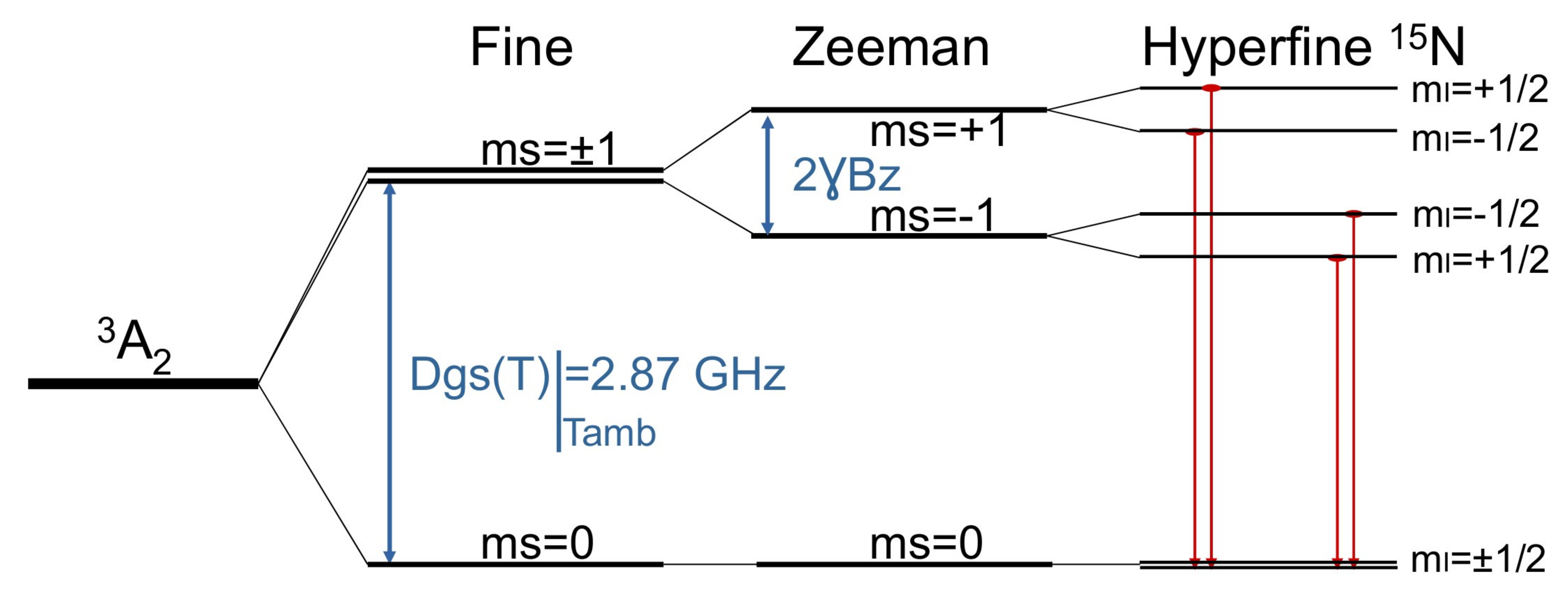} 
		
		\includegraphics[scale=0.2]{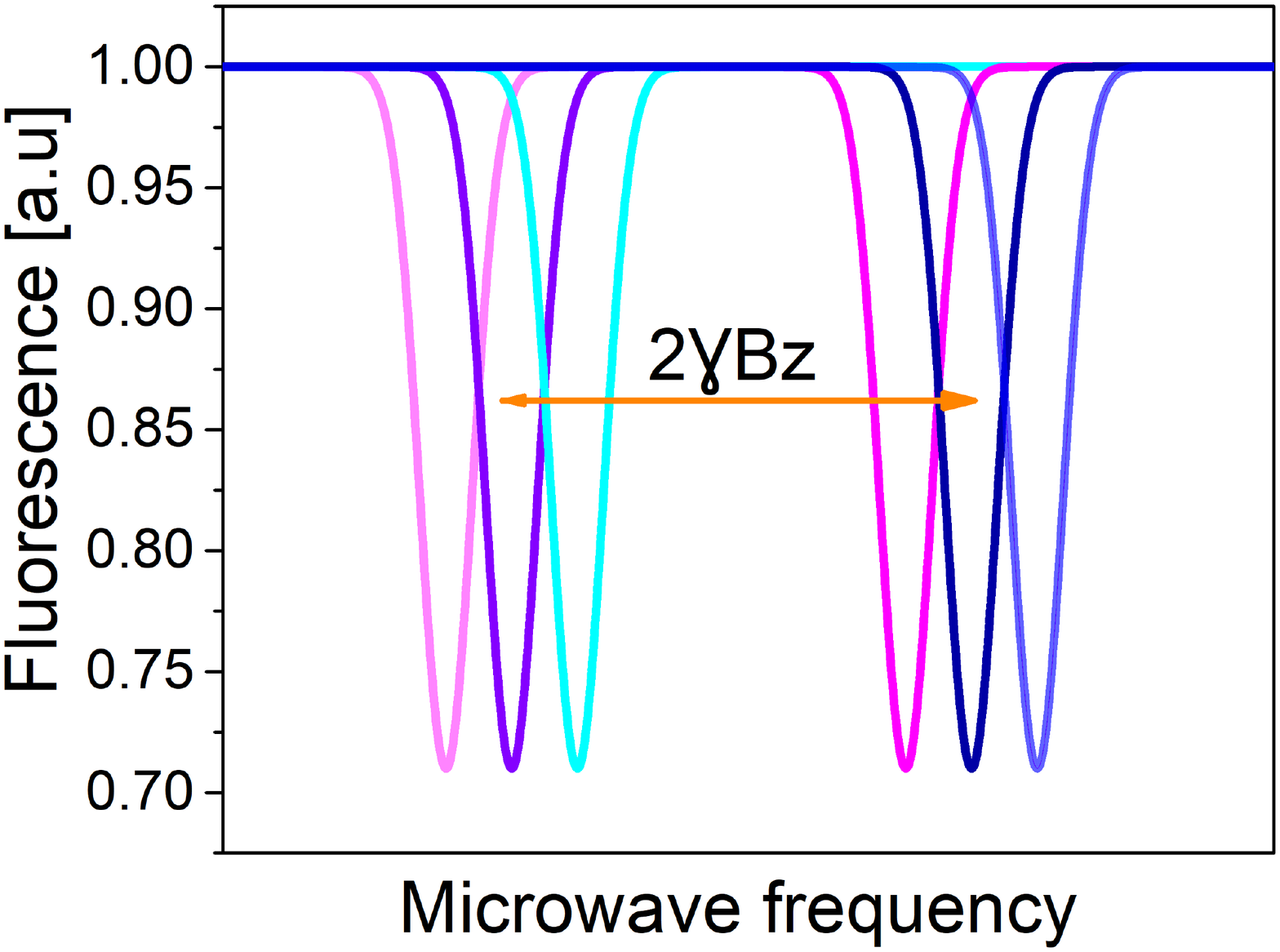} \quad             \includegraphics[scale=0.2]{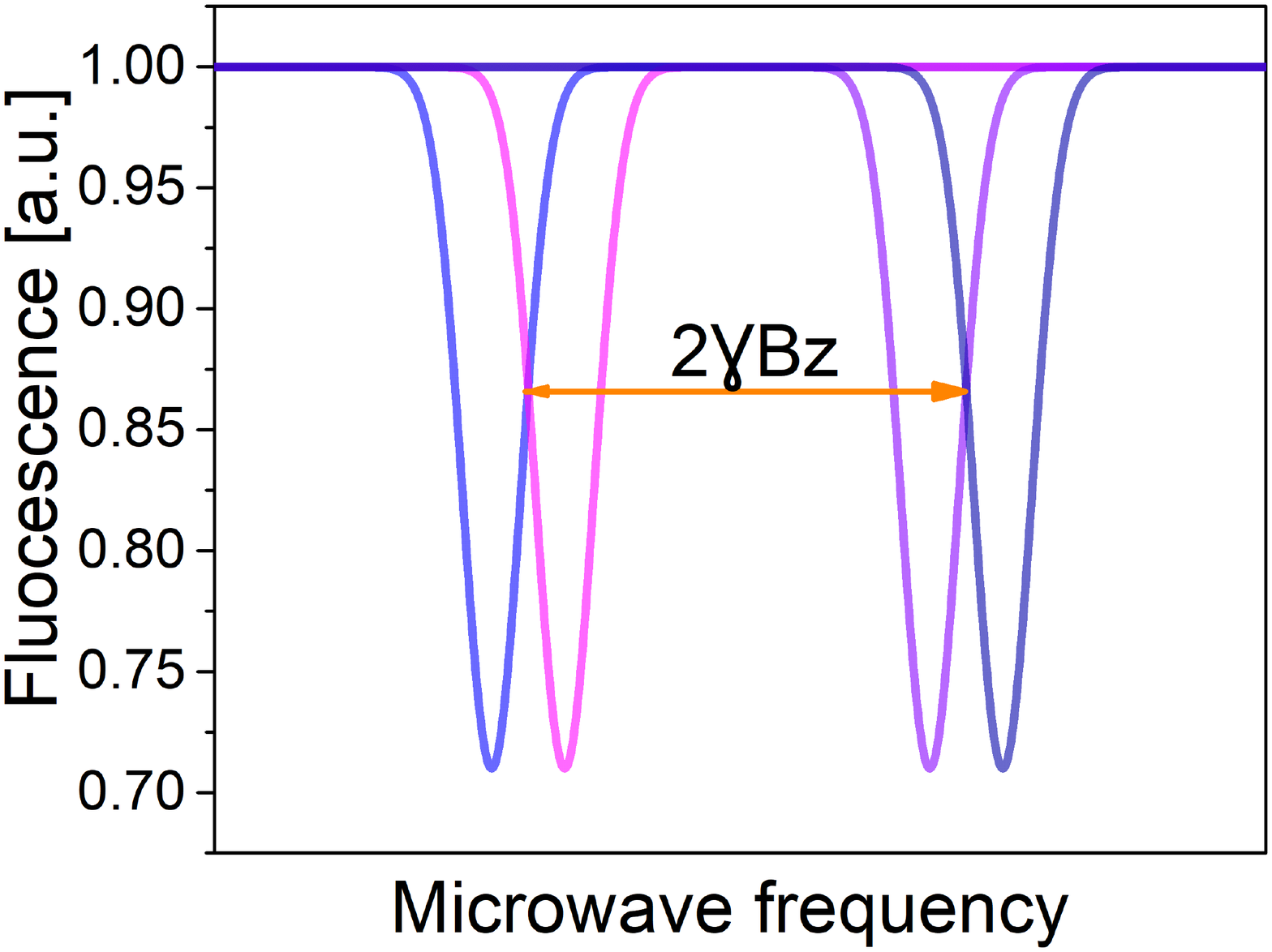} 
		\caption{NV ground-state $^3$A$_2$ scheme. Above: a) $^{14}$N hyperfine states and b)$^{15}$N hyperfine states. Below: schematic ODMR spectra. The spectra are shown considering Zeeman splitting and hyperfine splitting.}
		\label{fig:14N_15N}
	\end{center}
\end{figure}

A static magnetic field produces the well-known Zeeman effect \cite{balasubramanian2008nanoscale}, that it is described by:
\begin{equation}
\frac{\hat{V}_{gs}}{h}= \underbrace{\frac{\mu_B g^{//}_{gs}}{h} \hat{S}_z B_z + \frac{\mu_B g^{\perp}_{gs}}{h} (\hat{S}_x B_x+ \hat{S}_y B_y)}_{Zeeman\ interaction} +\underbrace{ \frac{\mu_N g_{N}}{h} \hat{I}\vec{\textbf{\textit{B}}}}_{nuclear\ Zeeman\ interaction}
\end{equation} 
where $\mu_B$ is the Bohr magneton, $\mu_N$ is the nuclear magneton, $g^{//}_{gs}$ and $g^{\perp}_{gs}$ are the components of the ground state electronic g-factor tensor and $g_N$ is the isotropic nuclear g-factor. In the presence of relatively weak magnetic fields, it is possible to approximate the almost diagonal g-factor tensor in a diagonal form, with constant $g_e$ = 2.003 \cite{doherty2013nitrogen}.
As reported in \textbf{Table \ref{tab:Coupling_coefficinet2}}, the interaction of the magnetic field with the nucleus is 2000 times smaller and, consequently, it is typically neglected \cite{pham2011magnetic}.
The presence of external fields eliminates the energy degeneracy of the levels $|m_s= \pm 1>$, whose splitting become $\gamma_e B_z$, where  $\gamma_e = \frac{\mu_B g_e}{h}$ (see \textbf{Figure \ref{fig:14N_15N}}).

\begin{figure}[h]\begin{center}
		\includegraphics[scale=0.36]{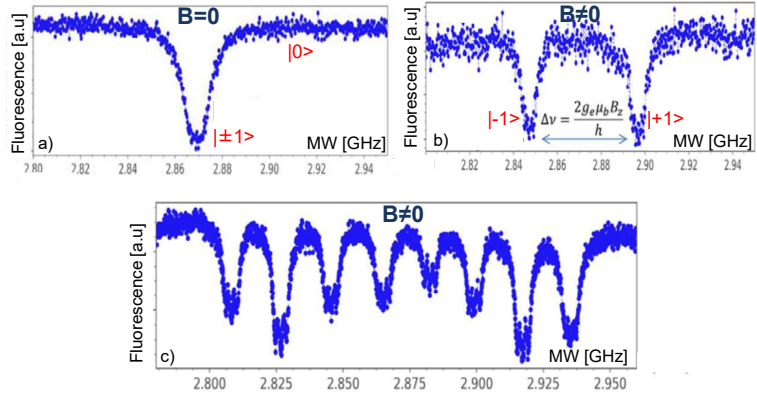}
	\end{center}
	\caption{ODMR spectra in the absence a) and in the presence b) of an external bias magnetic field. The magnetic field lifts the degeneracy of the $|m_s=\pm1>$ states and results in two separate dips in the ODMR spectrum. c) An example ODMR spectrum (excited at 532 nm) with a magnetic field in an arbitrary direction for an ensemble NV centers in diamond. Each of the four NV alignments has a different magnetic field projection along its quantization axis, leading to eight ODMR peaks (two for each NV alignment). For each dip a coupling with the nuclear spin of the $^{14}$N atom generates additional three hyperfine levels \cite{moreva2020biosensing}.}
	\label{fig:OBMR_B}
\end{figure}

If, instead of a single NV center, an ensemble of NV centers is considered, up to eight magnetic resonance dips can be observed, due to the four possible orientation of the NV axis in the diamond's crystalline matrix (see \textbf{Figure \ref{fig:OBMR_B}}). For certain directions of the magnetic field, some resonances can be degenerate.\\
A NV-based magnetometer can be realized, for example, by applying a bias field along the NV axis, removing the degeneracy, so that changes in the magnetic field projection along this axis affect the resonance frequencies almost linearly. Another option is to use all four NV alignments; although the eight ODMR frequencies have more complicated dependence on $\vec{B}$, this option yields information about the direction of magnetic field \cite{schloss2018simultaneous}.\\
The use of NV center as a magnetic field sensor firstly was proposed in \cite{taylor2008high,degen2008scanning} and demonstrated with single NV \cite{balasubramanian2008nanoscale,maze2008nanoscale} and NV ensembles \cite{acosta2009diamonds} in 2008.
\subsection{Electric field sensing}
The Hamiltonian describing the interaction with the electric field was derived from molecular orbit theory by Doherty et al.\cite{doherty2012theory} and it can be written in the following form:
\begin{equation}
\frac{\hat{V}_{gs}}{h}= d^{//}_{gs} (E_z+F_z)[\hat{S}_z^2-\frac{\hat{S}^2}{3}] + d^{\perp}_{gs} (E_x+F_x)(\hat{S}_y^2-\hat{S}_x^2) + d^{\perp}_{gs} (E_y+F_y)(\hat{S}_x \hat{S}_y + \hat{S}_y \hat{S}_x)  
\label{eq:electric}
\end{equation}
where $d^{//}_{gs}$ and $d^{\perp}_{gs}$ are respectively the axial and non-axial Stark shift components of the permanent electric dipole moment $d^{\perp}_{gs}$ in the ground triplet state \cite{sreenivasan2013luminescent}, $\vec{\textbf{\textit{E}}}$ is the electric field and $\vec{\textbf{\textit{F}}}$ is the mechanical strain.\\ 
According to Equation (\textbf{\ref{eq:electric}}) the effect of the electric field $\vec{\textit{\textbf{E}}}$ plays the same role as mechanical strain $\vec{F}$ \cite{dolde2011electric,van1990electric}. 
The strain depends on the diamond material: in single-crystal samples, the mechanical strain field is substantially negligible; while, in polycrystalline ones, a relatively high strain field is induced by the growth conditions, leading to a splitting of the spin state $|m_s= \pm 1>$ even in absence of external fields.\\
The frequency shift caused by the electric field is much smaller than the shift produced by the presence of a magnetic field (see \textbf{Table \ref{tab:Coupling_coefficinet2}}). For this reason, in order to reliably measure this second-order effect caused by the Stark shift, it is necessary to decouple it from the Zeeman shift.

Briefly, the fine structure Hamiltonian of the NV ground state, describing the energy levels of the electronic spin states due to the spin ($\hat{S}$) interaction with the static magnetic ($\vec{B}$), electric ($\vec{E}$), and strain ($\vec{F}$) fields, can be written in terms of the natural spin-triplet basis $\{|m_s =0>$ , $|m_s=+1>$, $|m_s=−1>\}$ in the following matrix form:
\[
\hat{H}_{gs}=
\begin{pmatrix}
0 & -\mu_B g_e \frac{B_x - i B_y}{\sqrt[]{2}} & -\mu_B g_e \frac{B_x + iB_y}{\sqrt[]{2}} \\
-\mu_B g_e \frac{B_x + iB_y}{\sqrt[]{2}} & h D+ \mu_B g_e B_z & - h d^{\perp}_{gs}(P_x- i P_y) \\
-\mu_B g_e \frac{B_x - i B_y}{\sqrt[]{2}} & - h d^{\perp}_{gs}(P_x+ i P_y) & h D- \mu_B g_e B_z
\end{pmatrix}
\] 
where it is possible to observe that the natural-spin basis vectors are eigenstates of the Hamiltonian only in the presence of both the magnetic and electric field aligned with the NV axis.
In this condition, $D= D_{gs}+ d^{//}_{gs}P_z$ describes the frequency shift of the resonance lines resulting from the zero-field splitting and from the Stark effect associated with the component of the vector $\vec{\textbf{\textit{P}}}=\vec{\textbf{\textit{E}}}+\vec{\textbf{\textit{F}}}$.
Otherwise, external fields not aligned to NV symmetry axis produce a non-diagonal matrix, and therefore energy levels of undefined spin.
In particular, the presence of additional transverse strain and electric-field components $P^{\perp}$ modifies the ground-state structure. \\
The Hamiltonian assumes a quasidiagonal form considering a new spin basis $\{|0>$, $|+>$, $|−>\}$, obtained by a field-dependent mixing of the $|m_s=+1>$ and $|m_s=−1>$ spin states according to the following unitary operator:
\[
\hat{U}=
\begin{pmatrix}
1 &0 & 0 \\
0 & e^{i \frac{\phi}{2}} sin(\frac{\theta}{2})  & e^{-i \frac{\phi}{2}} sin(\frac{\theta}{2})  \\
0 & e^{i \frac{\phi}{2}} cos(\frac{\theta}{2})  & -e^{-i \frac{\phi}{2} }sin(\frac{\theta}{2}) 
\end{pmatrix}
\]
where $tan(\phi)=P_x/P_y$ and $tan(\theta)=(d^{\perp}_{gs} P^{\perp}) / (\mu_B g_e B_z)$ are the field-dependent phases defining the spin state mixing. 
The Hamiltonian takes the following form in the $\{|0>$,$|+>$,$|−>\}$ basis:
\[
\hat{H'}_{gs}=\hat{U}\hat{H}_{gs} \hat{U^{\dagger}}=
\begin{pmatrix}
0 & c_1\mu_B g_e B^{\perp} &  c_2\mu_B g_e B^{\perp} \\
c^*_1\mu_B g_e B^{\perp} & h D+ W & 0 \\
c^*_2\mu_B g_e B^{\perp} & 0 &h D- W
\end{pmatrix}
\]
with \begin{equation}
W=\sqrt{(h d^{\perp}_{gs} P^{\perp})^2+(\mu_B g_e B_z)^2}
\end{equation}
The complex constants $c_1$ and $c_2$ represent the phase of the matrix elements and $B^{\perp}$ is the transverse component of the magnetic field with respect to the NV axis. If $B^{\perp} \approx 0$, the non-diagonal terms can be neglected and the Hamiltonian can be regarded as diagonal in the basis $\{|0>$, $|+>$, $|−>\}$. The energy difference between the $|0>$ and the $|\pm>$ states is $h D\pm W$, corresponding to ODMR resonances separated by 2W/h depending on the strengths of the magnetic, electric, and strain fields, as well as their orientations with respect to the axes of the NV center. 
Since the states $|\pm>$ are a coherent superposition of the states $|m_s=\pm 1>$, we underline that the ODMR resonance is observed also in this case as a reduction in the fluorescence emission at the new MW resonance frequencies.

\subsection{Temperature sensing}
Another interesting feature of the NV complex is the temperature dependence of its spin levels \cite{acosta2010temperature}.
Indeed, the microscopic origin of $D_{gs}$, also called the zero field splitting (ZFS) parameter, is due to spin-spin interactions in the NV's orbital structures, and the value depends on the lattice length, which is strongly correlated to the local temperature. When the local temperature increases the diamond lattice spacing of the NV center increases as well, lowering the spin-spin interaction and reducing the ZFS parameter $D_{gs}$.
Under ambient conditions $D_{gs} \simeq 2.87\,GHz$ and the temperature dependence is $dD /dT \simeq -74 kHz/K$ \cite{acosta2010temperature}. In general, the ZFS parameter shows a non-linear dependence, and its value increases when temperature decreases \cite{chen2011temperature}.

\begin{table}[htbp]
	\begin{center}
	\begin{tabular}{l l}
		\hline 
		Property          & Coupling coefficient      \\
		\hline 
		Magnetic field    &   $\gamma_e =	\frac{\mu_B g_e}{h} \simeq $ 28 GHz T$^{-1}$  \\&    $\gamma_N =	\frac{\mu_N g_N}{h} \simeq$ 15 MHz T$^{-1}$    \\
		
		Electric field    & $d_{//,gs} \simeq$ 3.5 mHz V$^{-1}$m \\ & $d_{\perp,gs} \simeq$  0.17   Hz V$^{-1}$m      \\
		
		Temperature       &  $\partial D_{gs}/ \partial T \simeq$ -74 kHz K$^{-1}$          \\
		\hline
	\end{tabular}
	\caption{Coupling coefficient of the NV center with the external fields and temperature.}
	\label{tab:Coupling_coefficinet2}
\end{center}
\end{table} 

\begin{figure}[!h]\begin{center}
		\includegraphics[scale=0.045]{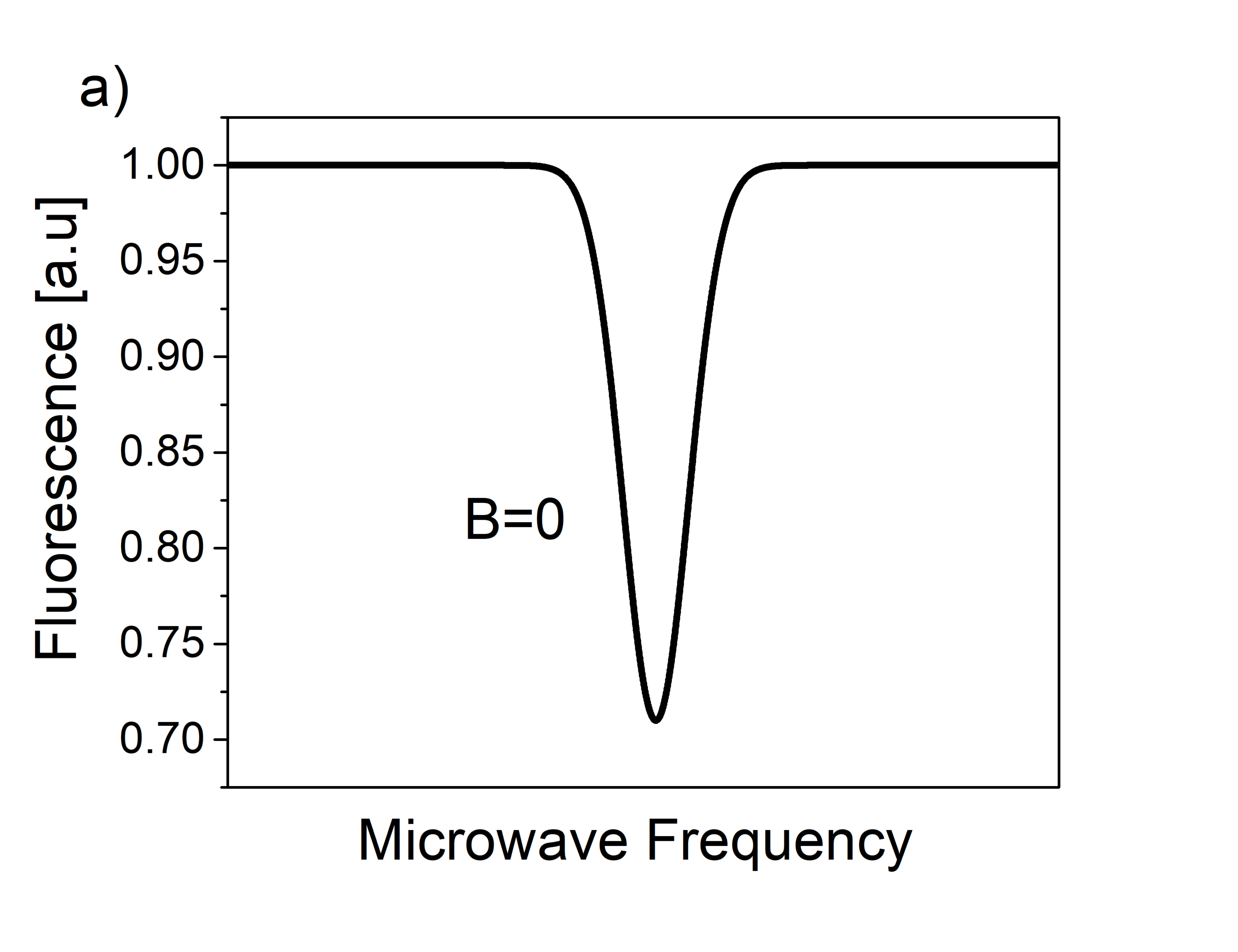} \quad \includegraphics[scale=0.045]{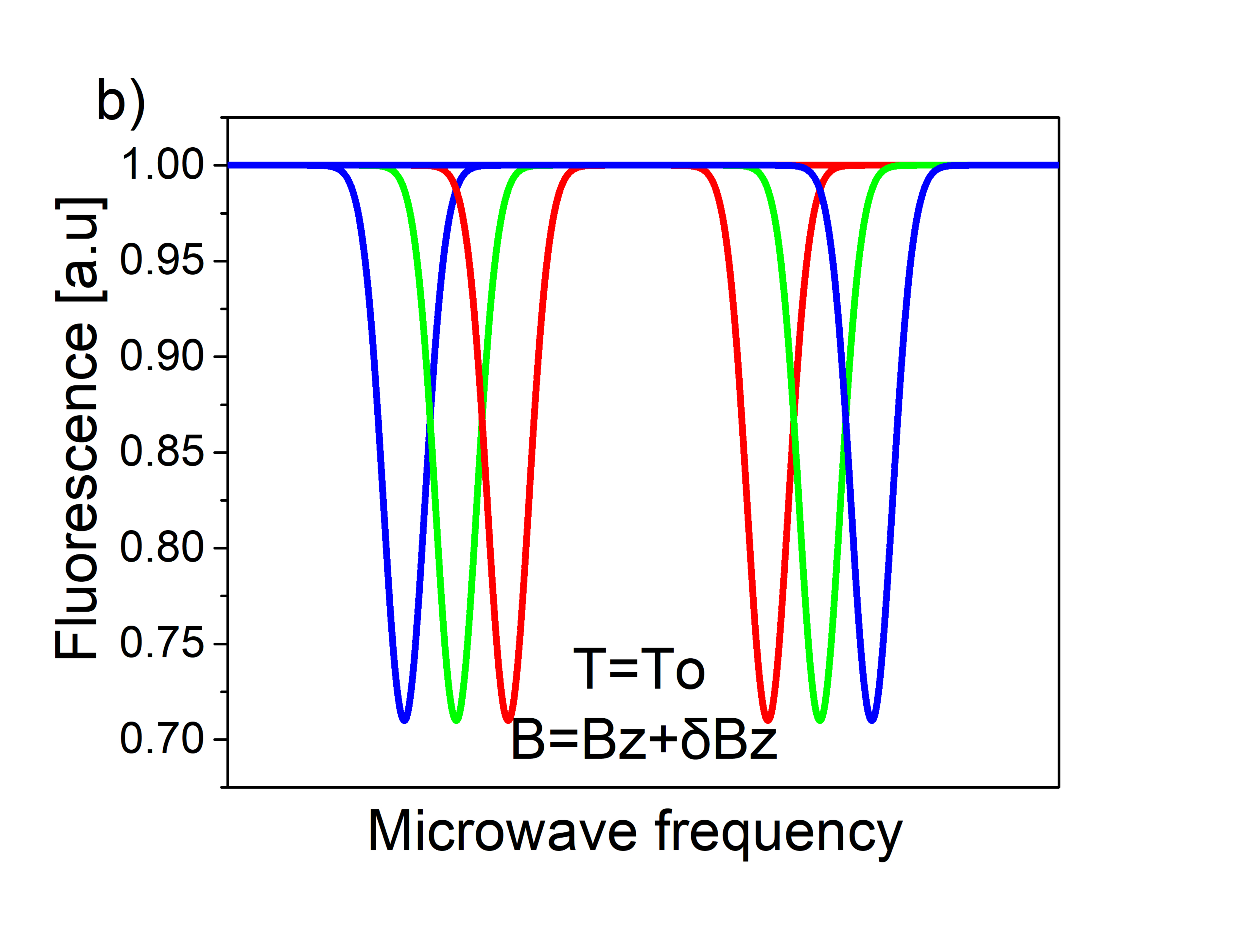} \quad \includegraphics[scale=0.045]{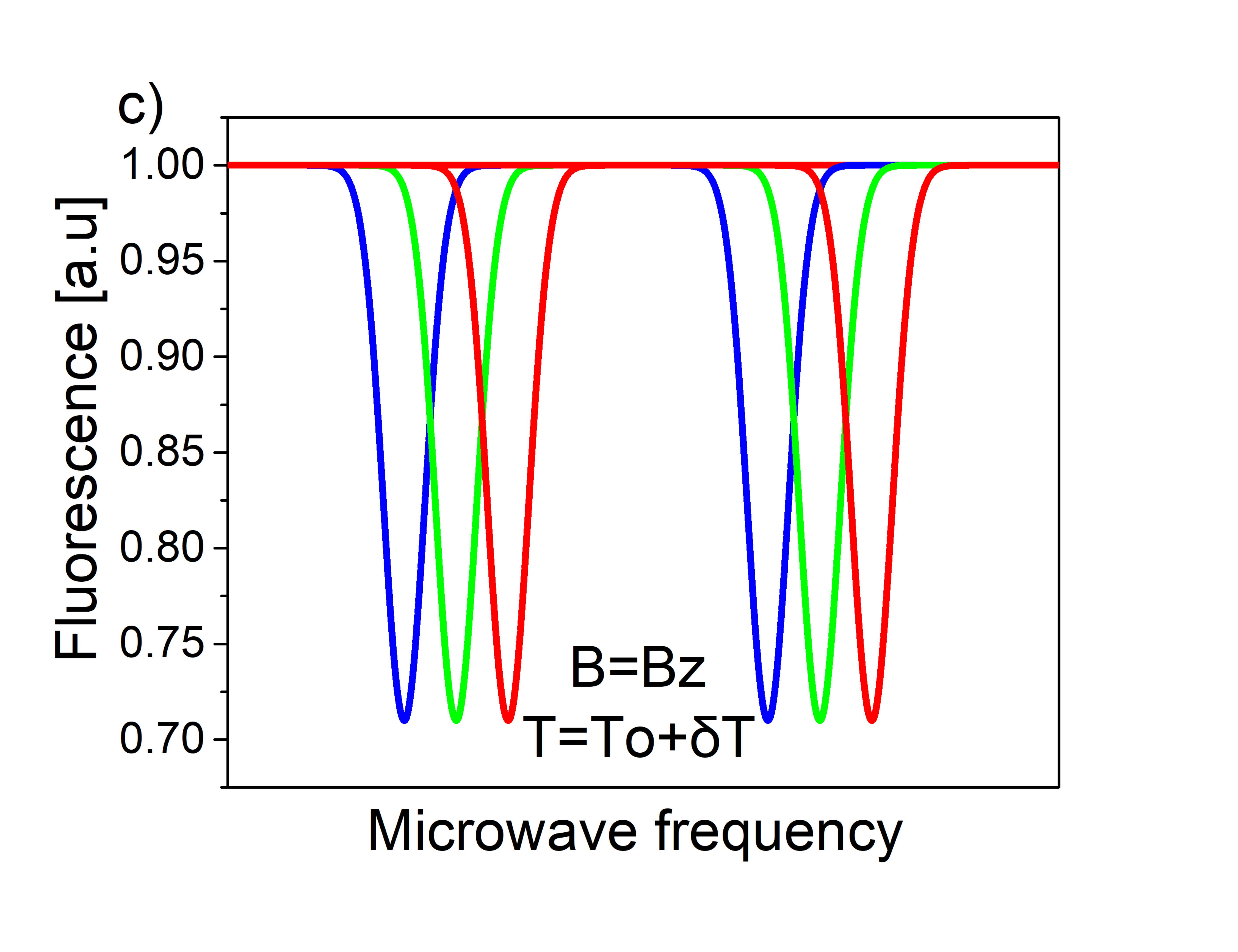} 
	\end{center}
	\caption{Example of magnetic and thermal shifts of the spin resonance, in ODMR spectra. Dips with equal colors correspond to paired resonances. The colors represent the timeline of the dips. The initial dips is red, then green and finally blue.}
	\label{fig:ODMR_BT}
\end{figure}

To realize a NV-based temperature sensor, the most obvious solution is exploiting the $D_{gs}$ temperature dependence. 
This requires that no external field is present ($\vec{\textbf{\textit{B}}},\vec{\textbf{\textit{E}}},\vec{\textbf{\textit{F}}}=0$), i.e. $|m_s=\pm1>$ is degenerate. In this case, an increase in temperature leads to a decrease in the resonance frequency, associated with a shift of the degenerate levels $|m_s=\pm1>$ towards the level $|m_s=0>$. \\
Nevertheless, this is the simplest but not the optimal solution, since, even in the absence of applied fields, the sample may have an internal strain and may be affected by the Earth's magnetic field. 
Unless it is possible to find a diamond sample with negligible $\vec{\textbf{\textit{F}}}$ and to design an experimental set up able to reasonably compensate for the external magnetic field (e.g. Helmholtz coils), the dips would not be perfectly overlapped because of the non perfect degeneracy of $|m_s=\pm1>$, thus showing a larger full-width-at-half-maximum (FWHM) and therefore a lower resolution. \\
A better solution is to apply an external magnetic field in order to significantly separate the spin levels. However in this configuration, a single dip can shift for a temperature variation, but also for a variation of magnetic field. To decouple the two contributions it is sufficient to monitor both $|m_s=+1>$ and $|m_s=-1>$ spin states at the same time, using simultaneous driving of the microwaves in ODMR technique \cite{tzeng2015time}.
As it can be seen in the \textbf{Figure \ref{fig:ODMR_BT}}, by simultaneously monitoring the initial dips (red curves), it is in principle possible to understand if there are variations in the magnetic field (the dips move in opposite directions) or in temperature (the dips move in the same direction).

\begin{figure}[!h]\begin{center}
		\includegraphics[scale=0.105]{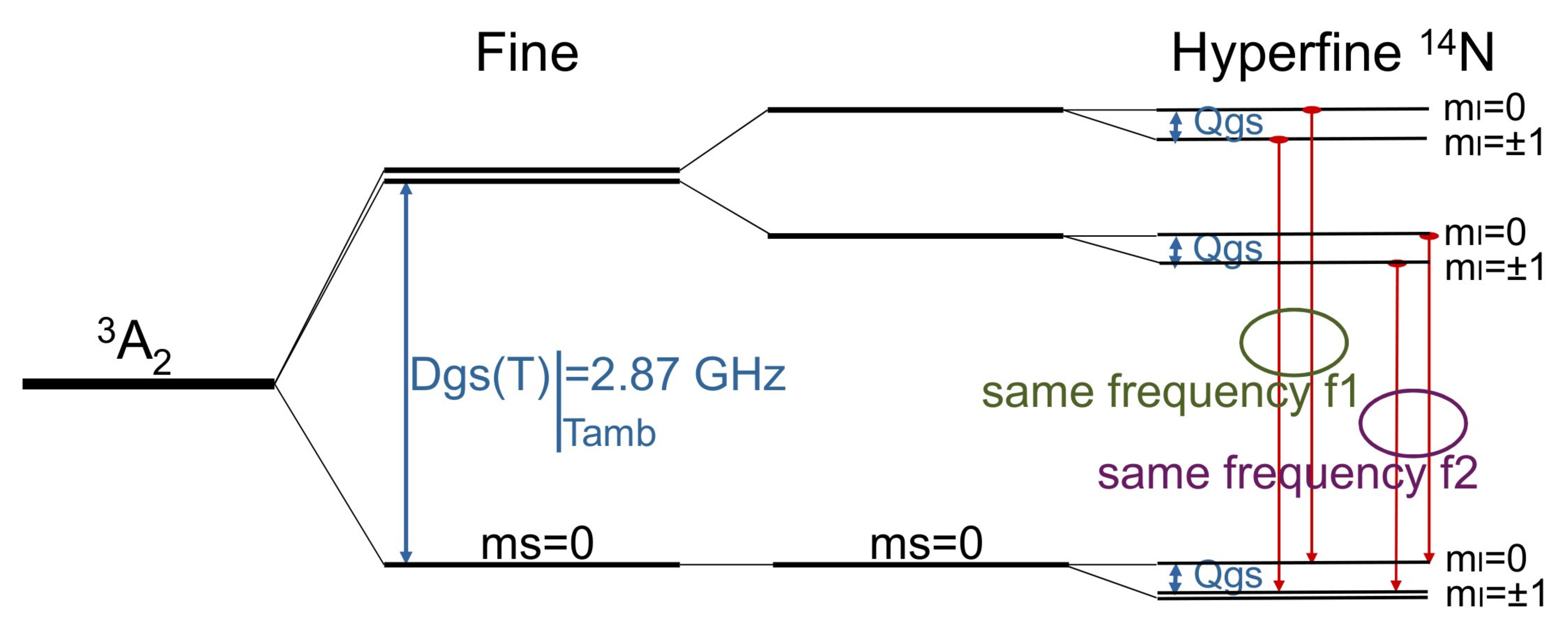} 
	\end{center}
	\caption{NV ground-state $^3$A$_2$ scheme, in presence of intense transverse magnetic field $B^{\perp}$.}
	\label{fig:Moreva}
\end{figure}

In this perspective, an improved technique exploiting the application of an intermediate transverse bias magnetic field $B^{\perp}\,$ has been implemented \cite{PhysRevApplied.13.054057}. Similarly to the case just discussed, the application of $B^{\perp}\,$ removes the degeneration of $|m_s=\pm1>$ and therefore improves the FWHM. The intensity and the transverse direction of that field creates a quantum superposition of states, which is insensitive to magnetic fields but sensitive to temperature \cite{moreva2019practical}. 
In this configuration, the expectation value of the spin along any direction is small, implying the degeneracy of the hyperfine structure between the levels $|m_I=\pm1>$ (except for the quadrupole contribution $Q_{gs}$, which seperates $|m_I=0>$ from $|m_I=\pm1>$). 
In \textbf{Figure \ref{fig:Moreva}} the corresponding scheme of the spin energy levels are reported (only the $^{14}$N isotope is considered as it is the most common). In this situation, the ODMR spectrum reduces to two dips \cite{moreva2019practical} (instead of 6), providing a substantial improvement in the signal-to-noise ratio. This particular orientation of the magnetic field ensures the protection of the measurements from the noise of other possible magnetic fields. 
In fact, the NV spin is non-sensitive to the magnetic field fluctuation, because the contribution of the magnetic component only appears at the second order in the Hamiltonian.
\section{Bio-sensing}
As mentioned above, NV sensors are particuarly suitable for biological sensing.
Before describing the experiments focusing on the NV-based sensor it is necessary to specify the type of biological specimens of interest, the expected magnitude of the electromagnetic field produced by these specimens and the principal parameters such as sensitivity and spatial/temporal resolution, required for the NV-based sensors.\\
This section, after reviewing some of the devices typically used for biosensing, analyzes in detail neuronal and cardiac cells. 
Higher sensitivity and resolution of electromagnetic fields is considered necessary to expand the understanding of the fundamental processes regulating the interaction of these cells.

\subsection{From the conventional electrophysiological techniques to NV sensors}
The electrical activity of excitable cells can be investigated by means of the conventional patch-clamp technique \cite{neher1992patch} or through Micro Electrode Array (MEA) \cite{stett2003biological} recordings.
Single-cell recordings, performed under voltage- or current-clamp configuration, respectively allow to monitor ion currents or the membrane resting potential, postsynaptic responses and action potential firing activity \cite{recording1995edited}. 
Besides having an extreme versatility (monitoring overall electrical events from the whole cell,  from microdomains of the cell membrane or even from single channel proteins), patch-clamp has a high temporal resolution and high sensitivity: all these features make this electrophysiological approach the gold standard for measuring the electrical activity. 
Though, patch-clamp is rather invasive, as it damages the cell membrane  through the recording electrode: this implies that only one recording is feasible for each cell \cite{mathias1990limitations}.
On the contrary, MEA is a non-invasive  approach, used to measure the membrane potential variations from many cells simultaneously. The MEA probe  is structured as an array of sensing electrodes, of variable geometry dimensions and material, which are immersed in a glassy (insulating) double layer. 
Commonly, sensing electrodes are made of titanium or indium tin oxide (ITO), and have a diameter that can vary between 10 and 30 $\mu m$ \cite{fejtl2006micro,spira2013multi}. 
By means of MEAs, it is possible to monitor the electrical activity of a neuronal network as a whole, and measuring its changes along with its maturation, even though informations on the biophysical properties of ion channel cannot be directly inferred. 
This specific measurement need, requiring non-invasive and iterative detection for biological applications has prompted the study and realization of different devices.
In the following we discuss and compare the most promising ones. \\
Promising devices for the detection of weak magnetic fields, in addition to the NV-based sensors, are the superconducting quantum interference device (SQUID) sensors \cite{vengalattore2007high,faley2006new,baudenbacher2003monolithic} and chip-scale atomic magnetometers (CSAMs) \cite{knappe2010cross}.
Until now, the measurement of very weak magnetic fields was the domain of SQUIDs sensors. These sensors have reached sensitivity levels of (0.9–1.4) fT/Hz$^{1/2}$ with a pick-up coil area of the order of 1 cm$^2\,$ \cite{faley2006new}. However, SQUIDs require cryogenic cooling, which, in addition to implying significant cost and maintenance complexity, requires positioning the sensor a few centimeters from the sample. 
An alternative is offered by the CSMAs, that are based on microfabricated alkali vapor cells integrated with small optical components such as diode lasers and fiber optics. These devices have reached sensitivities below 5 fT/Hz$^{1/2}$ at sensor volume 8 mm$^3$\cite{knappe2010cross}. However, despite the exceptional sensitivity, the minimum working distance between sensor and magnetic source for CSAM or SQUIDS remains at least few mm, that makes them unsuitable for monitoring individual cell signals or small tissues, being the amplitude of the magnetic field decreasing quadratically with distance.\\
The sensors for the detection of electric fields, emerging in the last few decades are single-electron transistors (SETs) \cite{nakajima2016application}, that are a promising candidate for achieving higher detection sensitivity due to the Coulomb oscillations. However, the existence of a SET- based biosensor has emerged only in recent years \cite{jalil2017sensing}, probably because of their difficulty of the room-temperature operation.\\
Finally, in recent years there has been a growing interest in the use of temperature sensors capable of operating on a nanometric scale. It has been known that local temperature variations at the intracellular level play a fundamental role in cellular activities related to body temperature homeostasis and energy balance \cite{takei2014nanoparticle}. Particular attention is paid to the possibility of measuring local temperature variations of cell organelles (i.e. nucleus, mitochondria, etc.) or ion channels. For example, different simulation model \cite{chen1995hydrodynamic,el2015mechanical} shows a hypothetical variation in temperature at the level of the ion channels, due to the flow of the ions from the inside to the outside of the plasma membrane, during the genesis of the action potential. Due to the difficulty of this local measurement, no one has ever measured this thermal variation. 
Interestingly, temperature changes may drastically alter the neuronal firing frequency, as demonstrated by Guatteo et al.\cite{guatteo2005temperature}.
Currently fluorescence probes are powerful method used to study intracellular temperature variation thanks their high spatio-temporal resolution. The probes typically used for this measurement are organic or inorganic fluorescent probes, such as fluorescent proteins, organic dyes \cite{donner2012mapping,yang2011quantum,kim2006micro,vetrone2010temperature}, quantum dots (QDs) \cite{maestro2010cdse,maestro2014quantum} and many others. 
Organic proteins are biocompatible probes, rather stable and very easy to chemically target. But there are different problems related the use of these probes: these are often autofluorescent and to avoid the phenomenon it is necessary to add specific quenchers; they cannot be used for a long time, in fact these sensors suffer from photobleaching and unstable photoluminescence. In the best case scenario, the probe degradation consists of fluorescence suppression, in the worst case scenario it releases an electron that binds to nearby molecules making them toxic. These probes are organic and by their nature they are also subject to even weak pH variations, for this reason it is fundamental a strict control of the cell environment \cite{jensen2012use,bernas2005loss}. 
The inorganic probes such as quantum dots (QDs) have the advantage of being stable in fluorescence, have a high sensitivity to temperature variations and their nanometric size allows obtaining a spatial resolution useful for cellular measurements. Although the size of these sensors would allow spatial resolution limited by the diffraction limit only, their chemical composition is found to be non-biocompatible in most of the cases.
Other temperature sensors are based on up converting nanoparticles (UCNPs)  \cite{takei2014nanoparticle,brites2016lanthanides}: nanoscale particles (diameter 1-100 nm) that exhibit photon upconversion, i.e. when stimulated by incident photons they are able to emit fluorescence’s of shorter wavelength. They are usually composed of rare-earth based lanthanide or actinide-doped transition metals. Their core-shell structure allows sensor compatibility, however, sensitivity is not high.\\
Extremely interesting devices able to realize all these measurements (magnetic, electrical and temperature sensing) eventually at the same time, are one based on the NV center in diamond. The advantages of these sensors are manifold: they have stable photoluminescence in the visible and near-infrared range, their chemical composition ensures resistance to photobleaching and diamond is an inert and therefore biocompatible material \cite{yu2005bright}, so cell/neurons can be grown directly on its surface \cite{schirhagl2014nitrogen,barry2016optical,specht2004ordered} or nanodiamonds can be injected inside them, allowing for sub-cellular spatial resolutions \cite{kucsko2013nanometre} with a non-invasive techniques. Finally, NV sensors can operate at room temperature and, in more detail, their dynamical range of temperature sensing extends further 500 K for both bulk \cite{toyli2012measurement} and nanoscale \cite{plakhotnik2014all} diamonds.\\
In the remainder of this section, the modelization of the target neural and cardiac signals for NV-based biosensing will be reviewed.

\subsection{NV center as sensor for neuronal signals}
\begin{figure}[!h]\begin{center}
		a)\includegraphics[scale=0.4]{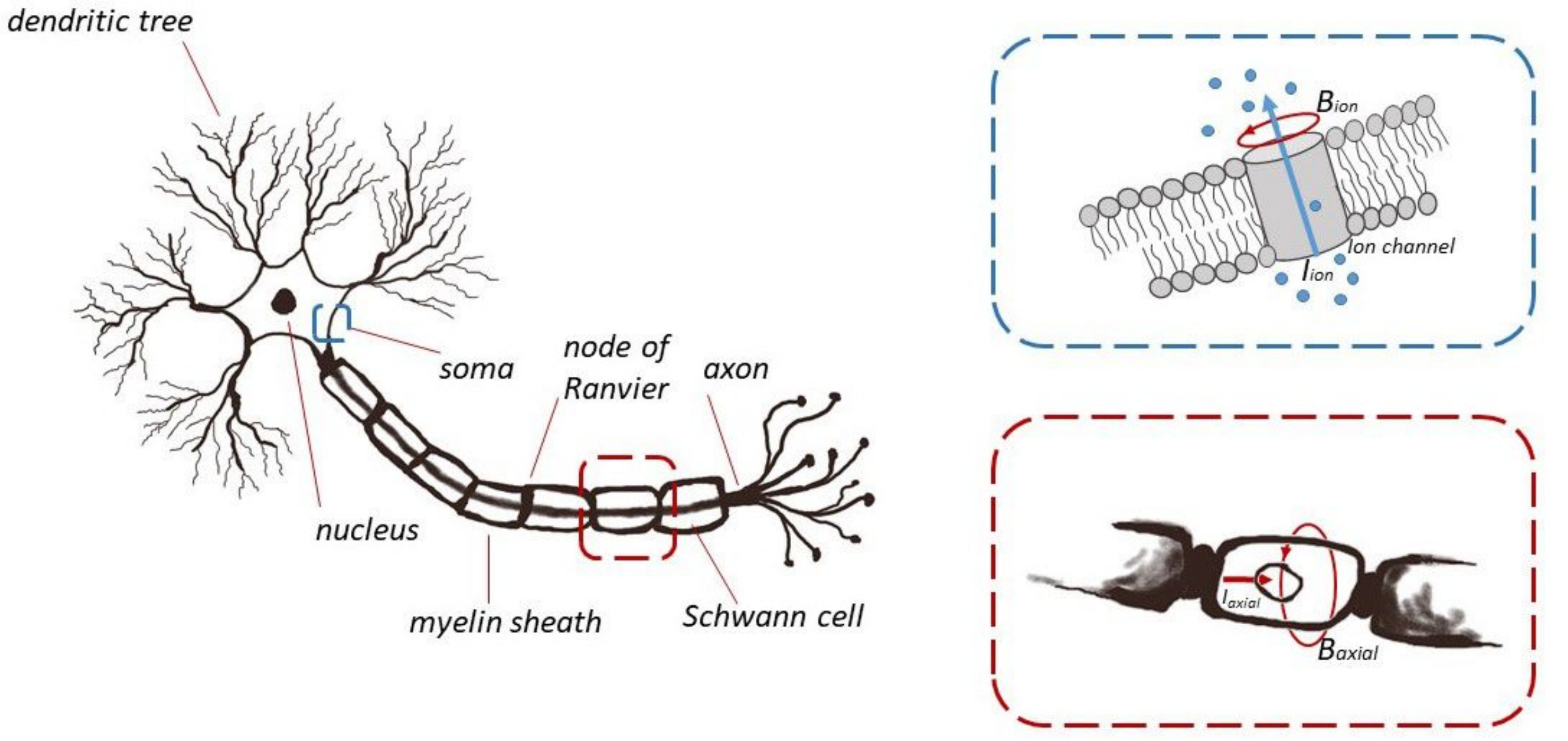}  \quad	b)\includegraphics[scale=0.14]{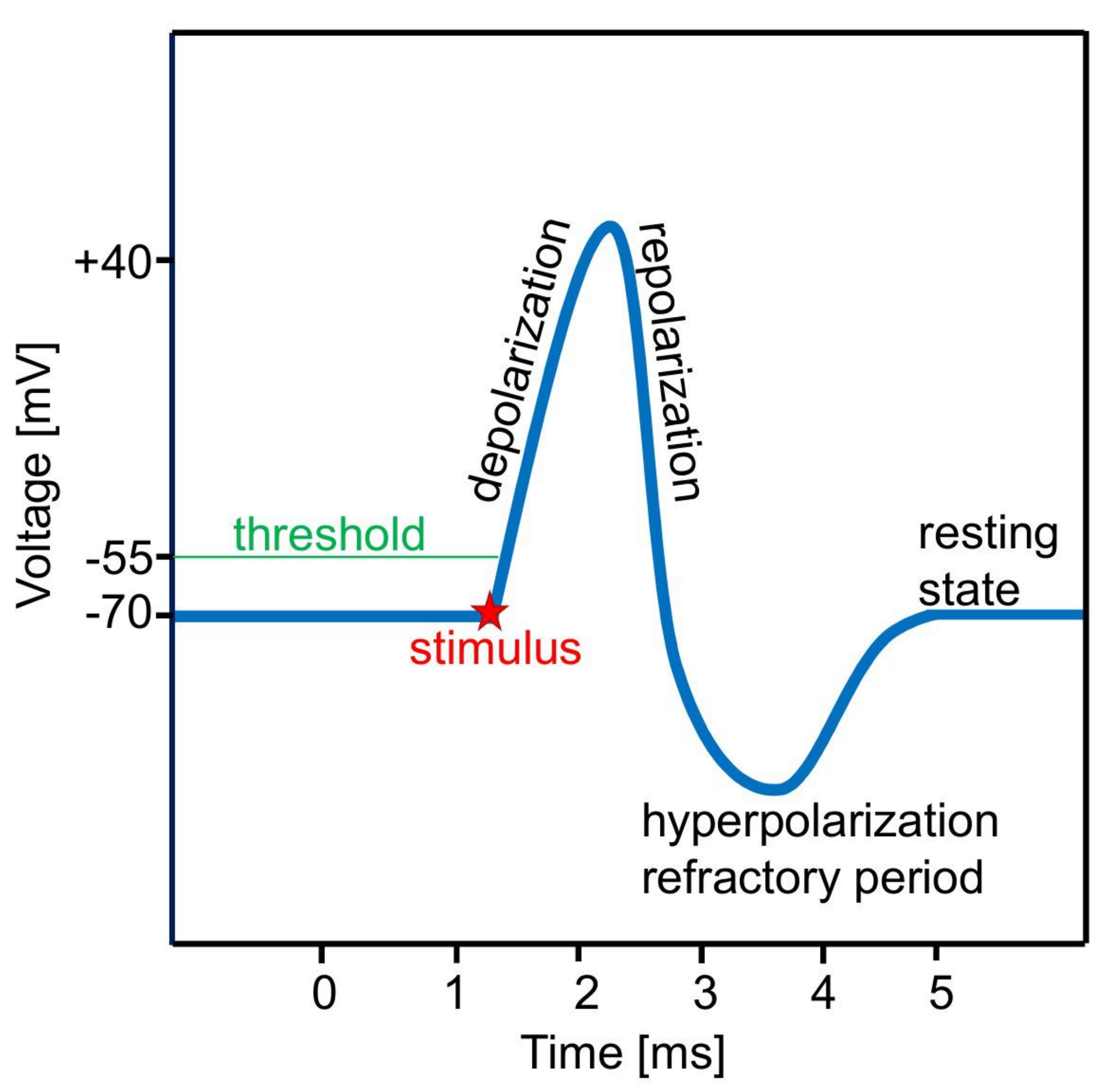} 
	\end{center}
	\caption{a) Single neuron simplified sketch. In the upper box a zoom of the neuronal membrane is reported, where the ionic current and the corresponding magnetic field are schematized. In the lower box the axial current and the relative magnetic field are shown. b) Schematic representation of neuronal action potential (AP). Resting membrane potential (-$V_m$) is -70 mV. When $V_m$ is driven and exceeds the threshold (following an initial stimulus), a rapid membrane depolarization occurs. In this phase the Na$^+$ channels open, allowing sodium to enter in the neuron and bringing $V_m$ to approximately +35 mV. Then the repolarization phase begins, caused by Na$^+$ channels inactivation and opening of K$^+$ channels. This outward current drives the membrane potential close to -93 mV (hyperpolarization). Finally, the Na$^+$/K$^+$ ATP-ase restores the initial conditions. During the depolarization, the influx of positive charges produces local internal and external longitudinal currents, which are responsible for the AP propagation in the axon adjacent area. The propagation directionality is guaranteed by the AP refractory period: although the local currents propagate in both directions, a new AP cannot be triggered in refractory membrane area.}
	\label{fig:Neuronal_action_potential}
\end{figure}

In the last decades, neuroscience has attracted great interest beyond the scientific community. Because of the increase in life expectation, cases of neurodegenerative diseases such as Parkinson's, Alzheimer's, Huntington's disease and many others are constantly growing. Currently, these diseases are incurable, even symptoms mitigation is difficult because of late diagnosis when most of the neurons involved have been irreparably damaged. This reason strongly prompts to develop new increasingly precise and sensitive techniques, allowing a deeper understanding of neuronal circuits ranging from functioning of synaptic sites to the behavior of the entire neuronal network. Neurons are the functional units of the nervous system. They communicate via electrical signals, known as action potentials.\\
The action potential (AP) consists in the variation in time of the membrane potential $V_m$, where $V_m=\Phi e_{in}-\Phi e_{out}$ is the electrical potential difference between intra- and extra-cellular side of the cell membrane. The AP characteristic waveform is shown in \textbf{Figure \ref{fig:Neuronal_action_potential}b}. The AP pulse is caused by several ionic species (Na$^+$, K$^+$, Ca$^{2+}$), which flow through the neuronal membrane. 
\\
The two electrophysiological techniques mostly used to study cell excitability and synaptic transmission in a neuronal network are the patch-clamp and the MEA. In the last decade scientists have tried to study more and more specifically the propagation of the electrical signal from the cell body (or soma) to the whole dendritic tree. In other words, the goal would be to create a device that allows scanning the neuron point by point from the soma to the axon and the dendrites, following and characterizing the electrophysiological variations of the electrical signal during its propagation. 
The technology closer to this ambitious goal is the one of the CMOS-MEA, that allows having a much higher density of electrodes with respect to the traditional MEA technology. Numerous studies have managed to scan the path of the electrical signal in a neuronal network at the level of the single neuron \cite{kuhn2020advanced,muller2015high,bertotti2014cmos,bakkum2013tracking}.\\\\
Bakkum et al.\cite{bakkum2013tracking} recently have developed a high electrode density CMOS-MEA device capable of stimulating a specific area and simultaneously scanning the signal along some points from the soma to the axon.
Clearly, this technique is much more sensitive than MEA, but given the stochasticity of the cell's placement in space, it requires cells to be marked in order to follow their path. Recently several groups have correlated this technology to the technique of optogenetics. They tagged the genes of interest and activated them following an optical stimulation and simultaneously followed the signal thanks to the integration of the CMOS-MEA \cite{buzsaki2015tools,radivojevic2016electrical}.\\
However, these techniques do not allow following the entire dynamics of the action potential, but to have a scan of a region depending on the position of the electrodes with respect to the neuron with its axon and its dendritic body.\\\\
NV sensors may therefore have a huge impact on these applications: nanodiamonds can be targeted on the membrane surface or, alternatively, cells can be plated and cultured on a bulk diamond \cite{tomagra2019micro}. Indeed, taking advantage of diamond biocompatibility and the exceptional spatial resolution displayed by color centers in diamonds , it will be worth exploiting these properties for a timely reconstruction of the AP dynamics. Furthermore, the possibility of positioning them adjacent to the cell membrane has the advantage of experiencing stronger magnetic fields. However, since neuronal magnetic fields are extremely weak ($\simeq pT$), their detection appears to be challenging even for NV-based sensors, at least for mammalian cells, while measurements have been performed on giant neurons of invertebrates \cite{barry2016optical}.

To predict the electromagnetic fields intensity created by the AP, and therefore to understand what sensitivity of the NV sensors is needed to sense it, it is necessary to model how the AP develops and propagates.

Hodgkin-Huxley model \cite{hines1993neuron,hines1997neuron,santamaria2009hodgkin,arcas2003computation} allows estimating the ionic current flowing through the neuron membrane (when the ion channels are open). For the human neuron, the total estimated ionic current, sum of the single channels contribution $I_{ion}$ is:
\begin{center}
	$ I_{\perp}=\sum I_{ion} \simeq 2\ pA \ \mu m^{-2}$
\end{center}
and the current pulse typically lasts $\Delta t\,\simeq1\ ms$.
Each $I_{ion}$ generates a magnetic field (see \textbf{Figure \ref{fig:Neuronal_action_potential}a}), which can be estimated by means of the Biot-Savart law: $\oint_{C} \vec{\textbf{\textit{B}}_{\textbf{\textit{ion}}}} \cdot \vec{\textbf{\textit{l}}}= \mu_0 I_{ion}$. However, the resulting amplitude of these fields depends on the channels density, which largely varies depending on the axon area being considered. Furthermore, we note that the $\vec{\textbf{\textit{B}}}^{\textbf{\textit{tot}}}_{\textbf{\textit{ion}}}$ field, sum of the contributions of the field produced by the various channels, can be vanishingly small on average, because of the different fields directions. 
To this purpose, channel clustering may be very significant \cite{rasband2001developmental,sato2019stochastic}. Assuming a current of $100\,\,pA/ \mu m^2$ and considering that the NV sensor positioned at an average distance of few nanometers (by selective targeting the channel using functionalized NDs \cite{zhang2017anchored,rendler2017optical}), a magnetic field of about $0.1-5\,\,nT$ (or even higher) could probably be sensed. 
This hypothesis is now under experimental analysis \cite{nostro}. \\
Current flowing through the membrane is not limited to the charge flow through ion channels, as longitudinal currents, but one should also consider the flow along the neuron axis, that is responsible for the AP propagation. \\
These currents also generate a magnetic field, around the neuron (see \textbf{Figure \ref{fig:Neuronal_action_potential}a}). Both the axial current and the corresponding magnetic field have been estimated \cite{isakovic2018modeling,karadas2018feasibility,hall2012high}. In particular, Ref.\cite{isakovic2018modeling} goes beyond the simplification of the Hodgkin-Huxley model, introducing the spatial and temporal progression of the AP along the various neuronal compartments, into which they have divided the axon. The theoretical prediction is a maximum field $B_{axial} \simeq $ 3 pT on the external membrane near the Ranvier node and a field $B_{axial} \simeq $ 2.3 pT on the myelin sheath external surface in those regions where the axon is wrapped by it.\\
The maximum magnetic field was also calculated by Isakovic et al. in Ref.\cite{isakovic2018modeling} for the nerve composed of 100 axons, obtaining only $B_{axial} \simeq$ 6 pT. This is due to the cancellation of the magnetic field component, caused by different axons within the same nerve, bringing opposite directional currents.
This estimated magnetic fields, in reality, are compatible with the fields detected by magnetoencephalography (MEG). MEG is able to detect fields if the order of $10^{-15}$ T because of the distance from the source \cite{hamalainen1993magnetoencephalography}.

Considering these values, a NV sensor positioned on the neuron surface or a few micrometers from it, should have a temporal resolution of about 0.1 ms (in order to be able to trace the time variation), and spatial resolution of about 10 $\mu m^3$ (which would allows a good reconstruction of the AP propagation, being the axon length ranging from 0.1 $\mu m$ to 1 m). Thus, the NV sensor should have a minimum sensitivity of \cite{schoenfeld2011real}:
\begin{equation}
\eta = \delta B_{min} \sqrt{\Delta t} \simeq 3\ pT \sqrt{0.1\, ms} \simeq 30 \,fT\,Hz^{-1/2}
\end{equation}
The NV sensor optimal sensitivity is in principle limited by the quantum projection noise. This fundamental sensitivity limit for spin-based magnetometers is given by \cite{budker2007optical}:
\begin{equation}
\eta_{q}=  \frac{1}{\gamma_e} \frac{1}{\sqrt{nT^*_2}} 
\label{projection}
\end{equation}
Where $\gamma_e$ is the magnetic coupling coefficient (\textbf{Table \ref{tab:Coupling_coefficinet}}), \textit{n} represents the number of NV centers and $T^*_2$ their characteristic dephasing time. It is important to underline that the number of NV centers \textit{n} refers to the sensing volume. As mentioned, for the single PA detection the sensing volume should be around 10 $\mu$m$^3$, the size of the cell. \\
In the Ref.\cite{barry2016optical}, the estimation of the parameters $n\simeq3\cdot 10^6$ cm$^{-3}$ and $T^*_2\simeq$ 450 ns determines a spin projection noise value of $\eta_{q}\simeq$ 30 pT Hz$^{-1/2}$ for the sensing volume of 10 $\mu$m$^3$ (the experimental sensitivity reached is instead $\eta \simeq$ 15 pT Hz$^{-1/2}$ for the sensing volume of $5\cdot10^6$ $\mu$m$^3$). This value is still 1000 times larger than the sensitivity required for the detection of a single AP. However, as will be discussed in section \textbf{5}, it is possible to optimize both the above mentioned parameters to improve the performances.\\\\
Once the biomagnetic field $\vec{\textbf{\textit{B}}}(\vec{\textbf{\textit{x}}},t)$ has been measured, to reconstruct the unknown currents generating it, one should solve an inversion problem. In general, its solution is not unique, due to the existence of the so-called "magnetically silent" currents (i.e. the ones producing magnetic fields that almost cancel each others) and due to the fact that the magnetic field can be influenced by the electric field \cite{sarvas1987basic,michel2004eeg}. However, in the single axon case, it can be uniquely resolved. On the contrary, in the biological tissue case and in the 3D structures case, that cannot be traced back to standard models (such as a spherically symmetrical conductor or a horizontally layered medium), the solution is not unique. In some cases this is resolved by the knowledge of the electric field on the conductor surface \cite{sarvas1987basic}.
\subsection{NV center as sensor for cardiac signals}
\begin{figure}[h]\begin{center}
		a)\includegraphics[scale=0.11]{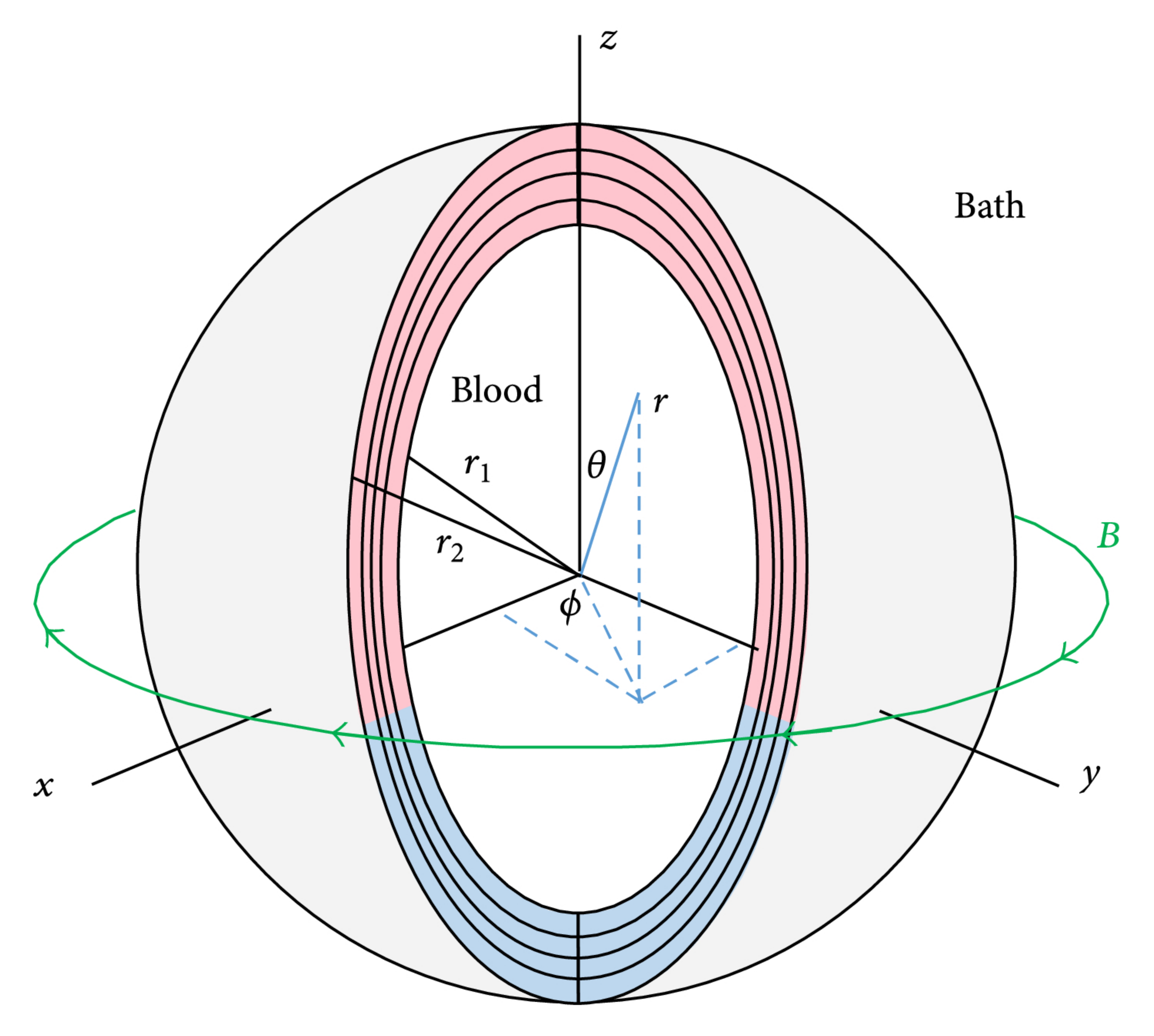}   \quad 	b)\includegraphics[scale=0.11]{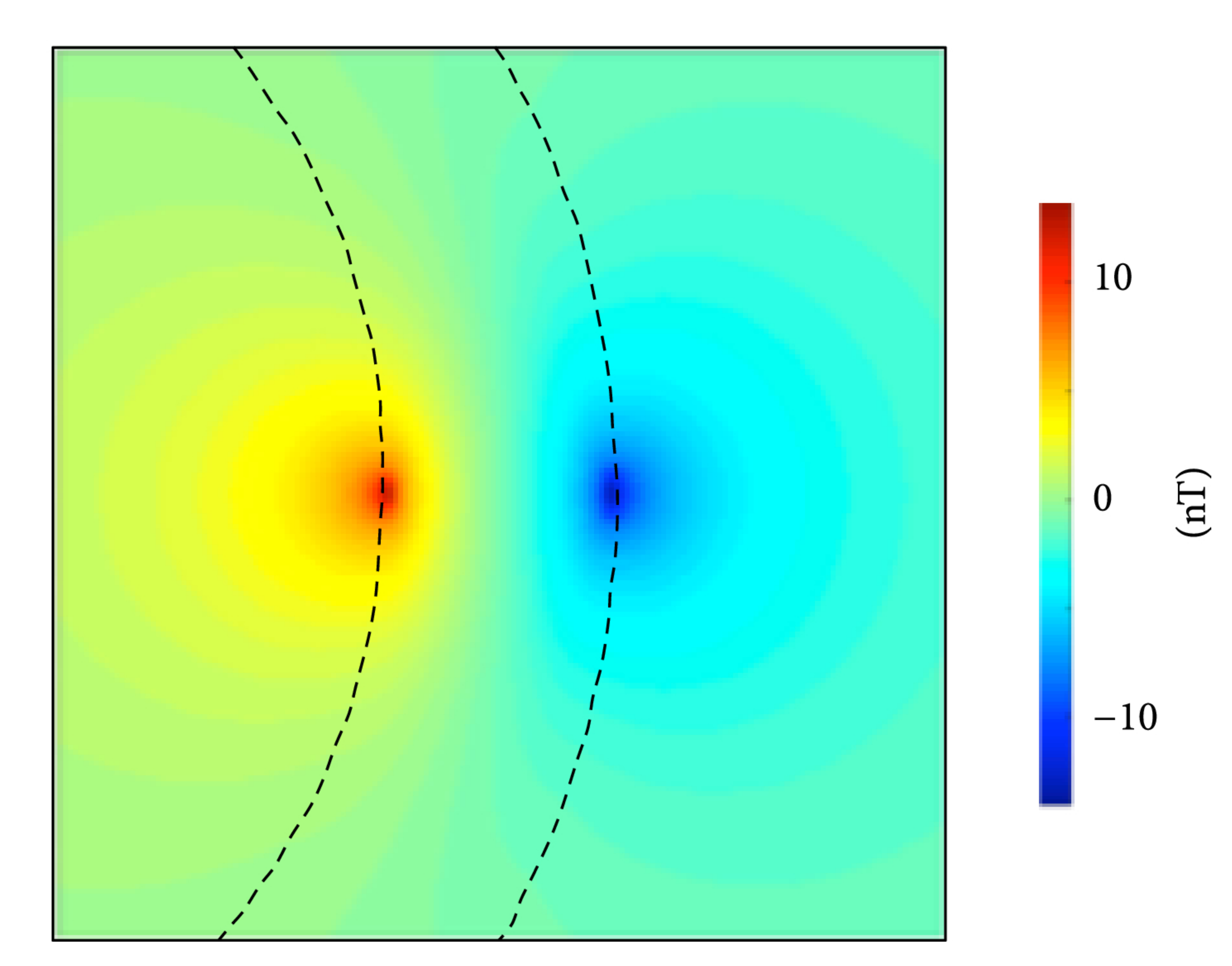} 
	\end{center}
	\caption{a) The model of a spherical heart, taken from the reference \cite{xu2017magnetic}. Part of the spherical shell has been cut out to show the heart wall. The black curves indicate the fiber orientation. The pink tissue has a transmembrane potential of 20 mV, and the blue tissue has a transmembrane potential of −80 mV. The green curve shows the magnetic field. The endocardial (inner) surface has radius r$_1$, and the epicardial (outer) surface has radius r$_2$. b) The magnetic field over a cross section of the heart. The dashed curves indicate the heart inner and outer surfaces. An area 40 mm by 40 mm is shown.}
	\label{fig:Cuore1}
\end{figure}

The human (and animal) heart generates the body's most intense electromagnetic field. In particular, by comparing measurements performed externally to the human body, the electric field generated by the heart, measured through the electrocardiogram (ECG) is about 60 times stronger than that the one of the brain, recorded by an electroencephalogram (EEG). In addition, the heart magnetic field detected by the magnetocardiogram (MCG) is about 5000 times higher than the neuronal magnetic field detected by magnetoencephalography (MEG): 0,05 nT (heart) vs 1 fT (neuron).
Thus, ODMR based on NV sensors can also find very significant applications in studying cardiac cells and tissues. To achieve a first qualitative estimation of the magnitude of the magnetic field in this case, one can start from a very simplified model: the spherical heart \cite{trayanova1993response}. Although this model is not physiologically accurate, it allows to extrapolate analytical solutions. \\
In a more recent work \cite{xu2017magnetic}, is proposed a further assumption concerning the origin of the currents. There are two currents sources in the heart: the first consists of intracellular currents, the second is given by the anisotropy of the tissue \cite{murdick2004comparative}. Regarding the first current contribution, the authors consider a spherical shell of cardiac tissue, which covers a blood cavity and is surrounded by an external bath of unlimited electrical conduction. The heart fibers propagate in the \textit{z} direction and a variation of the membrane potential $V_m$ is assumed following the activation of the action potential (AP), started at $\theta = 90$° (see \textbf{Figure \ref{fig:Cuore1}a}).\\
In this work the electric field is evaluated using the bidomain model \cite{trayanova1993response} and considering a situation of quasi-stationarity (although $V_m$ depends on time due to the action potential propagation, it is assumed that, given a certain $V_m(t_0)$, one can derive current and magnetic field in a quasistatic way). \\
Thus, the electric potential is obtained, using the continuity equations and the boundary conditions \cite{roth1991comparison,krassowska1994effective}, the current density distribution is obtained using Ohm's law and finally the magnetic field using Biot-Savart's law. 
Considering the anisotropic electrical conductance data \cite{roth1997electrical}, the $V_m$ values and typical heart dimensions \cite{trayanova1993response}, it turns out that the magnetic field is stronger near the internal and external surfaces tissue while it is weaker in the heart wall. The peak value of the magnetic field is around 14 nT (see \textbf{Figure \ref{fig:Cuore1}b}).

At the heart center, instead, the magnetic field reduces to $B=$ 2 nT \cite{mcbride2010measurements}. This is due to the fact that intracellular and extracellular currents are in opposite directions with almost the same magnitudes in the depths of the tissue and, therefore, the corresponding magnetic fields essentially cancel each other.\\ Considering a planar cardiac tissue sample, the spherical shell method is no longer valid. In this last case it has been found that the magnetic field reaches a peak value $B=$ 1 nT \cite{holzer2004high}.

The heart AP is about $\Delta t$ = 300÷500 ms long, however for some cardiac cells, such as ventricular or rapid response cells, the AP rapid rise occurs in 1 ms, as in the neuronal case. 
Considering a human heart, a NV sensor positioned on the heart surface should be sensitive to magnetic field $B$ = 14 nT, with a temporal resolution of about 0.1 ms (in order to be able to trace the time variation $\Delta t$ even in the case of the AP rapid rise), and a spatial resolution of about 10 $\mu$m$^3$ (which would allow a good reconstruction of the PA propagation, being the heart radius of about 40 mm \cite{trayanova1993response}). This corresponds to a minimum sensitivity:
\begin{equation}
\eta = \delta B_{min} \sqrt{\Delta t} \simeq 14\, nT \sqrt{0.1\, ms} \simeq 140\ pT\, Hz^{-1/2}
\end{equation}
This value can be considered a useful intermediate step for the application of the actual NV-based biosensing technologies, with the aim of reaching sensitivity that allows the  detection on neuronal signals.
\section{Methods and bio-applications}
To exploit NV-centers for biosensing, it is necessary to set up an optical microscope equipped with an ODMR apparatus featuring a non-resonant laser and a microwave antenna positioned near the diamond sample designed for a microwave source operating in the 2-4 GHz range, as described in section \textbf{2.2}.\\
The next sections are devoted to the presentation of several biosensing experiments exploiting NV centers in diamonds. 
Some of them are "proof-of-principle" in \textit{vitro} tests on a cell culture, while others are experiments carried on living organisms (in \textit{vivo} experiments).
The reviewed experiments will be divided in two categories describing respectively applications exploiting bulk diamond and experiments relying on nanodiamonds.
\subsection{Bulk diamond applications}
\begin{figure}[!h]\begin{center}
		\includegraphics[scale=0.3]{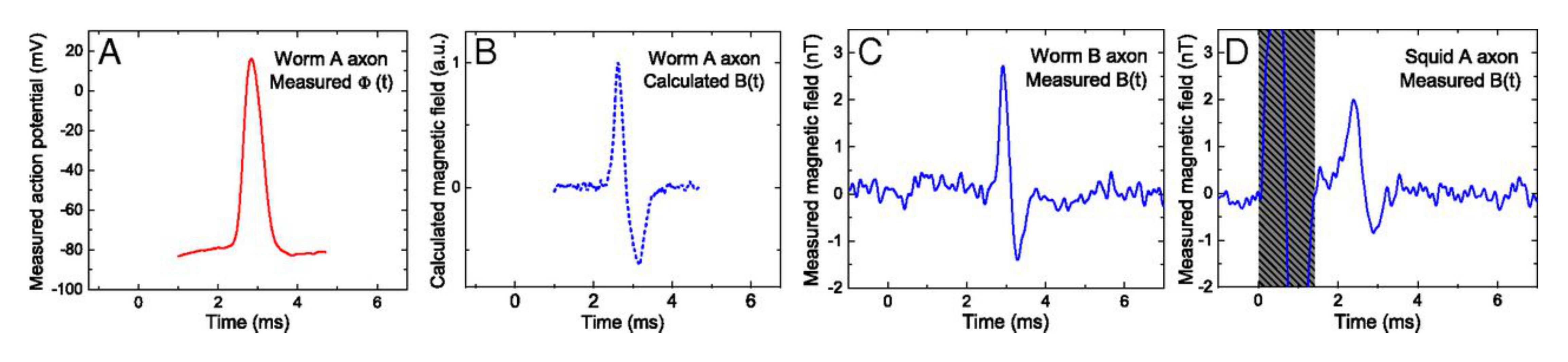}  
	\end{center}
	\caption{Measured AP voltage and magnetic field from excised single neurons, taken from the reference \cite{barry2016optical}. \\A) Measured time trace of intracellular axonal AP voltage $\Phi^{meas}_{in}(t)$ for giant axon from \textit{M. infundibulum} (worm). \\B) Calculated time trace of AP magnetic field $B(t)$ for \textit{M. infundibulum} extracted from data in A. \\C) Measured time trace of AP magnetic field $B(t)$ for \textit{M. infundibulum} giant axon with $N_{avg}$ = 600. \\D) Measured time trace of AP magnetic field $B(t)$ for \textit{L. pealeii} (squid) giant axon with $N_{avg}$ = 375. Gray box indicates magnetic artifact from stimulation current.}
	\label{fig:Barry1}
\end{figure}

\begin{figure}[!h] 	\begin{center}
		\includegraphics[scale=0.14]{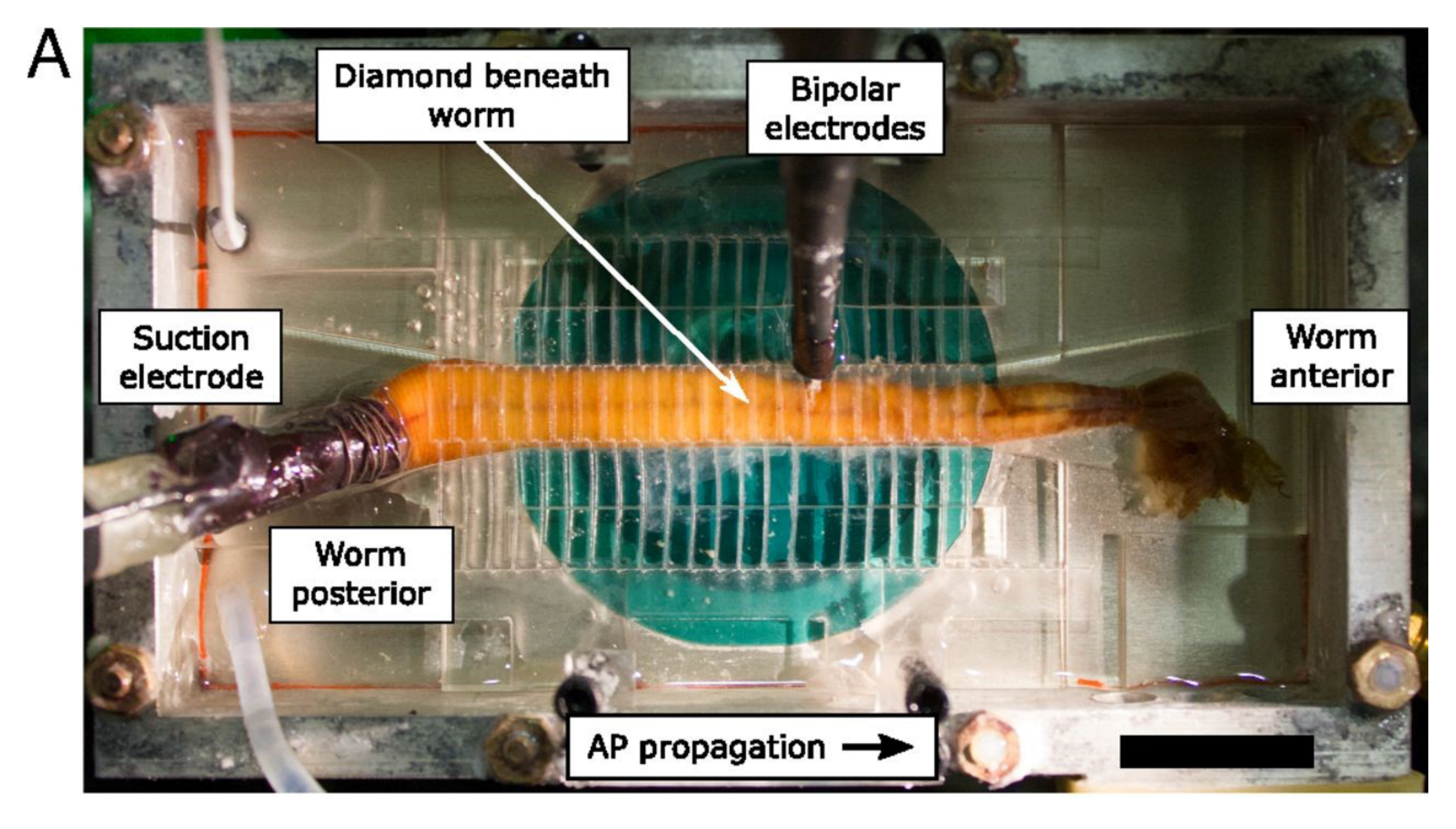}   \quad	\includegraphics[scale=0.2]{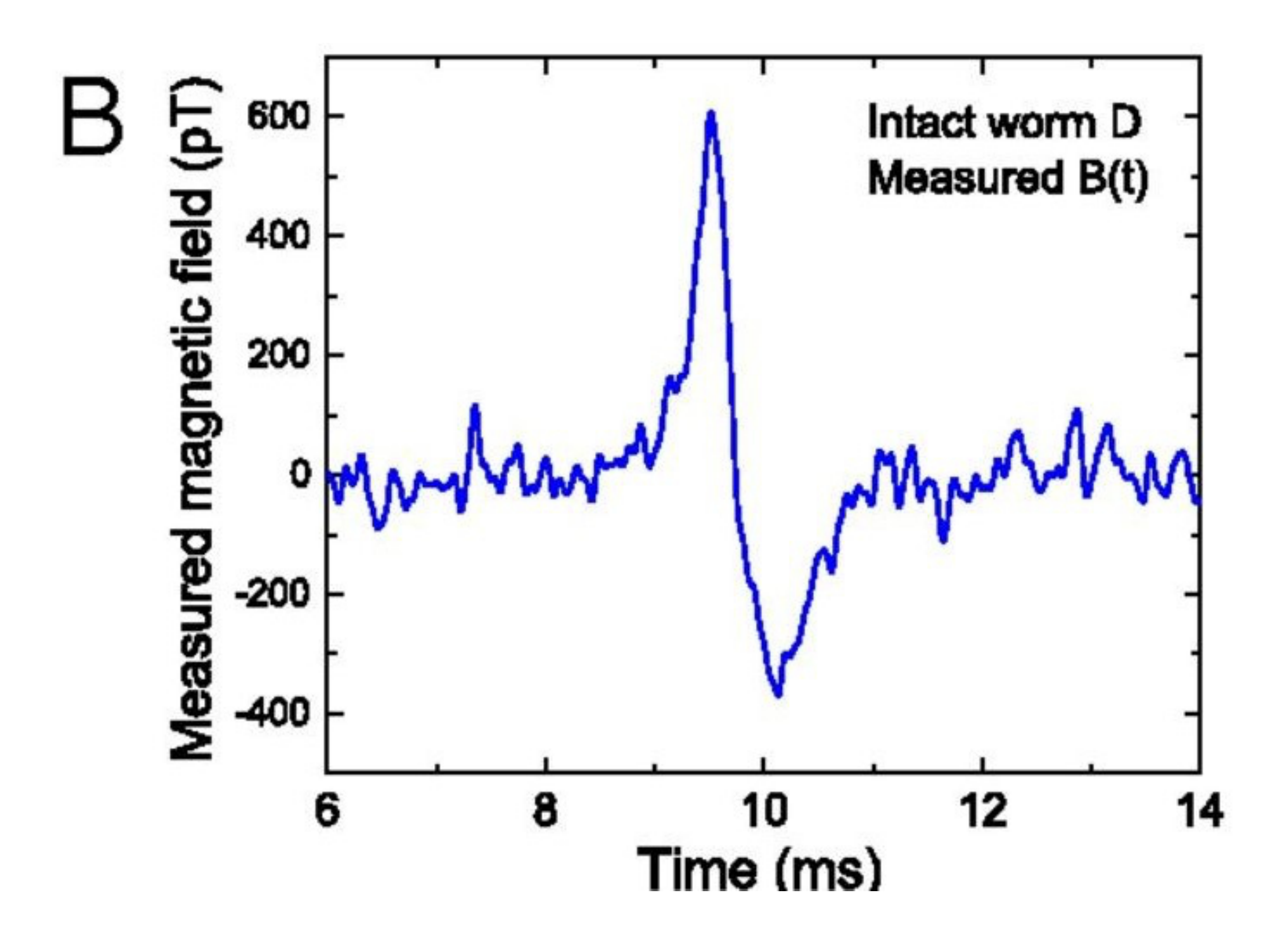} 
	\end{center}
	\caption{Single-neuron AP magnetic sensing exterior to live intact organism, taken from the reference \cite{barry2016optical}. \\
		A) Overhead view of intact living specimen of \textit{M. infundibulum} (worm) on top of NV diamond sensor. In configuration shown, animal is stimulated from posterior end by suction electrode, APs propagate toward worm’s anterior end, and bipolar electrodes confirm AP stimulation and propagation. (Scale bar 20 mm).\\ 
		B) Measured time trace of AP magnetic field $B(t)$ from live intact specimen of \textit{M. infundibulum} for N$_{avg}$ = 1,650 events.}
	\label{fig:Barry2}
\end{figure}

As discussed in the previous section, the electromagnetic fields produced by the excitable cells (as neurons, neuronal-like chromaffin cells, heart cells) even in mammals, are typically extremely weak (\textit{pT}). For this reason, several "proof-of-principle" experiments addressed measurements of fields produced by cells with peculiar electromagnetic properties.\\
Among the most suitable ones there are magnetotactic bacteria (MTB) \cite{proksch1995magnetic,qian2011magnetic,lam2010characterizing,dunin1998magnetic} containing magnetite ($^{III}Fe_2^{II}FeO_4$) or ferrite ($^{III}Fe_2^{II}FeS_4$) bacteria magnetic particles (BMP). The nanometer size of the BMPs is small to generate a single magnetic domain, but sufficient to create a permanent magnetic moment $\vec{\mu}_{BMP}$. This produces a cell magnetic moment $\vec{\mu}_{MTB}=\sum \vec{\mu}_{BMP}$, given by the sum of the BMP individual dipoles, which is exploited by the MTB to orient itself with respect to the earth's magnetic field \cite{komeili2012molecular,faivre2008magnetotactic}.\\
Among the various uses in the biomedical field, Sage et al. \cite{le2013optical} used \textit{Magnetospirillum magneticum AMB-1} for bio-magnetic imaging. The bacteria used in this work create magnetic nanoparticles with cubo-octahedral morphology and an average diameter of 50 nm. The experiment was performed both with bacteria dried on the surface of diamond chip implanted with NV centers, as well as with bacteria stored in phosphate-buffered saline (PBS) and laid on the chip surface (in \textit{vitro} experiment). \\The diamond sensor used to perform this experiment was a high-purity single-crystal diamond chip, with a 10 nm layer thickness of NV centers. The estimated surface density of nitrogen-vacancy centers was $3\cdot10^{17}$ cm$^{-3}$ in the case of experiment on bacteria in the liquid medium and $10^{18}$ cm$^{-3}$ for dry bacteria.\\
In the case of dry bacteria the objective was to demonstrate the possibility of measuring their static magnetic field, exploiting ODMR measurements at different bias magnetic field orientations ($B_{bias}$ = 3.7 mT). \\
In the case of live bacteria in the liquid medium, it was shown that it is possible to evaluate the magnetic field generated by the bacteria dipole $\vec{\mu}_{MTB}$ along the [111] crystallographic axis of the diamond, when also the bias magnetic field is oriented along it. Furthermore, cell viability was assessed immediately after magnetic imaging (lasting 4 minutes), using a standard fluorescence-based "live-dead" assay obtaining a viability of about 44$\%$. Cells mortality was attributed to the laser heating, since preliminary tests showed that 1 hour exposure to microwaves did not cause substantial cells mortality. Cells vitality was however partially preserved thanks to the strategy used to decouple laser light from the biological sample. Indeed in this set-up the laser impinged on diamond at an angle greater than the critical angle for the diamond–water interface, resulting in its total internal reflection within the diamond. \\
A wide field optical microscope was used for both MTB samples, with a field of view of $100\times30\,\,\mu m^2$ of the sample surface and a resolution of 400 nm. A CMOS camera was used to image the single magnetic nanoparticles inside the MTB. Their magnetic field was of the order of mT. Thanks to these measurements, the total magnetic moment $\vec{\mu}_{MTB}$ was determined by numerically fitting the modeled field distribution to the measured ones, with a mean value of $5\cdot10^{-17}$ m$^2$ A. \\
The magnetic field estimated from the ODMR measurements was compared with a scanning electron microscope (SEM) measurements. The position of nanoparticles revealed by the SEM was used to model the magnetic field they generated \cite{dunin1998magnetic,lam2010characterizing}. The two measurements were in excellent agreement and their values were compatible with the data reported in \cite{krichevsky2007trapping,moskowitz1993rock}. This highlights the potential of NV centers, able to perform sub-cellular magnetic field measurements at room temperature, allowing real-time imaging of magnetic dipole creation, single MTBs chain dynamics \cite{faivre2008magnetotactic} and magnetic particles formation in various organisms \cite{posfai2009magnetic,eder2012magnetic}.\\\\
Another "proof-of-principle" test was carried out by Davis et al.\cite{davis2018mapping} with the aim of measuring the magnetic field generated by iron oxide nanoparticles (IONs) incorporated in \textit{murine RAW 264,7} macrophages (a line established from a tumor induced by Abelson murine leukemia virus and often studied in relation to immune responses). The cells, after having phagocytized the iron ions (about 200 nm in size), were dried on the surface of a bulk diamond. High resolution magnetic imaging was performed exploiting ODMR measurements with a succession of bias magnetic field ($B_{bias}$= 10 mT) for each orientation of NV complex, for a total duration of 2 hours. Projection field maps were combined to form 3 orthogonal field maps, from which the cellular magnetic moment was obtained. For this experiment, the central dip of the hyperfine transition was used, which allowed to achieve a sensitivity of 17 nT at 1 $\mu$m in plane resolution, sufficient to reveal these magnetic nanoparticles. \\
To extend this technique to diagnostic imaging, Davis et al. performed NV magnetometry on liver specimens from a mouse model of hepatic iron overload, generated through intravenous administration of 900 nm IONs to C57bl/6 mice. To reduce the deposition of optical and thermal energy, the sample was illuminated for only 5 minutes with a duty cycle of 50\% and the ODMR technique was carried out with the bias magnetic field along only one of four NV axes. Furthermore, the laser beam was directed on the sample in total reflection mode. With these choices, time-lapse images of magnetic fields that coalesced within the macrophages after ION internalization were evaluated along that NV axes.
This experiment highlights the possibility of study the spatial distribution of iron deposits in the liver and other tissues. This has been a topic of interest in clinical literature as an indicator of disease state and the magnetic resonance imaging is becoming increasingly important in non-invasive quantification of tissue iron, overcoming the drawbacks of traditional techniques (liver biopsy) \cite{ghugre2011relaxivity}.\\\\
Barry et al. \cite{barry2016optical} studied individual neurons of marine worms (\textit{Myxicola infundibulum}) and squids (\textit{Loligo pealeii}).
The marine worm has a long axon \cite{nicol1948giant}, which stretches over its entire length (tens of mm and diameter of about 5 mm). The giant squid neuron (about 0.5 m long) did not extend over the entire length and is isolated following specific protocols \cite{song2013analysis}. An initial proof-of-principle test was performed on isolated neurons for both species.\\ 
The AP is stimulated by means of a current pulse, received by an electrode directly in contact with the neuron. The pulse was generated by a  current of about 10 mA, had a duration of about 1 ms and was repeated with a frequency of 0.4 Hz for the worm and 100 Hz for the squid. \\
The AP generation and its propagation was verified by micro-electrodes (see \textbf{Figure \ref{fig:Barry1}A}). 
From this axonal AP intracellular time trace, the shape of the associated magnetic field could be modeled \cite{swinney1980calculation,roth1985magnetic,wikswo1988magnetic} (see \textbf{Figure \ref{fig:Barry1}B}).
This was compared with the experimentally measured magnetic field, performed by means of the NV-based sensor in contact with the excised single neuron. Traces are shown in \textbf{Figure \ref{fig:Barry1}C} and\textbf{ $\,$\ref{fig:Barry1}D} respectively for the worm and the squid neuron. These measurements were performed using the ODMR technique at bias magnetic field $B_{bias}$ = 0.7 mT, oriented along two diamond axes and perpendicular to the axon axis (being the magnetic field generated by the AP pulse perpendicular to this last one). 
In Ref.\cite{barry2016optical}, Barry et al. carried on also a measurement on a living worm. The worm was directly fixed on the diamond and the distance between the neuron and the active NV layer was about 1.2 mm (see\textbf{ Figure \ref{fig:Barry2}A}). The magnetic field generated by the propagation of the AP pulse measured by ODMR technique is shown in \textbf{Figure \ref{fig:Barry2}B}. It is smaller than the one measured in the excised neuron, but its value is compatible with the increasing sensor distance. \\
The diamond sensor, exploited an electronic grade (N $<$ 5 ppb) single crystal chip, with a NV center layer of 13 $\mu m$. This layer had a NV centers density of $d=3 \cdot 10^{17}$ cm$^{-3}$ and a characteristic dephasing time $T_2^*$= 450 ns. The sensing volume is $V=5\cdot 10^{-6}$ cm$^{3}$, consequently the number of potentially stimulated centers was $n=15\cdot 10^{11}$. \\
Referring to the Equation (\textbf{\ref{projection}}), the fundamental sensitivity limit: $\eta_{q}\simeq$ 10 fT Hz$^{-1/2}$, while the sensitivity reached experimentally was $\eta \simeq$ 15 pT Hz$^{-1/2}$, allowing, anyway, a reliable measure of the magnetic fields generated by these animal species (of the order of nT). Further development should be needed for revealing those of human neurons (of the order of pT).
\subsection{Nanodiamonds}
The techniques for the creation of NV centers in diamond are well established also for nanodiamonds (NDs). Nanodiamonds-based sensors exploit colloidal suspensions of single diamond particles of minimum diameter of 4-5 nm, but on average the nanodiamonds typically used in experiment have a size of 50-100 nm. The nanometer size makes nanodiamonds-based sensor of extreme interest for bio-sensing application, as they are potentially usable in \textit{vivo} experiments. Nonetheless, they have also important drawbacks such as e.g. the increased sensitivity of NV spins to environmental noise. Indeed, while in a bulk diamond the coherence time $T_2^*$ is mainly influenced by the electronic impurities and nuclear spins in the surrounding, for nanodiamonds the coherence time is further reduced due to the surface spin noise. This should be taken into account in the estimation of the sensitivity limit (see Equation (\textbf{\ref{projection}})). \\
Despite this limitation, nanodiamonds have attracted interest also as a non-toxic alternative to quantum dots for biomedical imaging, as magnetic sensors and finally as drug transporters (thanks to the discovery of the possibility to functionalize the diamonds surface in various ways, exploiting the covalents carbon bonds). The great interest and the exceptional range of applications of NDs is boosting the development of novel fabrication techniques, even if the actual technologies are already able to provide very pure nanodiamonds with controlled surface chemistry at a relative low cost \cite{markham2011cvd,narayan2017novel}.

\subsubsection{Biocompatibility and functionalization studies}
To understand the perspective in bio-medical application, deep investigation of NDs biocompatibility is required. More specifically, it is important to understand their diffusion in tissues and their acute and long-term biological effect.
While bulk diamonds are non-toxic and inert, NDs interaction with cells should be carefully investigated \cite{mochalin2012properties,schrand2007diamond}. There is a huge variety of nanodiamond specimens, which differ in dimensions, functionalized surface and potential interaction with the biological sample. \\
Various experiments were therefore conducted to evaluate the cell viability, e.g. in HeLa cells \cite{neugart2007dynamics,mcguinness2011quantum,hsiao2016fluorescent} (a cell line deriving from tumoral human cells), in human neurons \cite{simpson2017non,guarina2018nanodiamonds}, in human trachea \cite{yuan2010pulmonary}, in the translucent \textit{Caenorhabditis elegans} worm \cite{mohan2010vivo} and intravenous infusions \cite{chow2011nanodiamond}. 
Briefly, the nanodiamonds of size between 50 and 100 nm have been found to be incorporated by the cells, without producing a significant damage.\\
In particular in Guarina et al. \cite{guarina2018nanodiamonds} an ODMR detection scheme with NV centers in nanodiamonds internalized in hippocampal neurons was performed in suitable conditions (3 mW of excitation power, -20 dBm of continuous-wave MW power). This experiment demonstrated that neuron functionality was not significantly affected by the implementation of the measurement protocol: their spontaneous firing (bursts synchronization) was preserved, as well as the amplitude of spontaneous inhibitory and excitatory events. Even thought some alteration both at the single-cell level and in neuronal networks was observed, this was principally attributed to the effects of nanoparticles aggregation.
The aim of the work was to assess the feasibility of in \textit{vitro} imaging and targetable drug delivery via nanodiamonds, but the same argument holds for the other sensing applications. 
Furthermore, if properly functionalized, the NDs can anchor themselves to the surface of the cell sample in the targeted areas.

\subsubsection{Nanodiamonds applications}
\begin{figure}[!h]\begin{center}
		\includegraphics[scale=0.3]{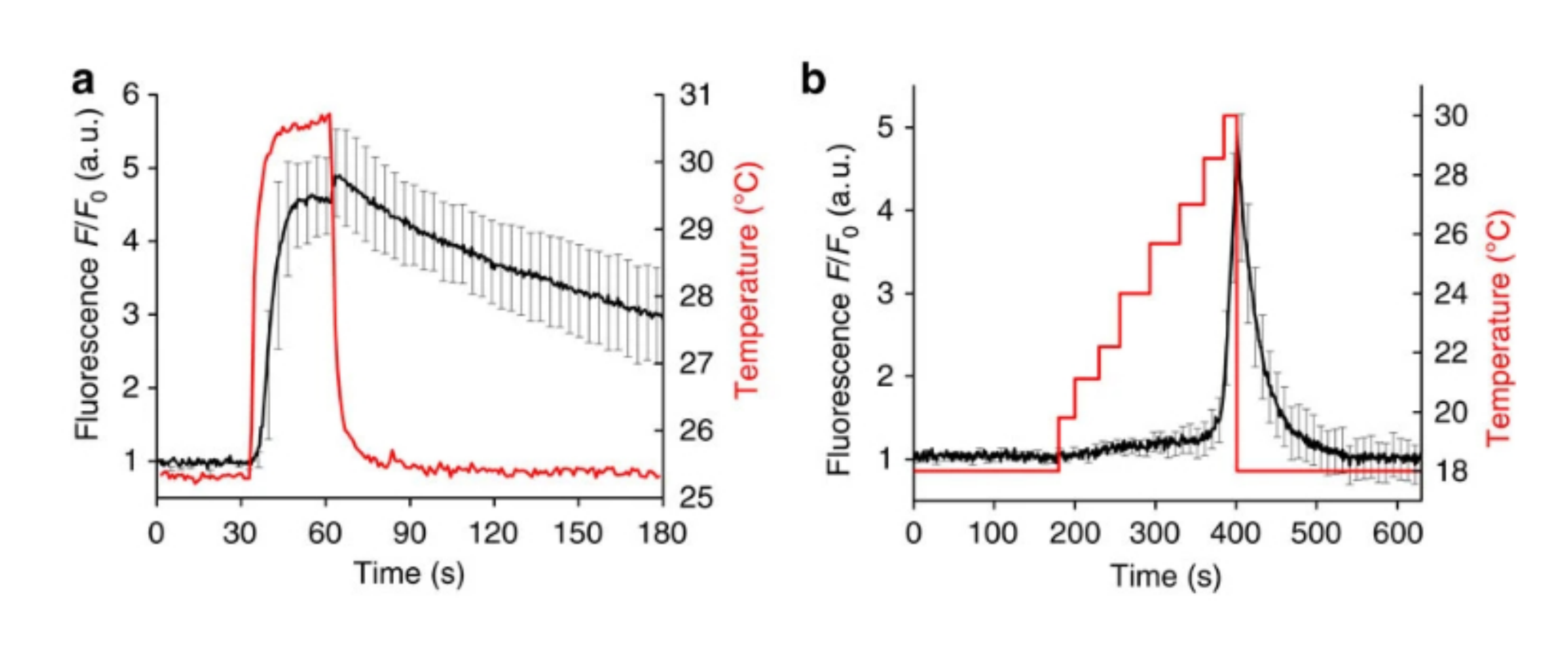} \quad  \includegraphics[scale=0.3]{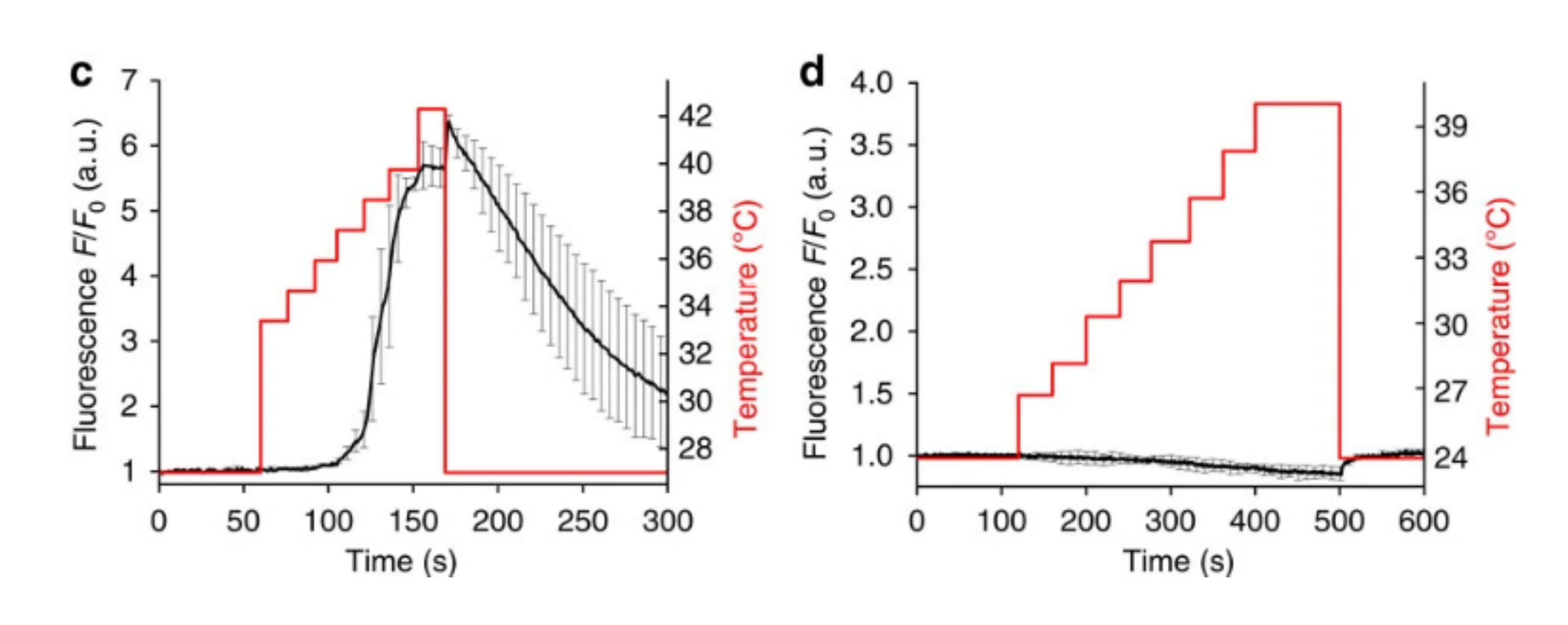} 
	\end{center}
	\caption{Activation of snake TRPA1 in cells expressing TRPA1-IRES-EGFP using femtosecond IR laser pulses, taken from the reference  \cite{ermakova2017thermogenetic}.\\
		a) R-GECO1.1 fluorescence (black line) reflects $Ca^{2+}$ dynamics in the cytoplasm with the 20 mW laser beam turned on at $t$=30 s and off at $t$=60 s. \\
		b,c) With the temperature of HEK293 cells expressing snake TRPA1 increased in a stepwise fashion using properly adjusted IR laser radiation, the activation thresholds of caTRPA1 b) and eolTRPA1 c) were determined. \\
		d) A similar heating of control cells does not induce $Ca^{2+}$ elevation. The black line is the fluorescence response. The red line is the temperature in the medium.}
	\label{fig:Fedotov}
\end{figure}

Once the biocompatibility of nanodiamonds is assessed, it is necessary to understand to which extent the sensing techniques developed for sensor based on NV in bulk diamond can be extended to nanodiamonds based sensors, functionalized and incorporated in the cells of interest.\\
A proof-of-principle demonstration of quantum control techniques to map the intracellular temperature of a neuronal network was performed by Simpson et al. \cite{simpson2017non}.
The NDs were dispersed in cell media in concentration of $6\,\,\mu g$/$ml$, sonicated for few minutes, and then applied to the primary cultures during a routine change of cell media.
The $170\,\, nm$ diameter NDs contained approximately 500 NV centers each.\\
Using ODMR techniques in combination with standard wide-field microscopy with a field of view of $80\times80\,\,\mu m^2$ was possible to observe NV resonance frequency in only 6 seconds. Specifically, in Ref.\cite{simpson2017non} the ODMR signal presented two fluorescence \textit{dips} (see section \textbf{2.4}) because of the strain. 
This effect is negligible in bulk diamonds while nanodiamonds crystal lattice suffers strong deformation inducing line splitting.
In that paper the two \textit{dips}, spaced by few MHz, were modeled as a single one with higher spectral broadening. 
By interpolating the ODMR graph with a \textit{Lorentzian} function, it was estimated the mean crystal field splitting $D_{gs}$= (2868.59$\pm$0.17) MHz.\\
To demonstrate the NV thermo-sensor performance in biological measurement, the temperature of the neuronal solution was reduced by $ 1.9$ °C. 
Repeating the ODMR analysis for a total acquisition time of $12\,\,s$, a resonance frequency shift was observed. 
The respective temperature variation was estimated using the temperature coupling coefficient $dD /dT \simeq$ -74 kHz K$^{-1}$ (see section \textbf{2.5)}. The distribution reported a mean temperature change of $(−1.36 \pm 0.08)$°C, consistent with the reduction in environmental temperature. \\
We underline that NDs allowing to create spatial maps of the temperature inside the cells will bring new insight on the understanding of cell activity.
There are many biological processes whose knowledge would be enriched by nanoscale thermometry, such as temperature increases following the opening of ion channels \cite{chen1995hydrodynamic}, or the correlation of temperature changes and the occurrence of neurological disorders and pathological conditions.\\\\
Another biological application was reported by Ermakova et al.\cite{ermakova2017thermogenetic}, using nanodiamonds with NV centers as thermo-sensors, exploits optically-induced thermal gradients for thermogenetic neural modulation \cite{lima2005remote,boyden2005millisecond}. 
This thermal gradient is generated at the transient receptor potential channels (TRP channels): a group of ion channels that are commonly present on the plasma membrane of numerous types of animal cells \cite{bernstein2012optogenetics}. A particular specialized form of these ion channels appears to be highly sensitive to temperature changes \cite{bath2014flymad}. Some species of snakes can use TRP channels to detect the thermal build-up caused by infrared IR radiation emitted by nearby prey, allowing them to estimate the direction and distance of the (IR) source \cite{newman1982infrared}.\\
To experimentally recreate this local temperature change and therefore study the TRPs response, Ermakova et al. used IR short pulsed laser. This method, with respect to conventional techniques as environmental heating \cite{hamada2008internal} or TRPs chemical agonists \cite{chen2016trp}, allows cellular spatial resolution and ultrahigh temporal resolution.
The precise temperature control was performed by varying the laser intensity, whose actual thermal impact was monitored by the nitrogen-vacancy complex. This quantum probe (whose dimension was about 300 nm) was integrated on the tip of an optical fiber, together with a microwave antenna. The optical fiber was positioned near the cell irradiated by the IR laser, allowing a measurement of its temperature by the ODMR technique. \\
In this experiment \cite{ermakova2017thermogenetic} it was initially evaluated the thermal stimulation via IR laser of two TRP channels of the snake. The TRP channels considered were the \textit{Crotalus atrox TRPA1} (caTRPA1) and the \textit{Elaphe obsoleta lindheimeri TRPA1} (eolTRPA1). Fluorescent proteins (caTRPA1-IRES-EGFP) had been added to the channels, allowing to monitor the opening and closing of the calcium channels. Thanks to the NV sensors and by slowly changing the cell temperature with properly tuning pulsed laser intensity, it was possible to obtain the threshold temperature, inducing opening of the calcium channels. The threshold temperatures were found to be $T_0$=(27.8 ± 0.6) °C for caTRPA1  (see \textbf{Figure \ref{fig:Fedotov}b}) and $T_0$=(38.5 ± 0.7) °C for eolTRPA1 (see \textbf{Figure \ref{fig:Fedotov}c}).\\
Once estimated the threshold temperature $T_0$, Ermakova et al. proved the technique on other biological samples: mouse neurons and zebrafish larvae, whose thermogenetic activation is induced by TRPA1 channels causing responses.\\
In the case of caTRPA1 channels, the cultured neurons were maintained at a temperature of 27 °C lower than the threshold temperature obtained before for this channel; in the case of eolTRPA1-expressing, neurons were kept at basal temperatures of 35.5 °C.
As expected, they found that the thermal increase induced by the IR laser activates the TRPs channels triggering the generation of the neuronal AP, measured through conventional electrophysiological techniques.\\
When a measurements on live samples is considered, the sample can no longer be kept at the desired temperature, therefore it is necessary to choose the TRP channel suitable for body temperature of the animal species analyzed. As for the zebrafish neurons, whose body temperature is found to be 26 °C, the eolTRPA1 channels may be suitable. As for the mammalian brain, the perfect TRP candidate has still to be found. For example, the mouse body temperature is too close to the threshold temperature of eolTRPA1 and it may be desensitized.\\
The results of the application of this technique in living zebrafish showed that it is possible to thermogenetically activate neurons using the IR laser. In particular, the technique demonstrated a spatial resolution of 60 $\mu$m (fiber size in which the IR laser was focused on the sample), allowing one or few neurons to be stimulated. As for the IR laser intensity, Emarkova et al. observed that 30 mW laser power induced the escape behavior exhibition of 93$\%$ of the larvae. 
NV-based temperature sensors allowed careful monitoring of the temperature reached by the cells with high spatial resolution and temperature sensitivity up to 0.1 °C. To preserve cellular integrity and to avoid cell ablation \cite{chen2016trp} is essential to heat-up the tissues by a few degrees only and for a time interval not exceeding a few minutes.
\\
Finally, another demonstration of the effectiveness of NV nanosensors for thermometry comes from Fujiwara et al.'s experiment \cite{fujiwara2020realtime}. A first test allowed to measure the temperature dynamics inside live \textit{C. elegans} adults worms during environmental temperature changes. The sensitivity reached was 1.4 °C Hz$^{-1/2}$. Having obtained this result, Fujiwara et al. successfully determined the temperature increase caused by the worm's thermogenesis under the
treatment of mitochondrial uncoupler stimuli.
\section{Techniques for improving ODMR sensitivity}
\begin{figure}[!h]\begin{center}
		\includegraphics[scale=0.28]{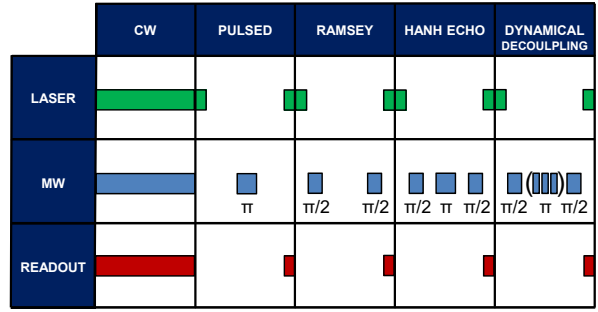}
	\end{center}
	\caption{Scheme of timing and duration of laser pulses, microwave pulses and reading sequences associated with the most common measurement protocols of the external fields for the NV complex.}
	\label{fig:tecniche}
\end{figure}

\begin{figure}[!h]\begin{center}
		\includegraphics[scale=0.3]{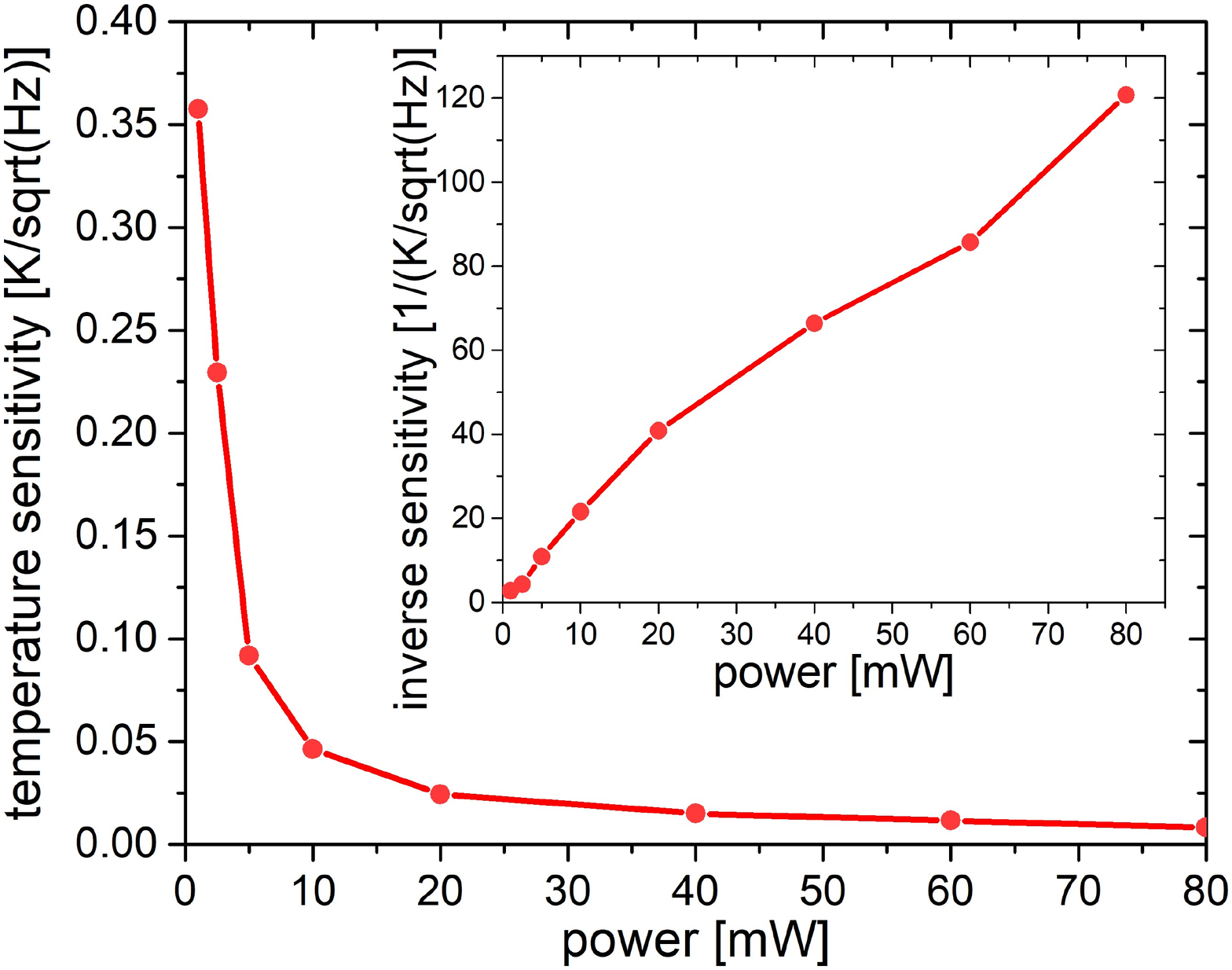}  
	\end{center}
	\caption{Temperature sensitivity versus the laser excitation power at 532 nm. The inset shows the inverse of the thermal sensitivity versus the excitation laser power.}
	\label{fig:sensVSpower}
\end{figure}

In this section we discuss some technological solutions to improve the sensitivity of NV-based sensors as well as the precautions to be taken when they are used as a bio-sensor.\\
Equation (\textbf{\ref{projection}}) provides the ultimate sensitivity limit reachable highlighting that the number $n$ of NV centers and their decoherence time $T_2^*$ play a key role. 
To increase $n$, while maintaining the same spatial resolution, it is necessary to have diamonds with an increased NV centers density. This can be achieved by enhancing the number of nitrogen implanted in the diamond and improving the N-to-NV conversion efficiency, minimizing the concentration of residual paramagnetic substitutional nitrogen \cite{acosta2009diamonds}. 
In parallel, to increase $T_2^*$, it is also recommended the production of ultra-pure diamonds, with reduced unwanted electronic impurities (e.g. the P1 centers) and nuclear spins impurities (e.g. the paramagnetic $^{13}C$ isotopes, whose natural abundance is about 1.1$\%$) \cite{balasubramanian2009ultralong,achard2020cvd,bernardi2017nanoscale}. 
It is important to note that the NV density increase will necessary worsen the decoherence time of the NVs themselves, because of their mutual interaction. Consequently, an optimal trade-offs between these parameters must be sought.\\\\
In addition to the NV-density and diamond sample engineering, the sensitivity can be improved by implementing specific experimental techniques, that are based on laser and microwave pulses of particular duration, synchronized appropriately \cite{schirhagl2014nitrogen,levine2019principles,barry2020sensitivity}.\\
For example, if an unknown electromagnetic field, responsible for the ODMR resonance frequency shift, is constant or slowly varying, it is possible to adopt the experimental \textit{pulsed} ODMR protocols \cite{dreau2011avoiding} or the \textit{Ramsey} method \cite{taylor2008high} instead of the continuous wave (\textit{CW}) ODMR \cite{shin2012room}. 
The \textit{CW} ODMR is the simplest and most widely employed magnetometry method with NV-based sensors, wherein the microwave driving and the optical polarization and readout (laser pumping) occur simultaneously. 
Although this technique is easy to be implemented, the relative ODMR spectrum dips are affected by the broadening induced by the continuous exposure of the laser beam and microwave field on the sample. 
With \textit{pulsed} ODMR techniques this broadening effect is substantially suppressed, allowing to obtain a narrower ODMR spectrum dips and therefore to improve the measurement sensitivity. This protocol uses temporally separated optical laser initializations, $\pi$ microwave control pulses, and laser readout pulses.
The $\pi$ pulses, whose name derives from the representation of the process on the \textit{Bloch} sphere, is an oscillating microwave field that brings the electronic state from the state $|m_s=0>$ to $|m_s=\pm 1>$.
\textit{Ramsey} ODMR spectroscopy, on the other hand, consists on in the application of two $\pi/2$ pulses, separated by a time $\tau$.
Also the $\pi/2$ pulse is an oscillating microwave field that brings the electronic state from the state $|m_s=0>$ to a balanced superposition of $|m_s=+1>$ and $|m_s=-1>$.
By varying the time $\tau$, the so-called "Ramsey fringes" are obtained, from which it is possible to extrapolate an estimation of the magnetic fields amplitude.
Also this technique allows sensitivity improvement with respect to the \textit{CW}: the decoupling of the MW from the laser power allows increasing the MW power improving the contrast, without degrading the FWHM.\\
In the case of time-varying electromagnetic fields, there are other even more complex microwave pulse sequences, capable of decoupling the measurement from surrounding spin environment \cite{bauch2019decoherence}. In this way the decoherence time of the NV centers increases and consequently it becomes possible to interrogate the quantum system for longer times, improving the measurement statistic and therefore the sensitivity. 
One of these experimental protocols is the \textit{Hahn Echo} sequence  \cite{grinolds2013nanoscale,mcguinness2011quantum}, which refocuses the dephasing NVs spin, applying an additional $\pi$ pulse in the middle of \textit{Ramsey} sequence. The characteristic time of the spin coherence decay, measured with this protocol, is called $T_2$ and it is typically one or two orders of magnitude longer than $T_2^*$.
Even more complex dynamic decoupling sequences, which apply multiple refocusing $\pi$ pulses further improving the decoherence time $T_2$ have been devised \cite{de2010universal,szankowski2017environmental,naydenov2011dynamical,knowles2014observing}. 
Among these, the most famous are the \textit{Carr-Purcell-Meiboom-Gill} (CPMG) \cite{de2010universal,szankowski2017environmental,naydenov2011dynamical,knowles2014observing}
and the \textit{XY8} sequences \cite{de2010universal,cao2020protecting}, which differ in the rotation axes (around which the spin rotates): the first method applies the pulses along the same axis, while the second chooses a different one for each $\pi$ pulses.
It is useful to underline that, although these techniques allow to extended the coherence time of the NV centers, they cannot go beyond the spin-lattice relaxation time $T_1$, that for an NV spin ensemble in bulk diamond is about 3 ms \cite{mrozek2015longitudinal}.\\
The \textbf{Figure \ref{fig:tecniche}} briefly summarizes the above mentioned pulse sequences.
\\
It is useful to underline that the sensitivity formula in Equation (\textbf{\ref{projection}}) describes an idealized measurement with a perfect readout mechanism.
On the contrary, typically the readout mechanism adds noise in the measurement, that can be described introducing, in the previous equation, the spin-readout fidelity factor $\mathcal{F}$ \cite{barry2020sensitivity}:
\begin{equation}
\eta=\frac{1}{\gamma_e \sqrt{nT^*_2}}\frac{1}{\mathcal{F}}
\end{equation}
Keeping the usual optical-readout, but improving the photon collection is expected to increase $\mathcal{F}$ (see ref.\cite{radtke2019nanoscale} for different methods to improve photon collection). 
Ancilla-assisted repetitive readout, which is based on mapping the NV spin state to the nuclear spin state, also improves $\mathcal{F}$ \cite{barry2020sensitivity}. Finally, in a more far perspective, quantum methods of noise reduction can be applied \cite{bar2012suppression, zwick2016maximizing,muller2020noise,arenz2016universal}.\\\\
When the ultimate goal is bio-sensing, some constraints rise limiting the implementation of the above described pulse sequences. 
One constraint is the frequency bandwidth. In fact, the dynamic decoupling techniques mentioned above are capable of measuring time-varying external field only if this time variation is of the order of the time interval separating the $\pi$ pulses. 
Furthermore, in order to control the system quantum state, the time between these pulses cannot exceed the coherence of the NV center. Consequently, the frequency of the signal to be measured must be of the order of the coherence time of the NV centers. In the biological case, the electromagnetic fields pulse lasts about 1 ms. This value is very far from $T_2$, marking a boundary for the use of these techniques in biological applications.
Another constraint is associated to the optical laser power.
The higher the laser power the better the sensitivity in measurements with ensembles, since it increases the percentage of excited centers and consequently the fluorescence signal. However precautions must be taken to avoid cells and proteins damaging. An efficient solution can be to direct the laser beam towards the diamond sample at an angle allowing total reflection (\textit{Brewster} angle). In this way only the fluorescence emitted by the NV centers travels through the cells, placed on the other diamond surface \cite{barry2016optical,le2013optical}.
In this case, however, precise control over sensing volume would be lost, deteriorating spatial resolution. In a standard configuration, where the laser impinges perpendicularly on the sample, it is necessary to limit the optical power reaching the cells to few mW. 
In this regard, \textbf{Figure \ref{fig:sensVSpower}} shows a sensitivity curve versus the laser optical power, obtained by adopting the technique described in Moreva et al.\cite{PhysRevApplied.13.054057}.
As anticipated in the introduction, the application of a transverse bias magnetic field $B^{\perp}_{bias} \simeq 3\,\,mT$, allows to improve the sensitivity of a the NV center based thermo-sensor with respect to other standard techniques in CW regime. In Ref.\cite{PhysRevApplied.13.054057}, the temperature sensitivity reached is $\eta \simeq 4.8\,\, mK/Hz^{1/2}$ in a sensing volume of $1\mu m^3$, obtained at a power level (80 mW) that can present biocompatibility problems. 
However, the sensitivity obtained is even beyond the one required to monitor biological mechanisms, usually requiring sensitivities of the order of 1 °C. Figure \ref{fig:sensVSpower} shows that it is possible to perform the temperature measurement with a lower laser power, finding an ideal compromise between the temperature sensitivity and laser intensity impinging on the cell sample. 
Indeed, with a power of a few mW it is already possible to discriminate biological processes with a sensitivity of the tenth of a degree.
\section{Conclusion}
Sensors based on the NV centers in artificial diamonds are one of the emerging quantum technologies of huge potential interest in biological applications, thanks to both their practicality and their technical performances. 
In fact, the capability to initialize and read out optically the spin state at room temperature, makes the use of this quantum sensor convenient and powerful even for biological applications.
Furthermore, the levels of sensitivity and spatial resolution achieved are extremely high, which in principle allows potential application towards the detection of very weak electromagnetic fields as the one generated by mammalian, and potentially human, cells.
Even if an eventual use of NV sensors for the detection of biological electric fields is more problematic due to its weak coupling constant, regarding the magnetic field sensing and especially temperature measurements astonishing results have already been achieved.  
Indeed, the thermal gradients generated by biological phenomena can be reliably observed thanks to the actual sensitivity of the NV-based sensors.
This is of the utmost importance because localized intracellular temperature gradients may affect neuronal functionality (including vesicular dynamics and neurotransmitter release) or  may provide indirect measurement of mitochondrial activity \cite{stanika2012comparative,humeau2018calcium}. 
Regarding the detection of bio-magnetic fields, the NV-based sensors have already shown good results with peculiar biological cells, presenting either an intrinsic magnetic field (magnetitotactic bacteria) or a generated magnetic field in axon of squids or long worms, much larger than the one generated in the human ones.
The improvement of these devices suggests the possibility of exploiting NV-based sensors also for the detection of weaker but more fascinating biological magnetic fields. 
In particular, an estimate of the cardiac magnetic field that is generated on the heart surface was here reported. This value is in the range of present measurement capability exploiting the NV center properties, exploiting optimized diamond sample engineering and the adoption of pulsed measurement protocols in order to improve the diamond coherence time. 
Furthermore, we have analyzed the magnetic field associated to human neuron activity. The weakness of these fields requires further improvements of the measurement technique in the case of the single AP, while measurement of clustered channels is likely a reasonable target for the actual technology. 
However, the considerable interest in the neuronal field detection as diagnostic and therapeutic tools for neurodegenerative diseases and aging effects, together with the recent years progress of these techniques (partially covered by this review), is expected to boost the technological developments and eventually the market success of quantum assisted biosensing based on NVs.

\medskip
\textbf{Acknowledgements} \par 
This work has received funding from the European Union’s PATHOS EU H2020 FET-OPEN grant no. 828946 and
Horizon 2020, from the EMPIR Participating States in the context of the projects EMPIR-17FUN06 ”SIQUST” and
from the project Piemonte Quantum Enabling Technologies (PiQuET) funded by the Piemonte Region.

\bibliographystyle{unsrt} 
\bibliography{review_final}

\begin{thebibliography}{100}

\bibitem{mathias1990limitations}
Richard~T Mathias, Ira~S Cohen, and Carlos Oliva.
\newblock Limitations of the whole cell patch clamp technique in the control of
  intracellular concentrations.
\newblock {\em Biophysical journal}, 58(3):759, 1990.

\bibitem{fejtl2006micro}
Michael Fejtl, Alfred Stett, Wilfried Nisch, Karl-Heinz Boven, and Andreas
  M{\"o}ller.
\newblock On micro-electrode array revival: its development, sophistication of
  recording, and stimulation.
\newblock In {\em Advances in network electrophysiology}, pages 24--37.
  Springer, 2006.

\bibitem{spira2013multi}
Micha~E Spira and Aviad Hai.
\newblock Multi-electrode array technologies for neuroscience and cardiology.
\newblock {\em Nature nanotechnology}, 8(2):83, 2013.

\bibitem{vengalattore2007high}
M~Vengalattore, JM~Higbie, SR~Leslie, J~Guzman, LE~Sadler, and DM~Stamper-Kurn.
\newblock High-resolution magnetometry with a spinor bose-einstein condensate.
\newblock {\em Physical review letters}, 98(20):200801, 2007.

\bibitem{faley2006new}
MI~Faley, U~Poppe, K~Urban, DN~Paulson, and RL~Fagaly.
\newblock A new generation of the hts multilayer dc-squid magnetometers and
  gradiometers.
\newblock In {\em Journal of Physics: Conference Series}, volume~43, page 1199.
  IOP Publishing, 2006.

\bibitem{baudenbacher2003monolithic}
F~Baudenbacher, LE~Fong, JR~Holzer, and M~Radparvar.
\newblock Monolithic low-transition-temperature superconducting magnetometers
  for high resolution imaging magnetic fields of room temperature samples.
\newblock {\em Applied Physics Letters}, 82(20):3487--3489, 2003.

\bibitem{knappe2010cross}
Svenja Knappe, Tilmann~H Sander, Olaf Kosch, Frank Wiekhorst, John Kitching,
  and Lutz Trahms.
\newblock Cross-validation of microfabricated atomic magnetometers with
  superconducting quantum interference devices for biomagnetic applications.
\newblock {\em Applied Physics Letters}, 97(13):133703, 2010.

\bibitem{nakajima2016application}
Anri Nakajima.
\newblock Application of single-electron transistor to biomolecule and ion
  sensors.
\newblock {\em Applied Sciences}, 6(4):94, 2016.

\bibitem{maestro2010cdse}
Laura~Martinez Maestro, Emma~Martin Rodriguez, Francisco~Sanz Rodriguez,
  MC~Iglesias-de la~Cruz, Angeles Juarranz, Rafik Naccache, Fiorenzo Vetrone,
  Daniel Jaque, John~A Capobianco, and Jose~Garcia Sole.
\newblock Cdse quantum dots for two-photon fluorescence thermal imaging.
\newblock {\em Nano letters}, 10(12):5109--5115, 2010.

\bibitem{brites2016lanthanides}
CDS Brites, A~Mill{\'a}n, and LD~Carlos.
\newblock Lanthanides in luminescent thermometry.
\newblock In {\em Handbook on the Physics and Chemistry of Rare Earths},
  volume~49, pages 339--427. Elsevier, 2016.

\bibitem{hines1993neuron}
Michael Hines.
\newblock Neuron—a program for simulation of nerve equations.
\newblock In {\em Neural systems: Analysis and modeling}, pages 127--136.
  Springer, 1993.

\bibitem{hines1997neuron}
Michael~L Hines and Nicholas~T Carnevale.
\newblock The neuron simulation environment.
\newblock {\em Neural computation}, 9(6):1179--1209, 1997.

\bibitem{santamaria2009hodgkin}
F~Santamaria and JM~Bower.
\newblock Hodgkin-huxley models.
\newblock In {\em Encyclopedia of Neuroscience}, pages 1173--1180. Elsevier
  Ltd, 2009.

\bibitem{arcas2003computation}
Blaise Ag{\"u}era~y Arcas, Adrienne~L Fairhall, and William Bialek.
\newblock Computation in a single neuron: Hodgkin and huxley revisited.
\newblock {\em Neural Computation}, 15(8):1715--1749, 2003.

\bibitem{recording1995edited}
Single-Channel Recording.
\newblock edited by b. sakmann and n. neher, 1995.

\bibitem{waxman1998demyelinating}
Stephen~G Waxman.
\newblock Demyelinating diseases—new pathological insights, new therapeutic
  targets, 1998.

\bibitem{kiyatkin2018brain}
Eugene~A Kiyatkin.
\newblock Brain temperature: from physiology and pharmacology to
  neuropathology.
\newblock In {\em Handbook of clinical neurology}, volume 157, pages 483--504.
  Elsevier, 2018.

\bibitem{maurer2013nanometer}
PC~Maurer, M~Kubo, MD~Lukin, HJ~Noh, G~Kucsko, NY~Yao, PK~Lo, and H~Park.
\newblock Nanometer scale quantum thermometry in a living cell.
\newblock 2013.

\bibitem{bai2016micro}
Tingting Bai and Ning Gu.
\newblock Micro/nanoscale thermometry for cellular thermal sensing.
\newblock {\em Small}, 12(34):4590--4610, 2016.

\bibitem{wu2016diamond}
Yuzhou Wu, Fedor Jelezko, Martin~B Plenio, and Tanja Weil.
\newblock Diamond quantum devices in biology.
\newblock {\em Angewandte Chemie International Edition}, 55(23):6586--6598,
  2016.

\bibitem{beveratos2001nonclassical}
Alexios Beveratos, Rosa Brouri, Thierry Gacoin, Jean-Philippe Poizat, and
  Philippe Grangier.
\newblock Nonclassical radiation from diamond nanocrystals.
\newblock {\em Physical Review A}, 64(6):061802, 2001.

\bibitem{beveratos2002room}
Alexios Beveratos, Sergei K{\"u}hn, Rosa Brouri, Thierry Gacoin, J-P Poizat,
  and Philippe Grangier.
\newblock Room temperature stable single-photon source.
\newblock {\em The European Physical Journal D-Atomic, Molecular, Optical and
  Plasma Physics}, 18(2):191--196, 2002.

\bibitem{zaitsev2013optical}
Alexander~M Zaitsev.
\newblock {\em Optical properties of diamond: a data handbook}.
\newblock Springer Science \& Business Media, 2013.

\bibitem{bernardi2020biocompatible}
Ettore Bernardi, Ekaterina Moreva, Paolo Traina, Giulia Petrini,
  Sviatoslav~Ditalia Tchernij, Jacopo Forneris, Zelijko Pastuovic, Ivo~Pietro
  Degiovanni, Paolo Olivero, and M~Genovese.
\newblock Biocompatible technique for nanoscale magnetic field sensing with
  nitrogen-vacancy centers.
\newblock {\em arXiv preprint arXiv:2005.13230}, 2020.

\bibitem{yu2005bright}
Shu-Jung Yu, Ming-Wei Kang, Huan-Cheng Chang, Kuan-Ming Chen, and Yueh-Chung
  Yu.
\newblock Bright fluorescent nanodiamonds: no photobleaching and low
  cytotoxicity.
\newblock {\em Journal of the American Chemical Society}, 127(50):17604--17605,
  2005.

\bibitem{barone2019long}
Frank~C Barone, Cezary Marcinkiewicz, Jie Li, Yi~Feng, Mark Sternberg, Peter~I
  Lelkes, David Rosenbaum-Halevi, Jonathan~A Gerstenhaber, and Giora~Z
  Feuerstein.
\newblock Long-term biocompatibility of fluorescent diamonds-(nv)-z\~{} 800 nm
  in rats: survival, morbidity, histopathology, particle distribution and
  excretion studies (part iv).
\newblock {\em International journal of nanomedicine}, 14:1163, 2019.

\bibitem{specht2004ordered}
Christian~G Specht, Oliver~A Williams, Richard~B Jackman, and Ralf Schoepfer.
\newblock Ordered growth of neurons on diamond.
\newblock {\em Biomaterials}, 25(18):4073--4078, 2004.

\bibitem{schirhagl2014nitrogen}
Romana Schirhagl, Kevin Chang, Michael Loretz, and Christian~L Degen.
\newblock Nitrogen-vacancy centers in diamond: nanoscale sensors for physics
  and biology.
\newblock {\em Annual review of physical chemistry}, 65:83--105, 2014.

\bibitem{kucsko2013nanometre}
Georg Kucsko, Peter~C Maurer, Norman~Ying Yao, MICHAEL Kubo, Hyun~Jong Noh,
  Po~Kam Lo, Hongkun Park, and Mikhail~D Lukin.
\newblock Nanometre-scale thermometry in a living cell.
\newblock {\em Nature}, 500(7460):54--58, 2013.

\bibitem{jeske2016laser}
Jan Jeske, Jared~H Cole, and Andrew~D Greentree.
\newblock Laser threshold magnetometry.
\newblock {\em New Journal of Physics}, 18(1):013015, 2016.

\bibitem{doherty2013nitrogen}
Marcus~W Doherty, Neil~B Manson, Paul Delaney, Fedor Jelezko, J{\"o}rg
  Wrachtrup, and Lloyd~CL Hollenberg.
\newblock The nitrogen-vacancy colour centre in diamond.
\newblock {\em Physics Reports}, 528(1):1--45, 2013.

\bibitem{wolf2017diamond}
Thomas Wolf, Philipp Neumann, and J{\"o}rg Wrachtrup.
\newblock Diamond magnetometer, December~26 2017.
\newblock US Patent 9,851,418.

\bibitem{gruber1997scanning}
A~Gruber, A~Dr{\"a}benstedt, C~Tietz, L~Fleury, J~Wrachtrup, and
  C~Von~Borczyskowski.
\newblock Scanning confocal optical microscopy and magnetic resonance on single
  defect centers.
\newblock {\em Science}, 276(5321):2012--2014, 1997.

\bibitem{doherty2012theory}
MW~Doherty, F~Dolde, H~Fedder, Fedor Jelezko, J~Wrachtrup, NB~Manson, and LCL
  Hollenberg.
\newblock Theory of the ground-state spin of the nv- center in diamond.
\newblock {\em Physical Review B}, 85(20):205203, 2012.

\bibitem{balasubramanian2008nanoscale}
Gopalakrishnan Balasubramanian, IY~Chan, Roman Kolesov, Mohannad Al-Hmoud,
  Julia Tisler, Chang Shin, Changdong Kim, Aleksander Wojcik, Philip~R Hemmer,
  Anke Krueger, et~al.
\newblock Nanoscale imaging magnetometry with diamond spins under ambient
  conditions.
\newblock {\em Nature}, 455(7213):648--651, 2008.

\bibitem{pham2011magnetic}
Linh~My Pham, David Le~Sage, Paul~L Stanwix, Tsun~Kwan Yeung, D~Glenn, Alexei
  Trifonov, Paola Cappellaro, Philip~R Hemmer, Mikhail~D Lukin, Hongkun Park,
  et~al.
\newblock Magnetic field imaging with nitrogen-vacancy ensembles.
\newblock {\em New Journal of Physics}, 13(4):045021, 2011.

\bibitem{schloss2018simultaneous}
Jennifer~M Schloss, John~F Barry, Matthew~J Turner, and Ronald~L Walsworth.
\newblock Simultaneous broadband vector magnetometry using solid-state spins.
\newblock {\em Physical Review Applied}, 10(3):034044, 2018.

\bibitem{taylor2008high}
JM~Taylor, P~Cappellaro, L~Childress, L~Jiang, D~Budker, PR~Hemmer, A~Yacoby,
  R~Walsworth, and MD~Lukin.
\newblock High-sensitivity diamond magnetometer with nanoscale resolution.
\newblock {\em Nature Physics}, 4(10):810--816, 2008.

\bibitem{degen2008scanning}
CL~Degen.
\newblock Scanning magnetic field microscope with a diamond single-spin sensor.
\newblock {\em Applied Physics Letters}, 92(24):243111, 2008.

\bibitem{maze2008nanoscale}
Jeronimo~R Maze, Paul~L Stanwix, James~S Hodges, Seungpyo Hong, Jacob~M Taylor,
  Paola Cappellaro, Liang Jiang, MV~Gurudev Dutt, Emre Togan, AS~Zibrov, et~al.
\newblock Nanoscale magnetic sensing with an individual electronic spin in
  diamond.
\newblock {\em Nature}, 455(7213):644--647, 2008.

\bibitem{dolde2011electric}
Florian Dolde, Helmut Fedder, Marcus~W Doherty, Tobias N{\"o}bauer, Florian
  Rempp, Gopalakrishnan Balasubramanian, Thomas Wolf, Friedemann Reinhard,
  Lloyd~CL Hollenberg, Fedor Jelezko, et~al.
\newblock Electric-field sensing using single diamond spins.
\newblock {\em Nature Physics}, 7(6):459--463, 2011.

\bibitem{acosta2010temperature}
Victor~M Acosta, Erik Bauch, Micah~P Ledbetter, Amir Waxman, L-S Bouchard, and
  Dmitry Budker.
\newblock Temperature dependence of the nitrogen-vacancy magnetic resonance in
  diamond.
\newblock {\em Physical review letters}, 104(7):070801, 2010.

\bibitem{guarina2018nanodiamonds}
Laura Guarina, C~Calorio, D~Gavello, E~Moreva, P~Traina, A~Battiato, S~Ditalia
  Tchernij, J~Forneris, M~Gai, F~Picollo, et~al.
\newblock Nanodiamonds-induced effects on neuronal firing of mouse hippocampal
  microcircuits.
\newblock {\em Scientific Reports}, 8(1):1--14, 2018.

\bibitem{le2013optical}
David Le~Sage, Koji Arai, David~R Glenn, Stephen~J DeVience, Linh~M Pham, Lilah
  Rahn-Lee, Mikhail~D Lukin, Amir Yacoby, Arash Komeili, and Ronald~L
  Walsworth.
\newblock Optical magnetic imaging of living cells.
\newblock {\em Nature}, 496(7446):486--489, 2013.

\bibitem{davis2018mapping}
Hunter~C Davis, Pradeep Ramesh, Aadyot Bhatnagar, Audrey Lee-Gosselin, John~F
  Barry, David~R Glenn, Ronald~L Walsworth, and Mikhail~G Shapiro.
\newblock Mapping the microscale origins of magnetic resonance image contrast
  with subcellular diamond magnetometry.
\newblock {\em Nature communications}, 9(1):1--9, 2018.

\bibitem{barry2020sensitivity}
John~F Barry, Jennifer~M Schloss, Erik Bauch, Matthew~J Turner, Connor~A Hart,
  Linh~M Pham, and Ronald~L Walsworth.
\newblock Sensitivity optimization for nv-diamond magnetometry.
\newblock {\em Reviews of Modern Physics}, 92(1):015004, 2020.

\bibitem{simpson2017non}
David~A Simpson, Emma Morrisroe, Julia~M McCoey, Alain~H Lombard, Dulini~C
  Mendis, Francois Treussart, Liam~T Hall, Steven Petrou, and Lloyd~CL
  Hollenberg.
\newblock Non-neurotoxic nanodiamond probes for intraneuronal temperature
  mapping.
\newblock {\em ACS nano}, 11(12):12077--12086, 2017.

\bibitem{ermakova2017thermogenetic}
Yulia~G Ermakova, Aleksandr~A Lanin, Ilya~V Fedotov, Matvey Roshchin, Ilya~V
  Kelmanson, Dmitry Kulik, Yulia~A Bogdanova, Arina~G Shokhina, Dmitry~S Bilan,
  Dmitry~B Staroverov, et~al.
\newblock Thermogenetic neurostimulation with single-cell resolution.
\newblock {\em Nature communications}, 8(1):1--15, 2017.

\bibitem{fujiwara2020realtime}
Masazumi Fujiwara, Simo Sun, Alexander Dohms, Yushi Nishimura, Ken Suto, Yuka
  Takezawa, Keisuke Oshimi, Li~Zhao, Nikola Sadzak, Yumi Umehara, Yoshio Teki,
  Naoki Komatsu, Oliver Benson, Yutaka Shikano, and Eriko Kage-Nakadai.
\newblock Real-time nanodiamond thermometry probing in-vivo thermogenic
  responses, 2020.

\bibitem{loubser1978electron}
JoHoNo Loubser and JoA van Wyk.
\newblock Electron spin resonance in the study of diamond.
\newblock {\em Reports on Progress in Physics}, 41(8):1201, 1978.

\bibitem{bauch2018ultralong}
Erik Bauch, Connor~A Hart, Jennifer~M Schloss, Matthew~J Turner, John~F Barry,
  Pauli Kehayias, Swati Singh, and Ronald~L Walsworth.
\newblock Ultralong dephasing times in solid-state spin ensembles via quantum
  control.
\newblock {\em Physical Review X}, 8(3):031025, 2018.

\bibitem{shin2014optically}
Chang~S Shin, Mark~C Butler, Hai-Jing Wang, Claudia~E Avalos, Scott~J Seltzer,
  Ren-Bao Liu, Alexander Pines, and Vikram~S Bajaj.
\newblock Optically detected nuclear quadrupolar interaction of n 14 in
  nitrogen-vacancy centers in diamond.
\newblock {\em Physical Review B}, 89(20):205202, 2014.

\bibitem{felton2009hyperfine}
S~Felton, AM~Edmonds, ME~Newton, PM~Martineau, D~Fisher, DJ~Twitchen, and
  JM~Baker.
\newblock Hyperfine interaction in the ground state of the negatively charged
  nitrogen vacancy center in diamond.
\newblock {\em Physical Review B}, 79(7):075203, 2009.

\bibitem{he1993paramagnetic}
Xing-Fei He, Neil~B Manson, and Peter~TH Fisk.
\newblock Paramagnetic resonance of photoexcited n-v defects in diamond. ii.
  hyperfine interaction with the n 14 nucleus.
\newblock {\em Physical Review B}, 47(14):8816, 1993.

\bibitem{acosta2009diamonds}
Victor~M Acosta, Erik Bauch, Micah~P Ledbetter, Charles Santori, K-MC Fu,
  Paul~E Barclay, Raymond~G Beausoleil, H{\'e}lo{\"\i}se Linget, Jean~Francois
  Roch, Francois Treussart, et~al.
\newblock Diamonds with a high density of nitrogen-vacancy centers for
  magnetometry applications.
\newblock {\em Physical Review B}, 80(11):115202, 2009.

\bibitem{moreva2020biosensing}
Ekaterina Moreva.
\newblock The biosensing with nv centers in diamond: Related challenges.
\newblock {\em International Journal of Quantum Information}, 18(01):1941023,
  2020.

\bibitem{sreenivasan2013luminescent}
Varun~KA Sreenivasan, Andrei~V Zvyagin, and Ewa~M Goldys.
\newblock Luminescent nanoparticles and their applications in the life
  sciences.
\newblock {\em Journal of Physics: Condensed Matter}, 25(19):194101, 2013.

\bibitem{van1990electric}
Eric Van~Oort and Max Glasbeek.
\newblock Electric-field-induced modulation of spin echoes of nv centers in
  diamond.
\newblock {\em Chemical Physics Letters}, 168(6):529--532, 1990.

\bibitem{chen2011temperature}
X-D Chen, C-H Dong, F-W Sun, C-L Zou, J-M Cui, Z-F Han, and G-C Guo.
\newblock Temperature dependent energy level shifts of nitrogen-vacancy centers
  in diamond.
\newblock {\em Applied Physics Letters}, 99(16):161903, 2011.

\bibitem{tzeng2015time}
Yan-Kai Tzeng, Pei-Chang Tsai, Hsiou-Yuan Liu, Oliver~Y Chen, Hsiang Hsu,
  Fu-Goul Yee, Ming-Shien Chang, and Huan-Cheng Chang.
\newblock Time-resolved luminescence nanothermometry with nitrogen-vacancy
  centers in nanodiamonds.
\newblock {\em Nano letters}, 15(6):3945--3952, 2015.

\bibitem{PhysRevApplied.13.054057}
E.~Moreva, E.~Bernardi, P.~Traina, A.~Sosso, S.~Ditalia Tchernij, J.~Forneris,
  F.~Picollo, G.~Brida, \ifmmode \check{Z}\else~\v{Z}\fi{}.
  Pastuovi\ifmmode~\acute{c}\else \'{c}\fi{}, I.~P. Degiovanni, P.~Olivero, and
  M.~Genovese.
\newblock Practical applications of quantum sensing: A simple method to enhance
  the sensitivity of nitrogen-vacancy-based temperature sensors.
\newblock {\em Phys. Rev. Applied}, 13:054057, May 2020.

\bibitem{moreva2019practical}
E~Moreva, E~Bernardi, P~Traina, A~Sosso, S~Ditalia Tchernij, J~Forneris,
  F~Picollo, G~Brida, Z~Pastuovic, IP~Degiovanni, et~al.
\newblock Practical applications of quantum sensing: a simple method to enhance
  sensitivity of nitrogen-vacancy-based temperature sensors.
\newblock {\em arXiv preprint arXiv:1912.10887}, 2019.

\bibitem{neher1992patch}
Erwin Neher and Bert Sakmann.
\newblock The patch clamp technique.
\newblock {\em Scientific American}, 266(3):44--51, 1992.

\bibitem{stett2003biological}
Alfred Stett, Ulrich Egert, Elke Guenther, Frank Hofmann, Thomas Meyer,
  Wilfried Nisch, and Hugo Haemmerle.
\newblock Biological application of microelectrode arrays in drug discovery and
  basic research.
\newblock {\em Analytical and bioanalytical chemistry}, 377(3):486--495, 2003.

\bibitem{jalil2017sensing}
Jubayer Jalil, Yong Zhu, Chandima Ekanayake, and Yong Ruan.
\newblock Sensing of single electrons using micro and nano technologies: a
  review.
\newblock {\em Nanotechnology}, 28(14):142002, 2017.

\bibitem{takei2014nanoparticle}
Yoshiaki Takei, Satoshi Arai, Atsushi Murata, Masao Takabayashi, Kotaro Oyama,
  Shin’ichi Ishiwata, Shinji Takeoka, and Madoka Suzuki.
\newblock A nanoparticle-based ratiometric and self-calibrated fluorescent
  thermometer for single living cells.
\newblock {\em ACS nano}, 8(1):198--206, 2014.

\bibitem{chen1995hydrodynamic}
Duan~P Chen, Robert~S Eisenberg, Joseph~W Jerome, and Chi-Wang Shu.
\newblock Hydrodynamic model of temperature change in open ionic channels.
\newblock {\em Biophysical Journal}, 69(6):2304, 1995.

\bibitem{el2015mechanical}
Ahmed El~Hady and Benjamin~B Machta.
\newblock Mechanical surface waves accompany action potential propagation.
\newblock {\em Nature communications}, 6:6697, 2015.

\bibitem{guatteo2005temperature}
Ezia Guatteo, Kenny~KH Chung, Tharushini~K Bowala, Giorgio Bernardi, Nicola~B
  Mercuri, and Janusz Lipski.
\newblock Temperature sensitivity of dopaminergic neurons of the substantia
  nigra pars compacta: involvement of transient receptor potential channels.
\newblock {\em Journal of neurophysiology}, 94(5):3069--3080, 2005.

\bibitem{donner2012mapping}
Jon~S Donner, Sebastian~A Thompson, Mark~P Kreuzer, Guillaume Baffou, and
  Romain Quidant.
\newblock Mapping intracellular temperature using green fluorescent protein.
\newblock {\em Nano letters}, 12(4):2107--2111, 2012.

\bibitem{yang2011quantum}
Jui-Ming Yang, Haw Yang, and Liwei Lin.
\newblock Quantum dot nano thermometers reveal heterogeneous local
  thermogenesis in living cells.
\newblock {\em ACS nano}, 5(6):5067--5071, 2011.

\bibitem{kim2006micro}
Soo~Ho Kim, Jermim Noh, Min~Ku Jeon, Ki~Woong Kim, Luke~P Lee, and Seong~Ihl
  Woo.
\newblock Micro-raman thermometry for measuring the temperature distribution
  inside the microchannel of a polymerase chain reaction chip.
\newblock {\em Journal of Micromechanics and Microengineering}, 16(3):526,
  2006.

\bibitem{vetrone2010temperature}
Fiorenzo Vetrone, Rafik Naccache, Alicia Zamarron, Angeles Juarranz de~la
  Fuente, Francisco Sanz-Rodriguez, Laura Martinez~Maestro, Emma
  Martin~Rodriguez, Daniel Jaque, Jose Garciia~Sole, and John~A Capobianco.
\newblock Temperature sensing using fluorescent nanothermometers.
\newblock {\em ACS nano}, 4(6):3254--3258, 2010.

\bibitem{maestro2014quantum}
Laura~Martinez Maestro, Qiming Zhang, Xiangping Li, Daniel Jaque, and Min Gu.
\newblock Quantum-dot based nanothermometry in optical plasmonic recording
  media.
\newblock {\em Applied Physics Letters}, 105(18):181110, 2014.

\bibitem{jensen2012use}
Ellen~C Jensen.
\newblock Use of fluorescent probes: their effect on cell biology and
  limitations.
\newblock {\em The Anatomical Record: Advances in Integrative Anatomy and
  Evolutionary Biology}, 295(12):2031--2036, 2012.

\bibitem{bernas2005loss}
Tytus Bernas, Bartlomiej~P Rajwa, Elikplimki~K Asem, and Joseph~Paul Robinson.
\newblock Loss of image quality in photobleaching during microscopic imaging of
  fluorescent probes bound to chromatin.
\newblock {\em Journal of biomedical optics}, 10(6):064015, 2005.

\bibitem{barry2016optical}
John~F Barry, Matthew~J Turner, Jennifer~M Schloss, David~R Glenn, Yuyu Song,
  Mikhail~D Lukin, Hongkun Park, and Ronald~L Walsworth.
\newblock Optical magnetic detection of single-neuron action potentials using
  quantum defects in diamond.
\newblock {\em Proceedings of the National Academy of Sciences},
  113(49):14133--14138, 2016.

\bibitem{toyli2012measurement}
DM~Toyli, DJ~Christle, A~Alkauskas, BB~Buckley, CG~Van~de Walle, and
  DD~Awschalom.
\newblock Measurement and control of single nitrogen-vacancy center spins above
  600 k.
\newblock {\em Physical Review X}, 2(3):031001, 2012.

\bibitem{plakhotnik2014all}
Taras Plakhotnik, Marcus~W Doherty, Jared~H Cole, Robert Chapman, and Neil~B
  Manson.
\newblock All-optical thermometry and thermal properties of the optically
  detected spin resonances of the nv--center in nanodiamond.
\newblock {\em Nano letters}, 14(9):4989--4996, 2014.

\bibitem{muller2015high}
Jan M{\"u}ller, Marco Ballini, Paolo Livi, Yihui Chen, Milos Radivojevic, Amir
  Shadmani, Vijay Viswam, Ian~L Jones, Michele Fiscella, Roland Diggelmann,
  et~al.
\newblock High-resolution cmos mea platform to study neurons at subcellular,
  cellular, and network levels.
\newblock {\em Lab on a Chip}, 15(13):2767--2780, 2015.

\bibitem{bertotti2014cmos}
Gabriel Bertotti, Dmytro Velychko, Norman Dodel, Stefan Keil, Dirk Wolansky,
  Bernd Tillak, Matthias Schreiter, Andreas Grall, Peter Jesinger, Sebastian
  R{\"o}hler, et~al.
\newblock A cmos-based sensor array for in-vitro neural tissue interfacing with
  4225 recording sites and 1024 stimulation sites.
\newblock In {\em 2014 IEEE Biomedical Circuits and Systems Conference (BioCAS)
  Proceedings}, pages 304--307. IEEE, 2014.

\bibitem{bakkum2013tracking}
Douglas~J Bakkum, Urs Frey, Milos Radivojevic, Thomas~L Russell, Jan
  M{\"u}ller, Michele Fiscella, Hirokazu Takahashi, and Andreas Hierlemann.
\newblock Tracking axonal action potential propagation on a high-density
  microelectrode array across hundreds of sites.
\newblock {\em Nature communications}, 4(1):1--12, 2013.

\bibitem{buzsaki2015tools}
Gy{\"o}rgy Buzs{\'a}ki, Eran Stark, Antal Ber{\'e}nyi, Dion Khodagholy, Daryl~R
  Kipke, Euisik Yoon, and Kensall~D Wise.
\newblock Tools for probing local circuits: high-density silicon probes
  combined with optogenetics.
\newblock {\em Neuron}, 86(1):92--105, 2015.

\bibitem{radivojevic2016electrical}
Milos Radivojevic, David J{\"a}ckel, Michael Altermatt, Jan M{\"u}ller, Vijay
  Viswam, Andreas Hierlemann, and Douglas~J Bakkum.
\newblock Electrical identification and selective microstimulation of neuronal
  compartments based on features of extracellular action potentials.
\newblock {\em Scientific reports}, 6:31332, 2016.

\bibitem{tomagra2019micro}
Giulia Tomagra, Pietro Apr{\`a}, Alfio Battiato, Cecilia~Colla Ruvolo, Alberto
  Pasquarelli, Andrea Marcantoni, Emilio Carbone, Valentina Carabelli, Paolo
  Olivero, and Federico Picollo.
\newblock Micro graphite-patterned diamond sensors: Towards the simultaneous in
  vitro detection of molecular release and action potentials generation from
  excitable cells.
\newblock {\em Carbon}, 152:424--433, 2019.

\bibitem{rasband2001developmental}
Matthew~N Rasband and James~S Trimmer.
\newblock Developmental clustering of ion channels at and near the node of
  ranvier.
\newblock {\em Developmental biology}, 236(1):5--16, 2001.

\bibitem{sato2019stochastic}
Daisuke Sato, Gonzalo Hern{\'a}ndez-Hern{\'a}ndez, Collin Matsumoto, Sendoa
  Tajada, Claudia~M Moreno, Rose~E Dixon, Samantha O’Dwyer, Manuel~F Navedo,
  James~S Trimmer, Colleen~E Clancy, et~al.
\newblock A stochastic model of ion channel cluster formation in the plasma
  membrane.
\newblock {\em Journal of General Physiology}, 151(9):1116--1134, 2019.

\bibitem{zhang2017anchored}
Bokai Zhang, Xi~Feng, Hang Yin, Zhenpeng Ge, Yanhuan Wang, Zhiqin Chu, Helena
  Raabova, Jan Vavra, Petr Cigler, Renbao Liu, et~al.
\newblock Anchored but not internalized: shape dependent endocytosis of
  nanodiamond.
\newblock {\em Scientific reports}, 7:46462, 2017.

\bibitem{rendler2017optical}
Torsten Rendler, Jitka Neburkova, Ondrej Zemek, Jan Kotek, Andrea Zappe, Zhiqin
  Chu, Petr Cigler, and J{\"o}rg Wrachtrup.
\newblock Optical imaging of localized chemical events using programmable
  diamond quantum nanosensors.
\newblock {\em Nature communications}, 8(1):1--9, 2017.

\bibitem{isakovic2018modeling}
Jasmina Isakovic, Ian Dobbs-Dixon, Dipesh Chaudhury, and Dinko Mitrecic.
\newblock Modeling of inhomogeneous electromagnetic fields in the nervous
  system: a novel paradigm in understanding cell interactions, disease etiology
  and therapy.
\newblock {\em Scientific reports}, 8(1):1--20, 2018.

\bibitem{karadas2018feasibility}
M{\"u}rsel Karadas, Adam~M Wojciechowski, Alexander Huck, Nils~Ole Dalby,
  Ulrik~Lund Andersen, and Axel Thielscher.
\newblock Feasibility and resolution limits of opto-magnetic imaging of neural
  network activity in brain slices using color centers in diamond.
\newblock {\em Scientific reports}, 8(1):1--14, 2018.

\bibitem{hall2012high}
LT~Hall, GCG Beart, EA~Thomas, DA~Simpson, LP~McGuinness, JH~Cole, JH~Manton,
  RE~Scholten, Fedor Jelezko, J{\"o}rg Wrachtrup, et~al.
\newblock High spatial and temporal resolution wide-field imaging of neuron
  activity using quantum nv-diamond.
\newblock {\em Scientific reports}, 2:401, 2012.

\bibitem{hamalainen1993magnetoencephalography}
Matti H{\"a}m{\"a}l{\"a}inen, Riitta Hari, Risto~J Ilmoniemi, Jukka Knuutila,
  and Olli~V Lounasmaa.
\newblock Magnetoencephalography—theory, instrumentation, and applications to
  noninvasive studies of the working human brain.
\newblock {\em Reviews of modern Physics}, 65(2):413, 1993.

\bibitem{schoenfeld2011real}
Rolf~Simon Schoenfeld and Wolfgang Harneit.
\newblock Real time magnetic field sensing and imaging using a single spin in
  diamond.
\newblock {\em Physical review letters}, 106(3):030802, 2011.

\bibitem{budker2007optical}
Dmitry Budker and Michael Romalis.
\newblock Optical magnetometry.
\newblock {\em Nature physics}, 3(4):227--234, 2007.

\bibitem{sarvas1987basic}
Jukka Sarvas.
\newblock Basic mathematical and electromagnetic concepts of the biomagnetic
  inverse problem.
\newblock {\em Physics in Medicine \& Biology}, 32(1):11, 1987.

\bibitem{michel2004eeg}
Christoph~M Michel, Micah~M Murray, G{\"o}ran Lantz, Sara Gonzalez, Laurent
  Spinelli, and Rolando~Grave de~Peralta.
\newblock Eeg source imaging.
\newblock {\em Clinical neurophysiology}, 115(10):2195--2222, 2004.

\bibitem{trayanova1993response}
Natalia~A Trayanova, Bradley~J Roth, and Lisa~J Malden.
\newblock The response of a spherical heart to a uniform electric field: a
  bidomain analysis of cardiac stimulation.
\newblock {\em IEEE transactions on biomedical engineering}, 40(9):899--908,
  1993.

\bibitem{xu2017magnetic}
Dan Xu and Bradley~J Roth.
\newblock The magnetic field produced by the heart and its influence on mri.
\newblock {\em Mathematical Problems in Engineering}, 2017, 2017.

\bibitem{murdick2004comparative}
Ryan~A Murdick and BJ~Roth.
\newblock A comparative model of two mechanisms from which a magnetic field
  arises in the heart.
\newblock {\em Journal of applied physics}, 95(9):5116--5122, 2004.

\bibitem{roth1991comparison}
Bradley~J Roth.
\newblock A comparison of two boundary conditions used with the bidomain model
  of cardiac tissue.
\newblock {\em Annals of biomedical engineering}, 19(6):669--678, 1991.

\bibitem{krassowska1994effective}
Wanda Krassowska and John~C Neu.
\newblock Effective boundary conditions for syncytial tissues.
\newblock {\em IEEE transactions on biomedical engineering}, 41(2):143--150,
  1994.

\bibitem{roth1997electrical}
Bradley~J Roth.
\newblock Electrical conductivity values used with the bidomain model of
  cardiac tissue.
\newblock {\em IEEE Transactions on Biomedical Engineering}, 44(4):326--328,
  1997.

\bibitem{mcbride2010measurements}
Krista~Kay McBride, Bradley~J Roth, VY~Sidorov, John~P Wikswo, and Franz~J
  Baudenbacher.
\newblock Measurements of transmembrane potential and magnetic field at the
  apex of the heart.
\newblock {\em Biophysical journal}, 99(10):3113--3118, 2010.

\bibitem{holzer2004high}
Jenny~R Holzer, Luis~E Fong, Veniamin~Y Sidorov, John~P Wikswo~Jr, and Franz
  Baudenbacher.
\newblock High resolution magnetic images of planar wave fronts reveal bidomain
  properties of cardiac tissue.
\newblock {\em Biophysical journal}, 87(6):4326--4332, 2004.

\bibitem{proksch1995magnetic}
Roger~B Proksch, TE~Sch{\"a}ffer, BM~Moskowitz, ED~Dahlberg, Dennis~A
  Bazylinski, and Richard~B Frankel.
\newblock Magnetic force microscopy of the submicron magnetic assembly in a
  magnetotactic bacterium.
\newblock {\em Applied Physics Letters}, 66(19):2582--2584, 1995.

\bibitem{qian2011magnetic}
Lisa Qian, Beena Kalisky, Amanda Hamilton, Bo~Dwyer, AC~Matin, and Kathryn
  Moler.
\newblock Magnetic characterization of individual magnetotactic bacteria.
\newblock In {\em APS Meeting Abstracts}, 2011.

\bibitem{lam2010characterizing}
Karen~P Lam, Adam~P Hitchcock, Martin Obst, John~R Lawrence, George~DW
  Swerhone, Gary~G Leppard, Tolek Tyliszczak, Chithra Karunakaran, Jian Wang,
  Konstantin Kaznatcheev, et~al.
\newblock Characterizing magnetism of individual magnetosomes by x-ray magnetic
  circular dichroism in a scanning transmission x-ray microscope.
\newblock {\em Chemical Geology}, 270(1-4):110--116, 2010.

\bibitem{dunin1998magnetic}
Rafal~E Dunin-Borkowski, Martha~R McCartney, Richard~B Frankel, Dennis~A
  Bazylinski, Mihaly Posfai, and Peter~R Buseck.
\newblock Magnetic microstructure of magnetotactic bacteria by electron
  holography.
\newblock {\em Science}, 282(5395):1868--1870, 1998.

\bibitem{komeili2012molecular}
Arash Komeili.
\newblock Molecular mechanisms of compartmentalization and biomineralization in
  magnetotactic bacteria.
\newblock {\em FEMS microbiology reviews}, 36(1):232--255, 2012.

\bibitem{faivre2008magnetotactic}
Damien Faivre and Dirk Schuler.
\newblock Magnetotactic bacteria and magnetosomes.
\newblock {\em Chemical Reviews}, 108(11):4875--4898, 2008.

\bibitem{krichevsky2007trapping}
A~Krichevsky, MJ~Smith, LJ~Whitman, MB~Johnson, TW~Clinton, LL~Perry,
  BM~Applegate, K~O’Connor, and LN~Csonka.
\newblock Trapping motile magnetotactic bacteria with a magnetic recording
  head.
\newblock {\em Journal of applied physics}, 101(1):014701, 2007.

\bibitem{moskowitz1993rock}
Bruce~M Moskowitz, Richard~B Frankel, and Dennis~A Bazylinski.
\newblock Rock magnetic criteria for the detection of biogenic magnetite.
\newblock {\em Earth and Planetary Science Letters}, 120(3-4):283--300, 1993.

\bibitem{posfai2009magnetic}
Mih{\'a}ly P{\'o}sfai and Rafal~E Dunin-Borkowski.
\newblock Magnetic nanocrystals in organisms.
\newblock {\em Elements}, 5(4):235--240, 2009.

\bibitem{eder2012magnetic}
Stephan~HK Eder, Herv{\'e} Cadiou, Airina Muhamad, Peter~A McNaughton, Joseph~L
  Kirschvink, and Michael Winklhofer.
\newblock Magnetic characterization of isolated candidate vertebrate
  magnetoreceptor cells.
\newblock {\em Proceedings of the National Academy of Sciences},
  109(30):12022--12027, 2012.

\bibitem{ghugre2011relaxivity}
Nilesh~R Ghugre and John~C Wood.
\newblock Relaxivity-iron calibration in hepatic iron overload: probing
  underlying biophysical mechanisms using a monte carlo model.
\newblock {\em Magnetic resonance in medicine}, 65(3):837--847, 2011.

\bibitem{nicol1948giant}
JAC Nicol.
\newblock The giant nerve-fibres in the central nervous system of myxicola
  (polychaeta, sabellidae).
\newblock {\em Quarterly Journal of Microscopical Science}, 3(5):1--45, 1948.

\bibitem{song2013analysis}
Yuyu Song and Scott~T Brady.
\newblock Analysis of microtubules in isolated axoplasm from the squid giant
  axon.
\newblock In {\em Methods in cell biology}, volume 115, pages 125--137.
  Elsevier, 2013.

\bibitem{swinney1980calculation}
KR~Swinney and JP~Wikswo~Jr.
\newblock A calculation of the magnetic field of a nerve action potential.
\newblock {\em Biophysical journal}, 32(2):719--731, 1980.

\bibitem{roth1985magnetic}
Bradley~J Roth and John~P Wikswo~Jr.
\newblock The magnetic field of a single axon. a comparison of theory and
  experiment.
\newblock {\em Biophysical journal}, 48(1):93--109, 1985.

\bibitem{wikswo1988magnetic}
John~P Wikswo~Jr and Bradley~J Roth.
\newblock Magnetic determination of the spatial extent of a single cortical
  current source: A theoretical analysis.
\newblock {\em Electroencephalography and clinical neurophysiology},
  69(3):266--276, 1988.

\bibitem{markham2011cvd}
ML~Markham, JM~Dodson, GA~Scarsbrook, DJ~Twitchen, G~Balasubramanian,
  F~Jelezko, and J~Wrachtrup.
\newblock Cvd diamond for spintronics.
\newblock {\em Diamond and Related Materials}, 20(2):134--139, 2011.

\bibitem{narayan2017novel}
Jagdish Narayan and Anagh Bhaumik.
\newblock Novel synthesis and properties of pure and nv-doped nanodiamonds and
  other nanostructures.
\newblock {\em Materials Research Letters}, 5(4):242--250, 2017.

\bibitem{mochalin2012properties}
Vadym~N Mochalin, Olga Shenderova, Dean Ho, and Yury Gogotsi.
\newblock The properties and applications of nanodiamonds.
\newblock {\em Nature nanotechnology}, 7(1):11, 2012.

\bibitem{schrand2007diamond}
Amanda~M Schrand, Houjin Huang, Cataleya Carlson, John~J Schlager, Eiji
  {\=O}sawa, Saber~M Hussain, and Liming Dai.
\newblock Are diamond nanoparticles cytotoxic?
\newblock {\em The journal of physical chemistry B}, 111(1):2--7, 2007.

\bibitem{neugart2007dynamics}
Felix Neugart, Andrea Zappe, Fedor Jelezko, C~Tietz, Jean~Paul Boudou, Anke
  Krueger, and J{\"o}rg Wrachtrup.
\newblock Dynamics of diamond nanoparticles in solution and cells.
\newblock {\em Nano letters}, 7(12):3588--3591, 2007.

\bibitem{mcguinness2011quantum}
Liam~P McGuinness, Yuling Yan, Alastair Stacey, David~A Simpson, Liam~T Hall,
  Dougal Maclaurin, Steven Prawer, P~Mulvaney, J~Wrachtrup, F~Caruso, et~al.
\newblock Quantum measurement and orientation tracking of fluorescent
  nanodiamonds inside living cells.
\newblock {\em Nature nanotechnology}, 6(6):358, 2011.

\bibitem{hsiao2016fluorescent}
Wesley Wei-Wen Hsiao, Yuen~Yung Hui, Pei-Chang Tsai, and Huan-Cheng Chang.
\newblock Fluorescent nanodiamond: a versatile tool for long-term cell
  tracking, super-resolution imaging, and nanoscale temperature sensing.
\newblock {\em Accounts of chemical research}, 49(3):400--407, 2016.

\bibitem{yuan2010pulmonary}
Yuan Yuan, Xiang Wang, Guang Jia, Jia-Hui Liu, Tiancheng Wang, Yiqun Gu,
  Sheng-Tao Yang, Sen Zhen, Haifang Wang, and Yuanfang Liu.
\newblock Pulmonary toxicity and translocation of nanodiamonds in mice.
\newblock {\em Diamond and Related Materials}, 19(4):291--299, 2010.

\bibitem{mohan2010vivo}
Nitin Mohan, Chao-Sheng Chen, Hsiao-Han Hsieh, Yi-Chun Wu, and Huan-Cheng
  Chang.
\newblock In vivo imaging and toxicity assessments of fluorescent nanodiamonds
  in caenorhabditis elegans.
\newblock {\em Nano letters}, 10(9):3692--3699, 2010.

\bibitem{chow2011nanodiamond}
Edward~K Chow, Xue-Qing Zhang, Mark Chen, Robert Lam, Erik Robinson, Houjin
  Huang, Daniel Schaffer, Eiji Osawa, Andrei Goga, and Dean Ho.
\newblock Nanodiamond therapeutic delivery agents mediate enhanced
  chemoresistant tumor treatment.
\newblock {\em Science translational medicine}, 3(73):73ra21--73ra21, 2011.

\bibitem{lima2005remote}
Susana~Q Lima and Gero Miesenb{\"o}ck.
\newblock Remote control of behavior through genetically targeted
  photostimulation of neurons.
\newblock {\em Cell}, 121(1):141--152, 2005.

\bibitem{boyden2005millisecond}
Edward~S Boyden, Feng Zhang, Ernst Bamberg, Georg Nagel, and Karl Deisseroth.
\newblock Millisecond-timescale, genetically targeted optical control of neural
  activity.
\newblock {\em Nature neuroscience}, 8(9):1263--1268, 2005.

\bibitem{bernstein2012optogenetics}
Jacob~G Bernstein, Paul~A Garrity, and Edward~S Boyden.
\newblock Optogenetics and thermogenetics: technologies for controlling the
  activity of targeted cells within intact neural circuits.
\newblock {\em Current opinion in neurobiology}, 22(1):61--71, 2012.

\bibitem{bath2014flymad}
Daniel~E Bath, John~R Stowers, Dorothea H{\"o}rmann, Andreas Poehlmann, Barry~J
  Dickson, and Andrew~D Straw.
\newblock Flymad: rapid thermogenetic control of neuronal activity in freely
  walking drosophila.
\newblock {\em Nature methods}, 11(7):756, 2014.

\bibitem{newman1982infrared}
Eric~A Newman and Peter~H Hartline.
\newblock The infrared" vision" of snakes.
\newblock {\em Scientific American}, 246(3):116--127, 1982.

\bibitem{hamada2008internal}
Fumika~N Hamada, Mark Rosenzweig, Kyeongjin Kang, Stefan~R Pulver, Alfredo
  Ghezzi, Timothy~J Jegla, and Paul~A Garrity.
\newblock An internal thermal sensor controlling temperature preference in
  drosophila.
\newblock {\em Nature}, 454(7201):217--220, 2008.

\bibitem{chen2016trp}
Shijia Chen, Cindy~N Chiu, Kimberly~L McArthur, Joseph~R Fetcho, and David~A
  Prober.
\newblock Trp channel mediated neuronal activation and ablation in freely
  behaving zebrafish.
\newblock {\em Nature methods}, 13(2):147--150, 2016.

\bibitem{balasubramanian2009ultralong}
Gopalakrishnan Balasubramanian, Philipp Neumann, Daniel Twitchen, Matthew
  Markham, Roman Kolesov, Norikazu Mizuochi, Junichi Isoya, Jocelyn Achard,
  Johannes Beck, Julia Tissler, et~al.
\newblock Ultralong spin coherence time in isotopically engineered diamond.
\newblock {\em Nature materials}, 8(5):383--387, 2009.

\bibitem{achard2020cvd}
Jocelyn Achard, Vincent Jacques, and Alexandre Tallaire.
\newblock Cvd diamond single crystals with nv centres: a review of material
  synthesis and technology for quantum sensing applications.
\newblock {\em Journal of Physics D: Applied Physics}, 2020.

\bibitem{bernardi2017nanoscale}
Ettore Bernardi, Richard Nelz, Selda Sonusen, and Elke Neu.
\newblock Nanoscale sensing using point defects in single-crystal diamond:
  recent progress on nitrogen vacancy center-based sensors.
\newblock {\em Crystals}, 7(5):124, 2017.

\bibitem{levine2019principles}
Edlyn~V Levine, Matthew~J Turner, Pauli Kehayias, Connor~A Hart, Nicholas
  Langellier, Raisa Trubko, David~R Glenn, Roger~R Fu, and Ronald~L Walsworth.
\newblock Principles and techniques of the quantum diamond microscope.
\newblock {\em Nanophotonics}, 8(11):1945--1973, 2019.

\bibitem{dreau2011avoiding}
A~Dr{\'e}au, M~Lesik, L~Rondin, P~Spinicelli, O~Arcizet, J-F Roch, and
  V~Jacques.
\newblock Avoiding power broadening in optically detected magnetic resonance of
  single nv defects for enhanced dc magnetic field sensitivity.
\newblock {\em Physical Review B}, 84(19):195204, 2011.

\bibitem{shin2012room}
Chang~S Shin, Claudia~E Avalos, Mark~C Butler, David~R Trease, Scott~J Seltzer,
  J~Peter~Mustonen, Daniel~J Kennedy, Victor~M Acosta, Dmitry Budker, Alexander
  Pines, et~al.
\newblock Room-temperature operation of a radiofrequency diamond magnetometer
  near the shot-noise limit.
\newblock {\em Journal of Applied Physics}, 112(12):124519, 2012.

\bibitem{bauch2019decoherence}
Erik Bauch, Swati Singh, Junghyun Lee, Connor~A Hart, Jennifer~M Schloss,
  Matthew~J Turner, John~F Barry, Linh Pham, Nir Bar-Gill, Susanne~F Yelin,
  et~al.
\newblock Decoherence of dipolar spin ensembles in diamond.
\newblock {\em arXiv preprint arXiv:1904.08763}, 2019.

\bibitem{grinolds2013nanoscale}
Michael~Sean Grinolds, Sungkun Hong, Patrick Maletinsky, Lan Luan, Mikhail~D
  Lukin, Ronald~Lee Walsworth, and Amir Yacoby.
\newblock Nanoscale magnetic imaging of a single electron spin under ambient
  conditions.
\newblock {\em Nature Physics}, 9(4):215--219, 2013.

\bibitem{de2010universal}
G~De~Lange, ZH~Wang, D~Riste, VV~Dobrovitski, and R~Hanson.
\newblock Universal dynamical decoupling of a single solid-state spin from a
  spin bath.
\newblock {\em Science}, 330(6000):60--63, 2010.

\bibitem{szankowski2017environmental}
Piotr Sza{\'n}kowski, Guy Ramon, Jan Krzywda, Damian Kwiatkowski, et~al.
\newblock Environmental noise spectroscopy with qubits subjected to dynamical
  decoupling.
\newblock {\em Journal of Physics: Condensed Matter}, 29(33):333001, 2017.

\bibitem{naydenov2011dynamical}
Boris Naydenov, Florian Dolde, Liam~T Hall, Chang Shin, Helmut Fedder, Lloyd~CL
  Hollenberg, Fedor Jelezko, and J{\"o}rg Wrachtrup.
\newblock Dynamical decoupling of a single-electron spin at room temperature.
\newblock {\em Physical Review B}, 83(8):081201, 2011.

\bibitem{knowles2014observing}
Helena~S Knowles, Dhiren~M Kara, and Mete Atat{\"u}re.
\newblock Observing bulk diamond spin coherence in high-purity nanodiamonds.
\newblock {\em Nature materials}, 13(1):21--25, 2014.

\bibitem{cao2020protecting}
Q-Y Cao, P-C Yang, M-S Gong, M~Yu, A~Retzker, Martin~B Plenio, C~M{\"u}ller,
  N~Tomek, B~Naydenov, LP~McGuinness, et~al.
\newblock Protecting quantum spin coherence of nanodiamonds in living cells.
\newblock {\em Physical Review Applied}, 13(2):024021, 2020.

\bibitem{mrozek2015longitudinal}
Mariusz Mr{\'o}zek, Daniel Rudnicki, Pauli Kehayias, Andrey Jarmola, Dmitry
  Budker, and Wojciech Gawlik.
\newblock Longitudinal spin relaxation in nitrogen-vacancy ensembles in
  diamond.
\newblock {\em EPJ Quantum Technology}, 2(1):22, 2015.

\bibitem{radtke2019nanoscale}
Mariusz Radtke, Ettore Bernardi, Abdallah Slablab, Richard Nelz, and Elke Neu.
\newblock Nanoscale sensing based on nitrogen vacancy centers in single crystal
  diamond and nanodiamonds: achievements and challenges.
\newblock {\em Nano Futures}, 3(4):042004, 2019.

\bibitem{bar2012suppression}
Nir Bar-Gill, Linh~My Pham, Chinmay Belthangady, David Le~Sage, Paola
  Cappellaro, JR~Maze, Mikhail~D Lukin, Amir Yacoby, and Ronald Walsworth.
\newblock Suppression of spin-bath dynamics for improved coherence of
  multi-spin-qubit systems.
\newblock {\em Nature communications}, 3(1):1--6, 2012.

\bibitem{zwick2016maximizing}
Analia Zwick, Gonzalo~A {\'A}lvarez, and Gershon Kurizki.
\newblock Maximizing information on the environment by dynamically controlled
  qubit probes.
\newblock {\em Physical Review Applied}, 5(1):014007, 2016.

\bibitem{muller2020noise}
Matthias~M M{\"u}ller, Stefano Gherardini, Nicola Dalla~Pozza, and Filippo
  Caruso.
\newblock Noise sensing via stochastic quantum zeno.
\newblock {\em Physics Letters A}, 384(13):126244, 2020.

\bibitem{arenz2016universal}
Christian Arenz, Daniel Burgarth, Paolo Facchi, Vittorio Giovannetti, Hiromichi
  Nakazato, Saverio Pascazio, and Kazuya Yuasa.
\newblock Universal control induced by noise.
\newblock {\em Physical Review A}, 93(6):062308, 2016.

\bibitem{stanika2012comparative}
Ruslan~I Stanika, Idalis Villanueva, Galina Kazanina, S~Brian Andrews, and
  Natalia~B Pivovarova.
\newblock Comparative impact of voltage-gated calcium channels and nmda
  receptors on mitochondria-mediated neuronal injury.
\newblock {\em Journal of Neuroscience}, 32(19):6642--6650, 2012.

\bibitem{humeau2018calcium}
Juliette Humeau, Jos{\'e}~Manuel Bravo-San~Pedro, Ilio Vitale, Lucia Nu{\~n}ez,
  Carlos Villalobos, Guido Kroemer, and Laura Senovilla.
\newblock Calcium signaling and cell cycle: progression or death.
\newblock {\em Cell calcium}, 70:3--15, 2018.

\end{thebibliography}

\end{document}